\documentclass{article}
\usepackage{arxiv}

\usepackage[utf8]{inputenc} 
\usepackage[T1]{fontenc}    
\usepackage[hidelinks]{hyperref}     
\usepackage{url}            
\usepackage{booktabs}       
\usepackage{amsfonts}       
\usepackage{nicefrac}       
\usepackage{microtype}      
\usepackage{lipsum}
\usepackage{bm}
\usepackage{graphicx}
\usepackage{subcaption}
\usepackage{amsmath}
\usepackage{placeins}
\usepackage{siunitx}
\graphicspath{ {./images/} }
\newcolumntype{C}[1]{>{\centering\arraybackslash}p{#1}}
\allowdisplaybreaks[4]

\title{Bimaterial Eshelby's inclusion problem for polyhedra}

\author{
  Chunlin Wu \\
  Shanghai Institute of Applied Mathematics \\and Mechanics
  ,Shanghai University\\
  Shanghai, 200044 \\
  \texttt{chunlinwu@shu.edu.cn} \\
\And
  Huiming Yin\thanks{Corresponding Author} \\
  Department of Civil Engineering and \\Engineering Mechanics,
  Columbia University\\
  New York, NY, 10027 \\
  \texttt{yin@civil.columbia.edu} \\
}

\begin{document}
\maketitle
\begin{abstract}
This paper presents the closed-form Eshelby's tensor for an arbitrarily oriented polyhedral inclusion in a bimaterial domain under general uniform eigenstrain. Existing bimaterial solutions are mainly restricted to special inclusion shapes or dilatational eigenstrains, because the bimaterial Green's function contains two Boussinesq's displacement potentials in addition to the harmonic and biharmonic potentials in Kelvin's solution. This paper derives the missing domain integrals of the two Boussinesq's potentials by reducing the volume integrals to surface and elementary line integrals. The formulae provide the complete elastic and thermoelastic bimaterial Eshelby's tensors, which are verified against analytical solutions for spherical and cuboidal inclusions parallel to the bimaterial interface, and finite element results of an inclined cuboid. Singularity analysis demonstrates that the interface-related contribution remains regular when the inclusion is separated from the bimaterial interface, while additional logarithmic singularities arise when an edge or vertex touches the interface without increasing the dominant singularity order. 
\end{abstract}

\keywords{Bimaterial Green's function \and Boussinesq's displacement potentials \and Eshelby's tensor \and Eigenstrain \and Singularity}

\section{Introduction}\label{sec:1 intro} 
Consider an ellipsoidal inhomogeneity embedded in an infinite domain. Eshelby \cite{Eshelby_1957, Eshelby_1959} proposed the equivalent inclusion method (EIM) to replace the inhomogeneity with the matrix material containing uniform eigenstrain. Eshelby's EIM transforms the original boundary value problem into determining eigen-fields through specific equivalent conditions, which lays a solid foundation for broad applications in micromechanics. For instance, EIM has been extended to steady-state heat conduction \cite{hatta1986}, viscoelastic \cite{KazemiLari2021}, magnetoelastic \cite{Yin2006}, thermoelastic analysis \cite{Wu2023}, and some recent progress on harmonic/transient heat transfer \cite{Wu2025, Wu2026}. Eshelby's tensor depends on the inclusion geometry and the elastic constants of the matrix. It establishes the relationship between eigen-fields and their induced local fields, and thus forms the foundation of EIM and homogenization theories. 

When an inclusion is embedded in an infinite domain, the elastic Green's function is composed of harmonic and biharmonic potentials. Thus, Eshelby's tensor can be obtained by superposing the corresponding domain integrals over the inclusion domain. Dyson \cite{Dyson1891} first derived the ellipsoidal integrals of harmonic potential with varying source density, while explicit formulae only exist for spheres and spheroids. Based on Dyson's work, Moschovidis and Mura \cite{Moschovidis1975} derived ellipsoidal integrals with polynomial-form source density, and Jin et al. \cite{Jin2016} derived a complete set of analytical solutions for ellipsoidal inclusion. Since the circular cylindrical inclusion can be considered a special case of the ellipsoidal inclusion (setting the third axis to infinity), Ma and Gao \cite{Ma2009} derived an explicit Eshelby's tensor involving strain-gradient theory. Alternatively, the classical ellipsoidal integrals have been reproduced using the convolution property of the Fourier space \cite{Michelitsch2003}. Although substantial developments have been achieved for ellipsoidal inclusions, the above works do not directly address general non-ellipsoidal or arbitrarily shaped inclusions. 

For non-ellipsoidal inclusions, numerous efforts have studied disturbances induced by eigenstrain, revealing interesting features of Eshelby's tensors, such as spatial variation at interior points and singularities at the vertices of polygonal/polyhedral inclusions. Chiu \cite{Chiu1977} studied the disturbance caused by initial strains in a cuboidal inclusion, and Waldvogel \cite{Waldvogel1979} derived explicit formulae of Newtonian (harmonic) potential for a polyhedral inclusion. To address the difficulty in the definition of integral limits, Rodin \cite{Rodin1996} proposed to construct transformed coordinates for polygons and polyhedra, which consist of the normal vector of the surface and normal/directional vectors of the edge. Using the transformed coordinates, Rodin derived the closed-form Eshelby's tensor for polygonal/polyhedral inclusions. Nozaki and Taya \cite{Nozaki1997, Nozaki2000} followed Mura's method \cite{Mura1987}, and the authors utilized a unit circular/spherical shell to derive closed-form Eshelby's tensors for polygonal/polyhedral convex inclusions. Building on Rodin's work, Gao and colleagues \cite{Gao2012, Liu2013} derived closed-form Eshelby's tensors for polyhedral and polygonal inclusions involving strain-gradient theory, respectively. Subsequently, Rosati's group \cite{Trotta2017, Trotta2018} revisited Rodin's framework and provided solutions that depend only on the vertices of the polygonal/polyhedral inclusion, thereby shortening Rodin's formulae. Recently, following Rodin's framework, we derived Eshelby's tensors for polygonal/polyhedral inclusions, where the eigenstrain exhibits polynomial-form variations \cite{Wu2021_jam_polygonal, Wu2021_polyhedral}. 

However, because the above papers used the infinite-domain Green's function, or Kelvin's solution \cite{thomson1848note}, those Eshelby's tensors, which are domain integrals with the corresponding Green's function, cannot take into account the boundary effects introduced by the bonded bimaterial interface. Instead, the Green's functions for half-space \cite{Mindlin1936, rongved1955force} or bimaterials \cite{Walpole1996, Wu2023} can be used for the domain integrals, but they are more complicated. In addition to the harmonic and biharmonic potentials in Kelvin's solution \cite{Mura1987}, those Green's functions also contain the first and second Boussinesq's displacement potentials \cite{Boussinesq1885}. Therefore, Eshelby's tensors for half-space or bimaterial scenarios require domain integrals of two Boussinesq's displacement potentials. Although these domain integrals can be evaluated numerically, closed-form formulae are fundamentally important because they not only eliminate quadrature errors and retain the exact singular features of the local fields but also provide a robust analytical benchmark for numerical methods. 

For ellipsoidal and spherical inclusions, related works are summarized as follows. Yu and colleagues \cite{Yu1991, Yu1992a, Yu1992b} derived a bimaterial Green's function and provided local elastic and thermoelastic fields caused by a spherical inclusion containing dilatational eigenstrain or prescribed thermal strain. Subsequently, Walpole \cite{Walpole1997} utilized the bimaterial Green's function \cite{Walpole1996} and established a formal representation of the elastic bimaterial Eshelby's tensor. Due to the difficulty of evaluating domain integrals of Boussinesq's displacement potentials, Walpole only demonstrated local elastic fields induced by a spherical inclusion with uniform eigenstrain. The complete explicit bimaterial Eshelby's tensor for spherical inclusions was later derived in our previous work \cite{Wu2023_rspa}.  Following Walpole's work, Li et al. \cite{Li2019} and Lyu et al. \cite{Lyu2022} subsequently derived Eshelby's tensors for an ellipsoidal inclusion with dilatational strain embedded in bimaterial domains under several interfacial conditions. Note that dilatational eigenstrain substantially simplifies the above problems. Because the corresponding eigenstrain is proportional to the identity tensor, Eshelby's tensor is reduced to its trace components. Consequently, the contributions from two Boussinesq's displacement potentials have been simplified as harmonic-potential-related integrals. Hence, although some of the above works solved ellipsoidal inclusions with dilatational eigenstrain, they did not derive the complete set of bimaterial Eshelby's tensors for ellipsoidal inclusions yet, which cannot be applied to general eigenstrain problems. Recently, Wu et al. \cite{Wu2023} derived the bimaterial thermoelastic Green's function and provided explicit bimaterial Eshelby's tensors for spherical inclusions containing uniform, linear, and quadratic eigenstrain.

Regarding non-ellipsoidal inclusions, Chiu \cite{Chiu1978,Chiu1980} first proposed the closed-form Eshelby's tensors when the cuboid is embedded in the half-space and half-plane, respectively. Although Chiu's works are only valid when cuboids are parallel to the surface/line of the half-space/-plane, they paved the way for subsequent EIM-based numerical implementations, see \cite{Liu2012,Wang2016}. However, the limitation on the orientation of cuboidal inclusions prevents its use in general applications. Subsequently, Ru \cite{Ru1999, Ru2001} utilized the technique of conformal mapping, and the author derived two-dimensional Eshelby's tensor for an arbitrarily shaped inclusion embedded in the half-plane and bimaterial domain, respectively. Although conformal mapping is a powerful tool for two-dimensional problems, its direct extension to three-dimensional problems is limited, especially for bimaterial domains. Jiang and Pan \cite{Jiang2004} and Zou and Pan \cite{Zou2012} derived two-dimensional bimaterial Eshelby's tensors for magnetoelastic and multiferroic problems, respectively. Recently, Wu and Yin \cite{Wu2023a} utilized the transformed coordinate \cite{Rodin1996} and derived bimaterial elastic/thermoelastic Eshelby's tensors for polygonal inclusions containing polynomial-form eigen-fields. Kuvshinov \cite{Kuvshinov2008} employed Gauss' theorem to derive closed-form expressions for polyhedral inclusions in a half-space and bimaterial domain containing dilatational eigenstrain under the hydrostatic case. Although the author \cite{Kuvshinov2008} discussed the possible extension to non-hydrostatic inclusions (Section 5 of \cite{Kuvshinov2008}), the complete form of Eshelby's tensor was not explicitly constructed. The domain integrals of Boussinesq's displacement potentials are essential for deriving the complete bimaterial Eshelby's tensor.

This paper aims to derive closed-form bimaterial Eshelby's tensors for arbitrarily shaped polyhedral inclusions with general uniform eigenstrain. The paper is organized as follows. Section 2 introduces the bimaterial Green's function and the components of elastic/thermoelastic Eshelby's tensors, which involve harmonic, biharmonic, and two Boussinesq's displacement potentials. Section 3 derives the closed-form domain integrals of the first and second Boussinesq's displacement potentials. Section 4 verifies Eshelby's tensors against classical results, including spherical and cuboidal inclusions parallel to the bimaterial interface. The stresses of an inclined cuboidal inclusion with anisotropic eigenstrain are validated against the finite element analysis. In addition, the EIM is applied to solve the inclined inhomogeneity, which shows the bimaterial interfacial disturbances. Section 5 conducts singularity analysis of bimaterial components and indicates that the logarithmic singularities can exist when the polyhedral inclusion touches the bimaterial interface. Finally, we provide some concluding remarks. 

\section{Bimaterial Green's functions and Eshelby's tensors}
\begin{figure}
    \centering
    \includegraphics[width=0.7\linewidth]{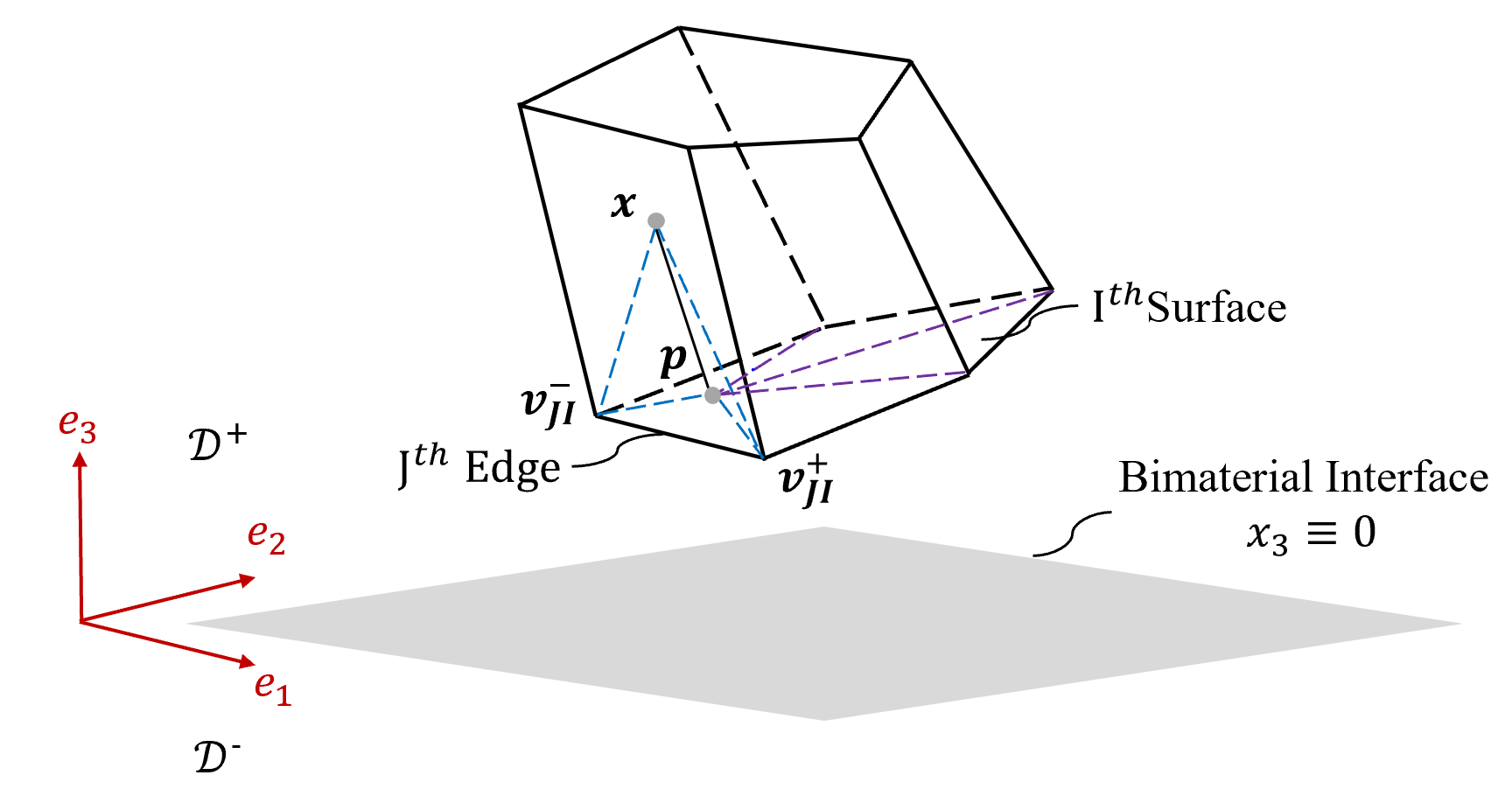}
    \caption{Schematic illustration of a polyhedral inclusion embedded in the upper half-space $\mathcal{D}^+$ of the bimaterial domain $\mathcal{D}$. The bimaterial interface is the $x_{1} - x_{2}$ plane, defined by $x_3 \equiv 0$. For the $J^{th}$ edge of the $I^{th}$ surface, its two vertices are denoted as $\textbf{v}_{JI}^{-}$ and $\textbf{v}_{JI}^+$. }
    \label{fig:fig1}
\end{figure}

Fig. \ref{fig:fig1} shows an infinite bimaterial domain $\mathcal{D}$ composed of two isotropic jointed dissimilar half-spaces, in which the upper and lower phases are denoted as $\mathcal{D}^+$ and $\mathcal{D}^-$, respectively. The bimaterial interface is the $x_1 - x_2$ plane, which is defined by $x_{3} \equiv 0$. In general, the upper $\mathcal{D}^+$ and lower $\mathcal{D}^-$ half-spaces exhibit different thermomechanical properties. Let the superscripts $(.)'$ and $(.)''$ denote properties belonging to the upper and lower phases, respectively. Specifically, (i) thermal conductivity: $K'$, $K''$; (ii) stiffness: $\textbf{C}'$, $\textbf{C}''$, and (iii) thermal modulus $\mathcal{A}'$, $\mathcal{A}''$. For isotropic material, the stiffness tensor $C_{ijkl} = \lambda \delta_{ij} \delta_{kl} + \mu (\delta_{il} \delta_{jk} + \delta_{ik} \delta_{jl})$, and the thermal modulus $\mathcal{A} = (3 \lambda + 2\mu) \gamma$, where $\lambda, \mu$ are Lam\`e constants and $\gamma$ is the thermal expansion coefficient. Without any loss of generality, the polyhedral inclusion is assumed to be located within the upper half-space $\mathcal{D}^+$, which contains prescribed uniformly distributed eigenstrain $\varepsilon_{ij}^*$. The polyhedral region is assumed to exhibit the same material properties as the upper half-space, which will be extended to inhomogeneity problems \cite{Mura1987} by the equivalent inclusion method in the case studies of Section 4.5. Note that this paper considers all components of eigenstrain and does not assume the dilatational case, which is distinct from the existing works \cite{Yu1991,Yu1992a,Yu1992b}. 

\subsection{Elastic bimaterial Green's function}
The elastic fundamental solution relates the unit excitation at the source point $\textbf{x}'$ to the displacement at the field point $\textbf{x}$. Based on Walpole's derivation and our recent work \cite{Wu2023}, the elastic fundamental solution can be written as (assuming the source is located within the upper half-space, $x_{3}' > 0$)

\begin{equation}
G_{ij}(\textbf{x}, \textbf{x}') = \frac{1}{4\pi \mu' }
\begin{cases}
\begin{aligned}
    & \left(\delta_{ij} \phi - \frac{\psi_{,ij}}{4(1-\nu')} \right) + A \overline{\phi} \delta_{ij}
    +  B (\delta_{i3} \delta_{jk} - \delta_{ik} \delta_{j3}) \overline{\alpha}_{,k} \\ & - C x_3 \left[ Q_{J} \overline{\psi}_{,ij3} + 4 (1-\nu') \delta_{j3} \overline{\phi}_{,i} + 2(1-2\nu') \delta_{i3} Q_{J} \overline{\phi}_{,j} - Q_{J} x_3 \overline{\phi}_{,ij} \right] \\ & - D Q_{I} Q_{J} \overline{\psi}_{,ji} - (G + B) Q_{J} \overline{\beta}_{,ij}
\end{aligned} & x_3 \geq 0 \\
\\
\begin{aligned}
    & \left(\delta_{ij} \phi - \frac{\psi_{,ij}}{4(1-\nu')} \right) + A \phi \delta_{ij} +  B (\delta_{i3} \delta_{jk} - \delta_{ik} \delta_{j3}) \alpha_{,k} \\ & - D \psi_{,ij} -  x_3 F \alpha_{,ij} - (G + B) Q_{I} \beta_{,ji} \end{aligned} & x_3 < 0 \\
\end{cases}
\label{eq:elastic_Green}
\end{equation}
where $A \sim G$ are material constants listed as follows \cite{yin2022inclusion}:  

\begin{equation}
    \begin{aligned}
    A = \frac{\mu' - \mu''}{\mu' + \mu''}, \quad B = \frac{2 \mu' (1 - 2 \nu') (\mu' - \mu'')}{(\mu' + \mu'')\left[\mu' + \mu''(3 - 4 \nu')\right]} \\
    C = \frac{\mu' - \mu''}{2(1 - \nu') \left[\mu' + (3 - 4\nu') \mu''\right]}, \quad D = \frac{3 - 4\nu'}{2} C \\
    F = \frac{2 \mu' \left[\mu'(1 - 2\nu'') - \mu'' (1 - 2\nu')\right]}{\left[ \mu' + \mu'' (3 - 4\nu') \right] \left[ \mu'' + \mu' (3 - 4 \nu'') \right]} \\ G = \frac{\mu' \left[\mu''(1 - 2\nu'')(3 - 4\nu') - \mu' (1 - 2\nu') (3 - 4\nu'')\right]}{\left[ \mu' + \mu'' (3 - 4\nu') \right] \left[ \mu'' + \mu' (3 - 4 \nu'') \right]}
    \end{aligned}
    \label{eq:mat_constant}
\end{equation}
and $\phi,  \overline{\phi}, \psi, \overline{\psi}$ are harmonic and biharmonic potentials, and $\alpha, \overline{\alpha}, \beta, \overline{\beta}$ are the first and second Boussinesq's displacement potentials \cite{yin2022inclusion}, which are listed as follows: 
\begin{equation}
    \begin{aligned}
    \phi = \frac{1}{|\textbf{x}' - \textbf{x}|}, \quad \overline{\phi} = \frac{1}{|\overline{\textbf{x}}' - \textbf{x}|}, \quad \psi = |\textbf{x}' - \textbf{x}|, \quad \overline{\psi} = |\overline{\textbf{x}}'-\textbf{x}| \\ 
    \alpha = \ln \left[|\textbf{x}' - \textbf{x}| + x_3' - x_3 \right], \quad \overline{\alpha} = \ln \left[ |\overline{\textbf{x}}' - \textbf{x}| + x_{3}' + x_3 \right]\\ 
    \beta = (x_3' - x_3) \alpha - \psi, \quad \overline{\beta} = (x_3' + x_3) \overline{\alpha} - \overline{\psi}
    \end{aligned}
    \label{eq:harmonic_potential}
\end{equation}
where $\overline{\textbf{x}}' = \{x_{1}', x_{2}', -x_{3}'\}$ is the image source point. Note that the dummy index rule does not apply to the capital index following Mura's notation \cite{Mura1987}, i.e., $Q_{J}$, and $\textbf{Q} = \{1, 1, -1\}$ mirrors the normal coordinate ($x_{3}$) across the fully bonded bimaterial interface.

\subsection{Bimaterial elastic Eshelby's tensors}
Based on Walpole's work \cite{Walpole1997}, the disturbed strain by the eigenstrain can be written as, 
\begin{equation}
    \varepsilon_{ij}' = \int_{\Omega} \frac{C_{mnkl}(\textbf{x}')}{2} \left[G_{im,n'j}(\textbf{x}, \textbf{x}') + G_{jm,n'i}(\textbf{x}, \textbf{x}') \right] \thinspace dV(\textbf{x}') \varepsilon_{kl}^* = S_{ijkl} \varepsilon_{kl}^*
    \label{eq:disturbed_strain}
\end{equation}
where $\varepsilon_{ij}'$ is the disturbed strain, and $S_{ijkl}$ is Eshelby's tensor. Conventional Eshelby's tensor is obtained by the domain integral of the harmonic and biharmonic potentials, $\phi, \psi$ and their imaged cases, 
\begin{equation}
    \Phi = \int_{\Omega} \phi \thinspace dV(\textbf{x}'), \quad \overline{\Phi} = \int_{\Omega} \overline{\phi} \thinspace dV(\textbf{x}'), \quad \Psi = \int_{\Omega} \psi \thinspace dV(\textbf{x}'), \quad \overline{\Psi} = \int_{\Omega} \overline{\psi} \thinspace dV(\textbf{x}')
    \label{eq:sym_harmonic}
\end{equation}
which have been addressed by Rodin $\cite{Rodin1996}$ and Trotta et al. \cite{Trotta2018}. However, the bimaterial Eshelby's tensor includes not only these integrals, but also the domain integrals of two Boussinesq's displacement potentials, i.e., $\alpha, \beta$, and their imaged cases. As previously mentioned, the challenge of domain integrals of Boussinesq's displacement potentials can be avoided in special cases: (i) for dilatational eigenstrain (hydrostatic) scenario, the elastic Green's function can be simplified, and Boussinesq's displacement potentials do not appear as independent ingredients in Eshelby's tensors; (ii) for cuboidal inclusions parallel to the bimaterial interface, since the base vector $\textbf{e}_{3}$ coincides or is perpendicular to the surface normal vectors, the domain integrals can be greatly simplified, which will be explained during the derivation in the following section. For a general uniform eigenstrain or polyhedral inclusions with arbitrary orientations, the domain integrals of Boussinesq's displacement potentials are necessary. Therefore, the following section derives closed-form domain integrals of those missing Boussinesq's displacement potential integrals, 
\begin{equation}
    \Theta = \int_{\Omega} \alpha \thinspace dV(\textbf{x}'), \quad \overline{\Theta} = \int_{\Omega} \overline{\alpha} \thinspace dV(\textbf{x}'), \quad \Lambda = \int_{\Omega} \beta \thinspace dV(\textbf{x}'), \quad \overline{\Lambda} = \int_{\Omega} \overline{\beta} \thinspace dV(\textbf{x}')
    \label{eq:boussinesq}
\end{equation}

Given the above integrals, when the field point is located in the upper phase ($x_{3} > 0$), the same as the source $(x_{3} > 0)$, Eshelby's tensors can be obtained as follows. For clarity, the Eshelby's tensor can be considered as the superposition of the conventional full-space contribution $S^C_{ijkl}$ and the interface-related contribution. The latter contains the harmonic, biharmonic, two Boussinesq's displacement potentials, along with their image counterparts. In particular, this decomposition will be further applied in the singularity analysis in Section 5. 
\begin{align}
    S_{ijkl} = S_{ijkl}^C + \frac{1}{8 \pi \mu'} \Big[ &- A \left[ 2 \lambda' \delta_{kl} Q_{I} \overline{\Phi}_{,ij} + \mu' ( \delta_{ik} Q_{L} \overline{\Phi}_{,lj} + \delta_{il} Q_{K} \overline{\Phi}_{,kj} + \delta_{jk} Q_{L} \overline{\Phi}_{,li} \right. \notag \\ & \left. \quad + \delta_{jl} Q_{K} \overline{\Phi}_{,ki} ) \right] - 2 C \left[ \lambda' (1 - 2\nu') \delta_{kl} (\delta_{j3} \overline{\Phi}_{,i3} + \delta_{i3} \overline{\Phi}_{,j3} + 2 x_{3} \overline{\Phi}_{,ij3}) \right. \notag \\ & \left. \quad - \mu' Q_K Q_L (\delta_{j3} \overline{\Psi}_{,ikl3} + \delta_{i3} \overline{\Psi}_{,jkl3} + 2 x_{3} \overline{\Psi}_{,ijkl3} ) \right. \notag \\ & \left. \quad - 2 \mu' (1 - \nu') \left\{ Q_L \delta_{k3} (\delta_{j3} \overline{\Phi}_{,il} + \delta_{i3} \overline{\Phi}_{,jl} + 2 x_{3} \overline{\Phi}_{,ijl} ) \right. \right. \notag \\ & \left. \left. \quad + Q_K \delta_{l3} (\delta_{j3} \overline{\Phi}_{,ik} + \delta_{i3} \overline{\Phi}_{,jk} ) + 2 x_{3} \overline{\Phi}_{,ijk} \right\} - 4 \mu' (1 - 2 \nu') Q_K Q_L \delta_{i3} \delta_{j3} \overline{\Phi}_{,kl} \right. \notag \\ & \left. \quad + 4 \mu' \nu' Q_K Q_L x_3 (\delta_{i3} \overline{\Phi}_{,jkl} + \delta_{j3} \overline{\Phi}_{,ikl}) + 2 \mu' Q_K Q_L x_{3}^2 \overline{\Phi}_{,ijkl} \right] \notag %
    \\
    & \quad -B \left[ 2 \lambda' \delta_{kl} (\delta_{i3} \overline{\Phi}_{,j3} + \delta_{j3} \overline{\Phi}_{,i3} - \overline{\Phi}_{,ij} ) - \mu' (Q_K + Q_L) (\delta_{i3} \overline{\Theta}_{,jkl} + \delta_{j3} \overline{\Theta}_{,ikl}) \right. \notag \\ & \left. \quad + 2 \mu' (Q_L \delta_{k3} \overline{\Theta}_{,ijl} + Q_K \delta_{l3} \overline{\Theta}_{,ijk}) \right] + 2 (Q_I + Q_J) \left[ D \lambda' \delta_{kl} \overline{\Phi}_{,ij} \right. \notag \\ & \left. \quad + D \mu' Q_K Q_L \overline{\Psi}_{,ijkl} + \mu' (G + B) \overline{\Lambda}_{,ijkl} \right] \Big]
    \label{eq:ElasticEshelbyTensors_u}
\end{align}
For a field point in the lower phase ($x_{3} < 0$), we obtain:
\begin{equation}
    \begin{split}
    S_{ijkl} = S_{ijkl}^C + \frac{1}{8 \pi \mu'} \Big[ & - A \left[ 2 \lambda' \delta_{kl} Q_{I} \Phi_{,ij} + \mu' ( \delta_{ik} Q_{L} \Phi_{,lj} + \delta_{il} Q_{K} \Phi_{,kj} + \delta_{jk} Q_{L} \Phi_{,li} \right. \\ & \left. \quad + \delta_{jl} Q_{K} \Phi_{,ki} ) \right] - 4 D \left[ \lambda' \delta_{kl} \Phi_{,ij} + \mu' \Psi_{,ijkl} \right] \\
    & \quad -2 B \left[ \lambda' \delta_{kl} \Phi_{,ij} + \mu' (\delta_{i3} \Theta_{,jkl} + \delta_{j3} \Theta_{,ikl} - \delta_{k3} \Theta_{,ijl} - \delta_{l3} \Theta_{,ijk} ) \right] \\ 
    & \quad + 4 F \mu' (\delta_{i3} \Theta_{,jkl} + \delta_{j3} \Theta_{,ikl} + 2 x_{3} \Theta_{,ijkl}) \\
    & \quad +  2  (G + B) (Q_I + Q_J) (\lambda' \delta_{kl} \Phi_{,ij} + \mu' \Lambda_{,ijkl})  \Big]
    \end{split}
     \label{eq:ElasticEshelbyTensors_l}
\end{equation}
where $S_{ijkl}^C$ is the conventional Eshelby's tensor for a single infinite domain as \cite{Mura1987}, 
\begin{equation}
    S_{ijkl}^C = \frac{1}{8 \pi (1 - \nu')} \left[ \Psi_{,klij} - 2 \nu' \delta_{kl} \Phi_{,ij} - (1 - \nu') (\delta_{il} \Phi_{,jk} + \delta_{ik} \Phi_{,jl} + \delta_{jl} \Phi_{,ik} + \delta_{jk} \Phi_{,il} ) \right]
\end{equation}

Note that when the inclusion is subjected to a dilatational eigenstrain, i.e., $\varepsilon_{kl}^* = \varepsilon^T \delta_{kl}$, the partial derivatives $\Theta_{,ikl}$ and $\Lambda_{,ijkl}$ in Eqs. (\ref{eq:ElasticEshelbyTensors_u}) and (\ref{eq:ElasticEshelbyTensors_l}) can be significantly simplified, as two Boussinesq's displacement potentials are harmonic functions, $\alpha_{,mm} = \beta_{,mm} = 0$, and other terms involving the first Boussinesq's displacement potential can be further reduced to the Newtonian potential using $\alpha_{,3} = - \phi$. Consequently, the two Boussinesq's displacement potentials do not need to be handled as independent integral ingredients in the dilatational case, which can be seen in previous works \cite{Lyu2022, Kuvshinov2008}. However, the complete domain integrals of the two Boussinesq's displacement potentials are essential for a general eigenstrain problem. 

The same two Boussinesq's displacement potentials also appear in the bimaterial thermoelastic Green's function. Hence, the closed-form domain integrals of $\alpha$ and $\beta$ in form of $\Theta$ and $\Lambda$ are required to construct the thermoelastic Eshelby's tensor as well. For completeness, the thermoelastic Green's function and Eshelby's tensor are provided in \ref{sec:thermoelastic}.

\section{Domain integrals of two Boussinesq's displacement potentials}

The domain integrals of two displacement potentials are evaluated through a successive dimensional reduction. To facilitate the derivation process and define the integral limits, the polyhedral inclusion is decomposed into tetrahedral subdomains related to its surfaces and edges, and a local orthogonal coordinate (transformed coordinate) is built upon the normal vector of the surface and normal/directional vectors of the edge. Based on the transformed coordinate, the original expressions are constructed through partial derivatives of higher-order potentials, and the volume integrals can be converted into surface integrals using Gauss' theorem. Moreover, surface tangential vectors are created to reduce the remaining surface integrals into elementary line integrals along edges. The successive procedures avoid directly evaluating volume integrals of the two displacement potentials in the transformed coordinates, and the resulting formulae consist only of simple surface integrals and elementary line integrals.

\begin{figure}
    \centering
    \includegraphics[width=0.8\linewidth]{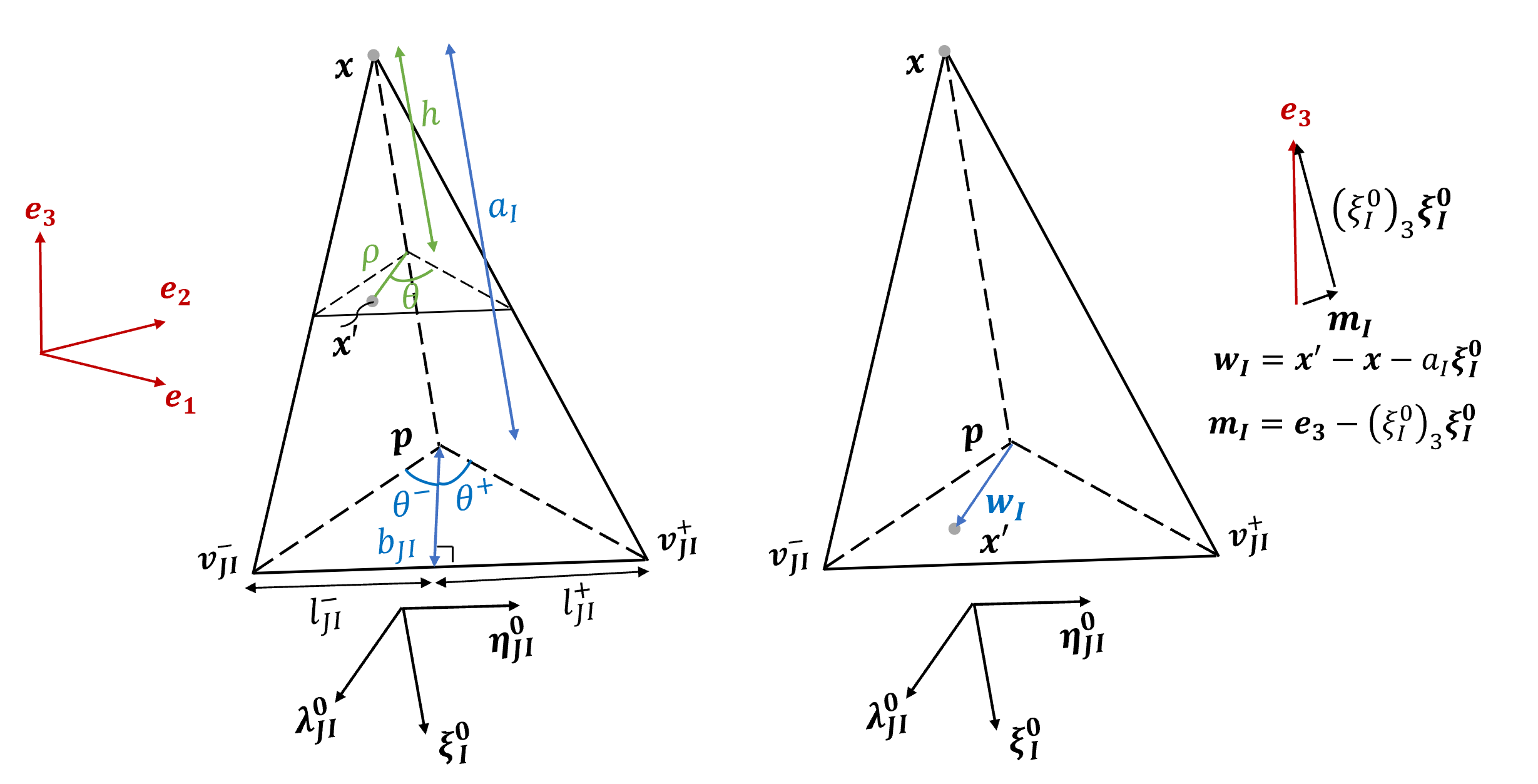}
    \caption{Schematic illustration of a tetrahedron composed of the field point $\textbf{x}$, its projection $\textbf{p}$ on the $\text{I}^\text{th}$ surface, and the $\text{J}^\text{th}$ edge with two vertices $\textbf{v}^\pm_{JI}$. In addition to the Cartesian coordinate $\textbf{e}_{1}, \textbf{e}_{2}, \textbf{e}_{3}$, a local cylindrical orthogonal coordinate, $\bm{\xi}_{I}^0, \bm{\lambda}_{JI}^0, \bm{\eta}^{0}_{JI}$, is constructed. When the source point $\textbf{x}'$ lies on the $\text{I}^\text{th}$ surface, two in-surface vectors $\bm{w}_{I}$ and $\bm{m}_{I}$ are defined.}
    \label{fig:fig2}
\end{figure}

\subsection{Transformed coordinate}
Consider a polyhedra containing $N_{I}$ surfaces, and each surface has $N_{JI}$ edges, which can be decomposed into tetrahedra \cite{Wu2021_polyhedral}. Fig. \ref{fig:fig2} shows a tetrahedron taken from the polyhedra in Fig. \ref{fig:fig1}, which is composed of the field point $\textbf{x}$, its projection $\textbf{p}$ on the $\text{I}^\text{th}$ surface, and the $\text{J}^\text{th}$ edge with two vertices $\textbf{v}^\pm_{JI}$. In addition to the Cartesian coordinate $\textbf{e}_{1}, \textbf{e}_{2}, \textbf{e}_{3}$, the transformed coordinate is constructed on the $\text{J}^\text{th}$ edge belonging to the $\text{I}^\text{th}$ surface that $\bm{\xi}_{I}^0$, $\bm{\lambda}_{JI}^0$, and $\bm{\eta}_{JI}^0$ represent the unit normal vector of the surface, edge, and the unit directional vector of the edge, respectively. 

The three base vectors form an orthogonal transformed coordinate, which helps to better define integral limits in the cylindrical coordinate. Because the displacement potentials contain the third component of the distance vector, for compactness, this section defines their directional cosines with respect to the $x_{3}$ axis as follows: 
\begin{equation}
    \chi_I = (\xi_{I}^0)_{3}, \quad \zeta_{JI} = (\lambda_{JI}^0)_{3}, \quad \gamma_{JI} = (\eta_{JI}^0)_{3}
\end{equation}
Based on the transformed coordinate, the following geometric parameters are defined as, 
\begin{equation}
    a_{I} = (\bm{v}_{JI}^+ - \textbf{x}) \cdot \bm{\xi}_{I}^0, \quad b_{JI} = (\bm{v}_{JI}^+ - \textbf{x}) \cdot \bm{\lambda}_{JI}^0, \quad l_{JI}^\pm = (\bm{v}_{JI}^+ - \textbf{x}) \cdot \bm{\eta}_{JI}^0
    \label{eq:vars}
\end{equation}
where $a_{I}$ refers to the signed distance between the field point $\textbf{x}$ and its projection $\textbf{p}$ on the $\text{I}^\text{th}$ surface; $b_{JI}$ represents the signed distance from the projection $\textbf{p}$ to the $\text{J}^\text{th}$ edge, and $l_{JI}^\pm$ specify the signed distances from the foot of the perpendicular to two vertices, respectively. Note that these geometric parameters are signed quantities, whose signs are determined by both the distance vector and the unit vectors. When the source point $\textbf{x}'$ is located on the $\text{I}^\text{th}$ surface, $a_I =(\bm{v}_{JI}^+ - \textbf{x}) \cdot \bm{\xi}_I^0 = (\textbf{x}' - \textbf{x}) \cdot \bm{\xi}_I^0$ is constant with respect to tangential differentiation on the surface. Hence, for ambient differentiation with respect to $\textbf{x}'$, the relation $\frac{\partial a_I}{\partial x_i'} = (\xi_I^0)_{i}$ is used in Eq. (\ref{eq:nabla_s_2}). When the source point $\textbf{x}'$ lies on the $\text{I}^\text{th}$ surface, two additional vectors are defined, 
\begin{equation}
    (w_{I})_{i} = x_i' - x_i - a_I (\xi_{I}^0)_{i} = r_i - a_I (\xi_I^0)_i, \quad (m_{I})_i = (e_{3})_{i} - \chi_I (\xi_{I}^0)_{i} = \delta_{i3} - \chi_I (\xi_{I}^0)_{i}
    \label{eq:def_w_m}
\end{equation}
where $r = \sqrt{r_i r_i}$, and it can be proved that $\bm{\xi}_{I}^0 \cdot \bm{w}_I = 0$ and $\bm{\xi}_{I}^0 \cdot \bm{m}_{I} = 0$. Therefore, two vectors $\bm{w}_I, \bm{m}_I$ are tangential to the $\text{I}^\text{th}$ surface. Note that the vector $\bm{m}_I$ is independent of both the field and source points, its partial derivatives with respect to $\textbf{x}, \textbf{x}'$ are therefore zero. The basis vector can be decomposed into its tangential and normal components with respect to the $\text{I}^\text{th}$ surface as $\textbf{e}_{3} = \bm{m}_{I} + \chi_I \bm{\xi}^0_{I}$. Some dot products of $\bm{w}_I$ and $\bm{m}_I$ are given, 
\begin{equation}
\begin{aligned}
    \bm{w}_I \cdot \bm{w}_I = [r_i - a_I (\xi_I^0)_i] [r_i - a_I (\xi_I^0)_i] = r^2 - a_I^2 \\ 
    \bm{w}_I \cdot \bm{m}_I = [r_i - a_I (\xi_I^0)_{i}] [\delta_{3i} - \chi_I (\xi_I^0)_{i}] =r_3 - \chi_I a_I \\
    \bm{m}_I \cdot \bm{m}_I = [\delta_{3i} - \chi_I (\xi_I^0)_{i}] [\delta_{3i} - \chi_I (\xi_I^0)_{i}] = 1 - \chi_I^2 
\end{aligned}
    \label{eq:dot_w_m}
\end{equation}

Although the bimaterial Green's function involves two branches of $\alpha$ and $\beta$, it is sufficient to evaluate the domain integrals $\Theta$ and $\Lambda$ for the original field point. The corresponding image terms can be acquired by setting the field point as $\textbf{x} = \overline{\textbf{x}} = \{x_{1}, x_{2}, -x_{3}\}$, which is elaborated in Section 3.4. Based on the transformed coordinate, the domain integral of the function $\mathcal{F}$ can be written as, 
\begin{equation}
    \int_{\Omega} \mathcal{F}(\textbf{x}, \textbf{x}') \thinspace dV(\textbf{x}') = \sum_{I = 1}^{N_I} \sum_{J = 1}^{N_{JI}} \int_{\tan^{-1} \left[ \frac{l_{JI}^-}{b_{JI}} \right]}^{\tan^{-1} \left[ \frac{l_{JI}^+}{b_{JI}} \right]} \int_{0}^{a_{I}} \int_{0}^{\frac{h b_{JI} \sqrt{1 + \tan^2 \theta}}{a_{I}}} \mathcal{F}(\rho, h, \theta) \thinspace \rho d \rho \thinspace d h \thinspace d\theta
    \label{eq:vol}
\end{equation}

For instance, it is intuitive to express Boussinesq's displacement potential as $\alpha = \ln[\sqrt{h^2 + \rho^2} + h \chi_I + \rho \cos \theta \zeta_{JI} + \rho \sin \theta \gamma_{JI}]$. However, it is challenging to conduct the analytical integral in such form, because the logarithmic term involves normal distance, planar radial coordinate, and Cartesian coordinate. The remainder of this section first converts the volume integral into surface integrals, and then further reduces the surface integrals to line integrals, so that the surface/volume integrals can be exactly evaluated.

\subsection{Domain integral of the first Boussinesq's displacement potential}
The details to transform the volume integral of $\alpha$ to surface and line integrals will be elaborated, which will be followed by the integral of $\beta$ in the next section. 
\subsubsection*{(1) Reduction of the volume integral of $\alpha$ into surface integral}
The first dimensional reduction follows from the divergence identity. Let $\textbf{r} = \textbf{x}' - \textbf{x}$ denote the distance between the source and field points. Because $\alpha$ depends only on $\textbf{r}$, its directional partial derivatives satisfy $r_{i} \alpha_{,i'} = 1$. Therefore, the first Boussinesq's displacement potential satisfies the following relation, 
\begin{equation}
    \nabla' \cdot \left[ \textbf{r} \alpha \right] = r_{i,i'} \alpha + r_{i} \alpha_{,i'} = 3 \alpha + 1
    \label{eq:vol_area_1}
\end{equation}
where $\nabla'$ refers to partial differentiation with respect to $\textbf{x}'$. Based on Gauss' theorem and Eq. (\ref{eq:vol_area_1}), $\alpha = \frac{1}{3} \left[ \nabla' \cdot (\textbf{r} \alpha) - 1 \right]$,  the volume integral of $\alpha$ can be converted into surface integrals and the volume of the polyhedra,
\begin{equation}
\begin{aligned}
   \int_{\Omega} \alpha \thinspace dV(\textbf{x}') & = \frac{1}{3}\int_{\partial \Omega} (r_{i} \alpha) n_{i}(\textbf{x}') \thinspace dS(\textbf{x}') - \frac{V_{\Omega}}{3} = \frac{1}{3} \sum_{I=1}^{N_I} a_{I}  \int_{S_{I}} \alpha \thinspace dS(\textbf{x}') - \frac{V_{\Omega}}{3} 
   %
\end{aligned}
    \label{eq:vol_area_2}
\end{equation}
where $r_{i} n_{i} = a_{I}$ when the source point lies on the $\text{I}^\text{th}$ surface $S_{I}$, and the volume of the polyhedra $V_{\Omega}$ can be evaluated with Eq. (\ref{eq:vol}) by setting the target function as the unit, 
\begin{equation}
    V_{\Omega} = \sum_{I=1}^{N_I} \sum_{J=1}^{N_{JI}} \frac{a_I b_{JI}}{6} \left(l_{JI}^+ - l_{JI}^-\right)
    \label{eq:vol_poly}
\end{equation}

Therefore, Eq. (\ref{eq:vol_area_2}) shows that the volume integral of $\alpha$ has been reduced to surface integrals of $\alpha$ over each surface and the volume of the polyhedra. 
\subsubsection*{(2) Reduction of the surface integral of $\alpha$ into line integral}
The next objective is to further reduce such surface integrals into line integrals along edges. 
Two useful surface identities are derived and utilized. Let $\nabla^s$ denote the surface gradient operator associated with the $\text{I}^\text{th}$ surface, 
\begin{equation}
    \nabla^s = (\textbf{I} - \bm{\xi}_I^0 \otimes \bm{\xi}_I^0) \cdot \nabla' = \left[ \delta_{ij} - (\xi_{I}^0)_{i} (\xi_{I}^0)_{j} \right] \frac{\partial}{\partial x_{j}'}
    \label{eq:surface_operator}
\end{equation}
where $\textbf{I}$ is the second-rank identity tensor. The surface divergence of the vector $\bm{w}_I$ can be simplified as,  
\begin{equation}
    \nabla^s \cdot \bm{w}_I = \left[ \delta_{ij} - (\xi_{I}^0)_{i} (\xi_{I}^0)_{j} \right] \left[ r_{i,j'} - \frac{\partial a_I}{\partial x_j'} (\xi_I^0)_{i} \right] = \left[ \delta_{ij} - (\xi_{I}^0)_{i} (\xi_{I}^0)_{j} \right] \left[ \delta_{ij} - (\xi_{I}^0)_{i} (\xi_{I}^0)_{j} \right] = 2
    \label{eq:nabla_s_2}
\end{equation}
where $r_{i, j'} = \delta_{ij}$ and $\frac{\partial a_I}{\partial x_j'} = (\xi_I^0)_j$ are used. It can be further derived that, 
\begin{equation}
    \begin{split}
    \bm{w}_{I} \cdot \nabla^s \alpha & = \left[r_{i} - a_{I} (\xi_{I}^0)_{i} \right] \left[ \delta_{ij} - (\xi_{I}^0)_{i} (\xi_{I}^0)_{j} \right] \alpha_{,j'} \\ 
    & = \left[ r_{j} - r_{i} (\xi_{I}^0)_{i} (\xi_{I}^0)_{j} - a_{I} (\xi_{I}^0)_{j} + a_{I} (\xi_{I}^0)_{i} (\xi_{I}^0)_{i} (\xi_{I}^0)_{j} \right] \alpha_{,j'} \\ 
    & = \left[ r_{j} - a_{I} (\xi_{I}^0)_{j} \right] \alpha_{,j'} = \bm{w}_{I} \cdot \nabla' \alpha
    \end{split}
    \label{eq:property_w_proof}
\end{equation}

The above results yield the following two surface identities, 
\begin{equation}
    \bm{w}_{I} \cdot \nabla^{s} \alpha = \bm{w}_I \cdot \nabla' \alpha, \quad \nabla^{s} \cdot \bm{w}_{I} = 2
    \label{eq:property_w}
\end{equation}

Using the properties in Eq. (\ref{eq:property_w}) and Green's theorem, the surface integral of $\alpha$ in Eq. (\ref{eq:vol_area_2}) can be simplified as, 

\begin{equation}
\begin{split}
    \Xi^{\alpha}_I = \int_{S_I} \alpha \thinspace dS(\textbf{x}') &= \frac{1}{2} \left[ \int_{S_{I}} \nabla^{s} \cdot (\bm{w}_{I} \, \alpha) \thinspace dS(\textbf{x}') - \int_{S_{I}} \nabla^s\alpha \cdot \bm{w}_{I} \thinspace dS(\textbf{x}') \right] \\ & = \frac{1}{2} \left[ \sum_{J=1}^{N_{JI}} b_{JI} \int_{\Gamma_{JI}} \alpha \thinspace d\textbf{x}' - \int_{S_I} \alpha_{,i'} (w_{I})_i \thinspace dS(\textbf{x}') \right]
\end{split}
\label{eq:surface_alpha_1}
\end{equation}
where $\Xi_I^\alpha$ refers to the integral of $\alpha$ over the $\text{I}^\text{th}$ surface; and $\Gamma_{JI}$ represents the $\text{J}^\text{th}$ edge of the $\text{I}^\text{th}$ surface. Since two vectors are orthogonal, $\bm{\lambda}^0_{JI} \cdot \bm{\xi}^0_I = 0$, the dot product of the unit outward normal vector ($\bm{\lambda}_{JI}^0$) and the other vector $\bm{w}_{I}$ is further simplified as, $\bm{\lambda}^0_{JI} \cdot \bm{w}_I = \bm{\lambda}_{JI}^0 \cdot \left( \textbf{r} - a_{I} \bm{\xi}_{I}^0 \right) = b_{JI}$, which is previously defined in Eq. (\ref{eq:vars}). In addition, the integrand in surface integral in Eq. (\ref{eq:surface_alpha_1}) can be written as, 
\begin{equation}
    \alpha_{,i'} (w_{I})_i = \frac{1}{r + r_{3}} \left( \frac{r_{j}}{r} + \delta_{j3} \right) \left( r_{j} - a_I (\xi_{I}^0)_{j} \right) = 1 - \frac{a_I}{r + r_{3}} \left( \frac{a_I}{r} + \chi_I \right)
    \label{eq:simp_alpha_grad}
\end{equation}

Substituting Eq. (\ref{eq:simp_alpha_grad}) into Eq. (\ref{eq:surface_alpha_1}), the surface integral of $\alpha$ can be further simplified as one line integral and two area integrals, 
\begin{equation}
\begin{split}
    \Xi^{\alpha}_I & = \frac{1}{2} \left[ \sum_{J=1}^{N_{JI}} b_{JI} \int_{\Gamma_{JI}} \alpha \thinspace d\textbf{x}' - \int_{S_I} \left( 1 - \frac{a_{I}^2}{r(r + r_{3})} - \frac{a_{I} \chi_I}{r + r_{3}} \right)  \thinspace dS(\textbf{x}')\right] \\ 
    & = \frac{1}{2} \left[ \sum_{J=1}^{N_{JI}} b_{JI} \int_{\Gamma_{JI}} \alpha \thinspace d\textbf{x}'-A_{I} + a_{I}^2 \int_{S_I} \frac{1}{r (r + r_{3})} \thinspace dS(\textbf{x}') + a_{I} \chi_I \int_{S_I} \frac{1}{r + r_{3}} \thinspace dS(\textbf{x}') \right]
\end{split}
    \label{eq:surfae_alpha_2}
\end{equation}
where $A_{I}$ refers to the surface area of the $\text{I}^\text{th}$ surface. Note that Eq. (\ref{eq:surfae_alpha_2}) still contains two surface integrals, and the remainder of this subsection utilizes Green's theorem to convert them into line integrals. Here we introduce a tangential vector $\bm{\mathcal{B}}_{I}= \chi_I \bm{w}_I - a_I \bm{m}_I$, whose surface divergence is equivalent to the linear combination of them because of: 
\begin{equation}
    \nabla^{s} \cdot \left( \frac{\bm{\mathcal{B}}_{I}}{r + r_{3}} \right) = \frac{1}{r + r_{3}} \left( \frac{a_{I}}{r} + \chi_I \right)
    \label{eq:surface_B}
\end{equation}

The left-hand side of Eq. (\ref{eq:surface_B}) can be written as, 
\begin{equation}
    \nabla^{s} \cdot \left( \frac{\bm{\mathcal{B}}_{I}}{r+r_{3}} \right) = \frac{\nabla^{s} \cdot \bm{\mathcal{B}}_{I}}{r + r_{3}} - \frac{\bm{\mathcal{B}}_{I} \cdot \nabla^{s} (r + r_{3})}{(r + r_{3})^2}
    \label{eq:lhs_tang}
\end{equation}

Based on Eqs. (\ref{eq:def_w_m}) and  (\ref{eq:dot_w_m}), $r = \sqrt{a_{I}^2 + (w_{I})_i (w_{I})_i}$ and $r_{3} = \bm{w}_{I} \cdot \bm{m}_{I} + \chi_I a_{I}$. Hence, the partial derivatives in Eq. (\ref{eq:lhs_tang}) are derived as below:
\begin{equation}
    \begin{aligned}
    \nabla^s( r + r_{3}) & = [\delta_{ij} - (\xi_I^0)_{i} (\xi_I^0)_{j}] [r_{,j'} + \delta_{3j}] = \frac{r_i - a_I (\xi_I^0)_{i}}{r} + \left( \delta_{i3} - (\xi_I^0)_{i} \chi_I \right) = \frac{\bm{w}_I}{r} + \bm{m}_{I} \\
    \nabla^{s} \cdot \bm{\mathcal{B}}_{I} & = \chi_I \nabla^{s} \cdot \bm{w}_{I} - a_{I} \nabla^{s} \cdot \bm{m}_{I} - \bm{m}_I \cdot \nabla^s a_I = 2 \chi_I
    \end{aligned}
\end{equation}
where $\nabla^{s} \cdot \bm{w}_{I} = 2$ in Eq. (\ref{eq:property_w}), $\nabla_{s} \cdot\bm{m}_I = [\delta_{ij} - (\xi_I^0)_i (\xi_I^0)_j] (m_I)_{i,j'} = 0$ ($(m_I)_{i,j'} = 0$), and $\nabla^s a_I = 0$ have been used. Therefore, Eq. (\ref{eq:lhs_tang}) can be further simplified as, 
\begin{equation}
\begin{split}
    & \nabla^{s} \cdot \left( \frac{\bm{\mathcal{B}}_{I}}{r+r_{3}} \right) = \frac{2 \chi_I}{r + r_{3}} - \frac{\left[ \chi_I \bm{w}_{I} - a_{I} \bm{m}_{I} \right] \left[ \frac{\bm{w}_{I}}{r} + \bm{m}_{I} \right]}{(r + r_{3})^2} \\ 
    & = \frac{2 \chi_I}{r + r_{3}} - \frac{1}{(r + r_{3})^2} \left[ \chi_I \frac{(w_{I})_i (w_{I})_i}{r} - a_{I} \frac{(w_{I})_{i} (m_{I})_{i}}{r} + \chi_I (w_{I})_{i} (m_{I})_{i} - a_{I} (m_{I})_{i} (m_{I})_{i} \right] \\ 
    %
    %
    & = \frac{2 \chi_I}{r + r_{3}} - \frac{(r + r_{3}) \left( \chi_I - \frac{a_I}{r} \right)}{(r + r_{3})^2} = \frac{\chi_I}{r+r_{3}} + \frac{a_{I}}{r (r + r_{3})}
\end{split}
\end{equation}
which shows that the surface divergence of the tangential vector $\mathcal{B}_{I}$ equals to the surface integrals in Eq. (\ref{eq:surfae_alpha_2}). Hence, substituting Eq. (\ref{eq:surface_B}) into Eq. (\ref{eq:surfae_alpha_2}), the surface integrals can be rewritten as line integrals using Green's theorem, 

\begin{equation}
    \begin{split}
    \Xi^{\alpha}_I & = \frac{1}{2} \left[ \sum_{J=1}^{N_{JI}} b_{JI} \int_{\Gamma_{JI}} \alpha \thinspace d\textbf{x}' - A_{I} + a_{I} \int_{S_I} \nabla^{s} \cdot \left( \frac{\bm{\mathcal{B}}_{I}}{r + r_{3}} \right) \thinspace d S(\textbf{x}') \right] \\ 
    & = \frac{-A_{I}}{2} + \frac{1}{2} \sum_{J=1}^{N_{JI}} \left[  b_{JI} \int_{\Gamma_{JI}} \alpha \thinspace d\textbf{x}' + a_{I} \int_{\Gamma_{JI}} \left( \frac{(\lambda_{JI}^0)_{i} \left[ \chi_I (w_{I})_{i} - a_{I} (m_{I})_{i} \right] }{r + r_{3}} \right) \thinspace d\textbf{x}' \right] \\ 
    & = \frac{-A_{I}}{2} + \frac{1}{2} \sum_{J=1}^{N_{JI}} \left[ b_{JI} \mathcal{H}_{JI} + a_{I} k_{JI} \mathcal{K}_{JI} \right]
    \end{split}
    \label{eq:surface_alpha_line}
\end{equation}
where $k_{JI} = \left[ \chi_I \textbf{r} - a_{I} \textbf{e}_{3} \right] \cdot \bm{\lambda_{JI}^0} = \chi_I b_{JI}  - a_{I} \zeta_{JI}$ ; $\mathcal{H}_{JI}$ and $\mathcal{K}_{JI}$ are two elementary line integrals, 

\begin{equation}
\begin{aligned}
    \mathcal{H}_{JI} = \int_{\Gamma_{JI}} \alpha \thinspace d\textbf{x}' = H_{JI}(l_{JI}^+) - H_{JI}(l_{JI}^-) \\ 
    \mathcal{K}_{JI} = \int_{\Gamma_{JI}} \frac{1}{r + r_{3}} \thinspace d\textbf{x}' = K_{JI}(l_{JI}^+) - K_{JI}(l_{JI}^-)
\end{aligned}
    \label{eq:def_H_K_int}
\end{equation}
where the detailed expression for $K_{JI}, H_{JI}$ and the derivation are provided in \ref{sec:KH}. Substituting Eq. (\ref{eq:surface_alpha_line}) into Eq. (\ref{eq:vol_area_2}), the domain integral of $\alpha$ only consists of its entire volume, area of each surface, and two elementary line integrals, 

\begin{equation}
    \Theta = \frac{1}{6} \sum_{I=1}^{N_I} a_{I} \left[ -A_{I} + \sum_{J=1}^{N_{JI}} \big(  b_{JI} \mathcal{H}_{JI} + a_{I} k_{JI} \mathcal{K}_{JI} \big) \right] - \frac{V_{\Omega}}{3} 
    \label{eq:alpha_final}
\end{equation}
Eq. (\ref{eq:alpha_final}) completes the dimensional reduction of the first Boussinesq's displacement. The domain integral can be expressed in terms of the polyhedral volume, surface areas, and two elementary edge primitives $H_{JI}$ and $K_{JI}$. 

\subsection{Domain integral of the second Boussinesq's displacement potential}
Following the procedure in the previous subsection, the domain integral of $\beta$ can be evaluated using vector manipulation and Gauss' and Green's theorems, reducing the volume integral to surface integrals and then to line integrals. 
\subsubsection*{(1) Reduction of the volume integral of $\beta$ into surface integral}
Specifically, the integration results consist only of two straightforward volume and area integrals, and four elementary line integrals. With some straightforward derivation, the second Boussinesq's displacement potential satisfies the following relation:
\begin{equation}
    \nabla' \cdot \left[ \textbf{r} \beta \right] = r_{i,i'} \beta + r_{i} \beta_{,i'} = 3 \beta + r_{i} \left[ \delta_{i3} \ln[r + r_{3}] + r_{3} \frac{r_{,i'} + \delta_{i3}}{r + r_{3}} - r_{,i'} \right] = 4 \beta + r_{3}
    \label{eq:beta_grad}
\end{equation}
which leads to $\beta = \frac{1}{4} \left[ \nabla' \cdot (\textbf{r} \beta) - r_{3} \right]$. Based on Gauss' theorem and Eq. (\ref{eq:beta_grad}), the volume integral of $\beta$ can be converted into surface integrals and the volume integral of a function $r_{3}$, 
\begin{equation}
    \int_{\Omega} \beta \thinspace d V(\textbf{x}') = \frac{1}{4} \int_{\partial \Omega} (r_{i} \beta) n_{i}(\textbf{x}') \thinspace d S(\textbf{x}') - \frac{1}{4} \int_{\Omega} r_{3} \thinspace d V(\textbf{x}') = \frac{1}{4} \sum_{I=1}^{N_I} a_{I} \int_{S_I} \beta \thinspace d S(\textbf{x}') - \frac{V_{\Omega3}}{4}
    \label{eq:beta_to_surf}
\end{equation}
where $V_{\Omega3}$ is the volume integral of $r_{3}$, which refers to the first volume moment of the relative coordinate $r_{3}$. Based on the transformed coordinate, the closed-form formulation for $V_{\Omega3}$ is derived with Eq. (\ref{eq:vol}), 
\begin{equation}
    V_{\Omega3} = \int_{\Omega} r_{3} \thinspace d\textbf{x}' = \sum_{I=1}^{N_{I}} \sum_{J=1}^{N_{JI}} \frac{a_{I} b_{JI}}{24} \left( l_{JI}^+ - l_{JI}^- \right) \left[ \gamma_{JI} \left( l_{JI}^+ + l_{JI}^- \right) + 2 b_{JI} \zeta_{JI} + 3 a_{I} \chi_I \right]
    \label{eq:grav_3}
\end{equation}

\subsubsection*{(2) Reduction of the surface integral of $\beta$ into line integral}
Using the identity of the surface gradient in Eq. (\ref{eq:property_w}), the surface integral in Eq. (\ref{eq:beta_to_surf}) can be written as, 
\begin{equation}
    \begin{split}
    \int_{S_I} \nabla^s \cdot (\bm{w}_{I} \beta) \thinspace d S(\textbf{x}') 
    & = 3 \int_{S_I} \beta \thinspace d S(\textbf{x}') + \int_{S_I} r_{3} \thinspace d S(\textbf{x}') - a_{I} \int_{S_I} \left[ \chi_I \alpha+ \frac{r_{3}}{r + r_{3}} \left( \chi_I  + \frac{a_{I}}{r} \right) - \frac{a_{I}}{r}\right] \thinspace d S(\textbf{x}')
    \end{split}
    \label{eq:beta_surface}
\end{equation}

Using the tangential vector $\bm{\mathcal{B}}_I$ and the identity in Eq. (\ref{eq:surface_B}), the surface integral of $\beta$ can be further simplified as, 
\begin{equation}
    \begin{split}
    \Xi^\beta_I = \int_{S_I} \beta \thinspace d S(\textbf{x}') & = \frac{1}{3}  \left[ \chi_I a_{I} \Xi^\alpha_I  - \int_{S_I} r_{3} \thinspace d S(\textbf{x}') - a_{I}^2 \int_{S_I} \frac{1}{r} \thinspace d S(\textbf{x}')  \right. \\ & \qquad \left. + a_{I} \int_{S_I} r_{3} \nabla^{s} \cdot \left( \frac{\bm{\mathcal{B}}_{I}}{r+r_{3}} \right) \thinspace d S(\textbf{x}') + \sum_{J=1}^{N_{JI}} b_{JI} \int_{\Gamma_{JI}} \beta \thinspace d \textbf{x}' \right]
    \end{split}
    \label{eq:beta_surface_final}
\end{equation}
where $\Xi_I^\beta$ refers to the integral of $\beta$ over the $\text{I}^\text{th}$ surface. As Eq. (\ref{eq:beta_surface_final}) shows, the surface integral of $\beta$ now consists of four surface integrals and one line integral, in which $\Xi^\alpha_I$ has been evaluated in Eq. (\ref{eq:surface_alpha_line}). For compact notations, the other four integrals are defined as, 

\begin{equation}
    \begin{aligned}
    \mathcal{A}_{I3} = \int_{S_I} r_{3} \thinspace d S(\textbf{x}'), \quad \Phi_{I} = \int_{S_I} \frac{1}{r} \thinspace d S(\textbf{x}'), \quad \mathcal{L}_{JI} = \int_{\Gamma_{JI}} \beta \thinspace d \textbf{x}', \quad \mathcal{D}_{I} = \int_{S_I} r_{3} \nabla^s \cdot \left( \frac{\bm{\mathcal{B}}_{I}}{r + r_{3}} \right) \thinspace d S(\textbf{x}')
    \end{aligned}
    \label{eq:def_int_beta}
\end{equation}
where $\mathcal{A}_{I3}$, $\Phi_I$, and $\mathcal{L}_{JI}$ can be obtained with straightforward derivation using a similar procedure for $\mathcal{K}_{JI}$ and $\mathcal{H}_{JI}$, which are elaborated in \ref{sec:three_integrals}. Note that $\Phi_I$ refers to the surface integral of the Newtonian potential, whose closed-form formulation has been reported in \cite{Wu2021_polyhedral}. For completeness, its result is provided in the \ref{sec:three_integrals}. 

However, the surface integral $\mathcal{D}_{I}$ requires additional treatment before Green's theorem can be applied. Unlike the previous surface integrals, $\mathcal{D}_I$ contains the directional component $r_{3}$ multiplying a surface divergence and therefore cannot be directly reduced to line integrals, which require additional treatment. Using integration by parts first separates the original integral into a boundary part and another surface integral, which involves $(\nabla^s r_{3}) \cdot \mathcal{B}_{I} / (r + r_{3})$. Following the concept in Eq. (\ref{eq:surface_B}), a tangential vector is constructed, so that the surface integral can be written as the surface divergence of it. A term-by-term justification of this derivation is provided in \ref{sec:proof_surface_grad_beta}. Thus, 

\begin{equation}
    \begin{split}
    \mathcal{D}_I & = \int_{S_I} \nabla^{s} \cdot \left( r_{3}\frac{\bm{\mathcal{B}}_{I}}{r + r_{3}} \right) \thinspace d S(\textbf{x}') - \int_{S_I} \frac{(\nabla^{s} r_{3}) \cdot\bm{\mathcal{B}}_{I}}{r+r_{3}} \thinspace dS(\textbf{x}') \\ 
    & = \int_{S_I} \nabla^{s} \cdot \left( r_{3}\frac{\bm{\mathcal{B}}_{I}}{r + r_{3}} \right) \thinspace dS(\textbf{x}') - \frac{1}{2} \int_{S_I} \nabla^{s} \cdot \left[ \frac{\chi_I r_{3} - a_{I}}{r+r_{3}} \bm{w}_{I} \right. \\ & \left.+ a_{I} \frac{(\chi_I^2 - 1) \bm{w}_{I} - a_{I} \chi_I \bm{m}_{I}}{r + r_{3}}  - a_{I} \frac{(\chi_I r_{3} - a_{I}) (\chi_I \bm{w}_{I} - a_{I} \bm{m}_{I})}{(r + r_{3})^2} \right] \thinspace dS(\textbf{x}') \\ 
    & = \sum_{J=1}^{N_{JI}} \left[ k_{JI} \mathcal{M}_{JI} - \frac{b_{JI}}{2} \mathcal{U}_{JI}  - \frac{a_{I}}{2} p_{JI} \mathcal{K}_{JI} \right. \left.+ \frac{a_{I}}{2} k_{JI} \mathcal{V}_{JI} \right]
    \end{split}
    \label{eq:beta_surface_manu}
\end{equation}
Therefore, all surface integrals in $\mathcal{D}_I$ are reduced to line integrals. 
\begin{equation}
    \mathcal{M}_{JI} = \int_{\Gamma_{JI}} \frac{r_{3}}{r + r_{3}} \thinspace d\textbf{x}', \quad \mathcal{U}_{JI} = \int_{\Gamma_{JI}} \frac{\chi_I r_{3} - a_{I}}{r + r_{3}} \thinspace d\textbf{x}' = \chi_I \mathcal{M}_{JI} - a_I \mathcal{K}_{JI}, \quad \mathcal{V}_{JI} = \int_{\Gamma_{JI}} \frac{\chi_{I} r_{3} - a_I}{(r + r_{3})^2}\thinspace d\textbf{x}'
    \label{eq:def_MUV}
\end{equation}
where $k_{JI} = \chi_I b_{JI} - a_I \zeta_{JI}$ is previously defined in Eq. (\ref{eq:surface_alpha_line}); and $p_{JI} = \bm{\lambda}_{JI}^0 \cdot \left[ (\chi_I^2 - 1) \bm{w}_{I} - a_I \chi_I \bm{m}_I \right] = (\chi_I^2 - 1) b_{JI} - a_I \chi_I \zeta_{JI}$. The integral $\mathcal{K}_{JI}$ has been evaluated in Eq. (\ref{eq:explicit_K}). The other two line integrals $\mathcal{M}_{JI}, \mathcal{V}_{JI}$ are derived in \ref{sec:M_V}. Therefore, the surface integral of $\beta$ can be written as, 
\begin{equation}
    \Xi_I^\beta = \frac{1}{3} \left( a_I \chi_{I} \Xi^\alpha_I - \mathcal{A}_{I3} - a_I^2 \Phi_{I} + a_I \mathcal{D}_I + \sum_{J=1}^{N_{JI}} b_{JI} \mathcal{L}_{JI} \right)
    \label{eq:surface_beta_final}
\end{equation}
Substituting Eq. (\ref{eq:surface_beta_final}) into Eq. (\ref{eq:beta_to_surf}), the volume integral of $\beta$ can be obtained as,
\begin{equation}
\begin{split}
    \Lambda = \frac{1}{12} \sum_{I=1}^{N_I} \Bigg\{ a_I^2 \chi_I \Xi_I^{\alpha} - a_I \mathcal{A}_{I3} - a_I^3 \Phi_I + a_I^2 \mathcal{D}_I + a_I \sum_{J=1}^{N_{JI}} b_{JI} \mathcal{L}_{JI} \Bigg\} - \frac{V_{\Omega3}}{4}
\end{split}
    \label{eq:beta_final}
\end{equation}
Compared with $\Theta$ in Eq. (\ref{eq:alpha_final}), the evaluation of Eq. (\ref{eq:beta_final}) requires additional treatment of the directional component $r_{3}$. Eq. (\ref{eq:beta_final}) completes the dimensional reduction of the second Boussinesq's displacement potential. The domain integral can be expressed in terms of the volume and area integrals of $r_{3}$, the surface integral of $\alpha$, and five edge primitives. 

\subsection{Partial differentiation of $\Theta, \Lambda, \overline{\Theta}, \overline{\Lambda}$}
Because Eshelby's tensors are composed of partial derivatives of $\Theta$ and $\Lambda$, the domain integrals should be differentiated with respect to $\textbf{x}$. For instance, as Eq. (\ref{eq:elastic_Green}) and Eq. (\ref{eq:disturbed_strain}) indicate, the evaluation of disturbed strain requires the third- and fourth-order partial derivatives of $\Theta$, and the fourth-order partial derivative of $\Lambda$, respectively. Instead of directly differentiating the volume-integral expressions in Eq. (\ref{eq:alpha_final}) and Eq. (\ref{eq:beta_final}), the translation invariance of the original Boussinesq displacement potentials is first applied to reduce the first-order differentiation to surface integrals. For instance, $\alpha$ and $\beta$ depend on the distance vector $\textbf{x}' - \textbf{x}$, which provides $\alpha_{,i} = -\alpha_{,i'}$ and $\beta_{,i} = -\beta_{,i'}$. This treatment allows the application of Gauss' theorem, which lowers the order of subsequent differentiation for both analytical and numerical implementations. Therefore, the first-order partial derivatives can be written as, 

\begin{equation}
\begin{aligned}
    \Theta_{,i} = -\int_{\Omega} \alpha_{,i'}(\textbf{x}', \textbf{x}) \thinspace dV(\textbf{x}') = \sum_{I=1}^{N_I} -(\xi_{I}^0)_{i} \int_{S_{I}} \alpha(\textbf{x}', \textbf{x}) \thinspace dS(\textbf{x}') = \sum_{I=1}^{N_I} -(\xi_I^0)_{i} \Xi_I^{\alpha} \\ 
    \Lambda_{,i} = -\int_{\Omega} \beta_{,i'}(\textbf{x}', \textbf{x}) \thinspace dV(\textbf{x}') = \sum_{I=1}^{N_I} -(\xi_{I}^0)_{i} \int_{S_{I}} \beta(\textbf{x}', \textbf{x}) \thinspace dS(\textbf{x}') = \sum_{I=1}^{N_I} -(\xi_I^0)_{i} \Xi_I^{\beta} 
\end{aligned}
\end{equation}
where the two surface integrals are provided in Eq. (\ref{eq:surface_alpha_line}) and Eq. (\ref{eq:surface_beta_final}), respectively. In addition, the image Boussinesq's displacement potentials can be obtained by mirroring the field point $\textbf{x}$ as $\overline{\textbf{x}} = (x_{1}, x_{2}, - x_{3})$, therefore, $\overline{\alpha}(\textbf{x}', \textbf{x}) = \alpha(\textbf{x}', \overline{\textbf{x}})$ and $\overline{\beta}(\textbf{x}', \textbf{x}) = \beta(\textbf{x}', \overline{\textbf{x}})$. Consequently, the partial derivatives of image potentials are, 

\begin{equation}
    \overline{\Theta}_{,i_1 i_2 ... i_{n}}(\textbf{x}) = \left( \prod_{n=1}^{N} Q_{I_{n}} \right) \Theta_{,i_1 i_2 ... i_{n}}(\overline{\textbf{x}}), \quad \overline{\Lambda}_{,i_1 i_2 ... i_{n}}(\textbf{x}) = \left( \prod_{n=1}^{N} Q_{I_{n}}\right) \Lambda_{,i_1 i_2 ... i_{n}}(\overline{\textbf{x}})
    \label{eq:image_diff}
\end{equation}
where $N$ refers to the times of partial differentiation; and each differentiation with the respect to $\textbf{x}_{i}$ introduces the corresponding factor $Q_I$. Specifically, the first-order partial derivatives can be obtained as, 

\begin{equation}
\begin{aligned}
    \overline{\Theta}_{,j} = \sum_{I=1}^{N_I} -Q_J (\xi_{I}^0)_{j} \int_{S_{I}} \alpha(\textbf{x}', \overline{\textbf{x}}) \thinspace dS(\textbf{x}'), \quad 
    \overline{\Lambda}_{,j} = \sum_{I=1}^{N_I} - Q_J (\xi_{I}^0)_{j} \int_{S_{I}} \beta(\textbf{x}', \overline{\textbf{x}}) \thinspace d S(\textbf{x}')
\end{aligned}
\end{equation}
For subsequent differentiation, it can be conducted through the partial differentiation chain rule, i.e., $\mathcal{P}_{JI}(a_I, b_{JI}, l_{JI}^+, l_{JI}^-)$,

\begin{equation}
    \frac{\partial \mathcal{P}_{JI}}{\partial x_m} = \frac{\partial \mathcal{P}}{\partial a_I} \frac{\partial a_I}{\partial x_m} + \frac{\partial \mathcal{P}}{\partial b_{JI}} \frac{\partial b_{JI}}{\partial x_m} + \left( \frac{\partial \mathcal{P}}{\partial l_{JI}^+} \frac{\partial l^+_{JI}}{\partial x_m} + \frac{\partial \mathcal{P}}{\partial l_{JI}^-} \frac{\partial l^-_{JI}}{\partial x_m} \right)
\end{equation}
where the partial derivatives of geometric variables $a_I, b_{JI}, l_{JI}^\pm$ can be derived from Eq. (\ref{eq:vars}), 

\begin{equation}
    \frac{\partial a_I}{\partial x_m} = -(\xi_I^0)_{m}, \quad \frac{\partial b_{JI}}{\partial x_m} = -(\lambda_{JI}^0)_{m}, \quad \frac{\partial l^\pm_{JI}}{\partial x_m} = -(\eta_{JI}^0)_{m}
\end{equation}
For instance, the second-order partial derivatives of $\Theta$ can be written as, (let $\mathcal{P}_{JI} = \frac{1}{2} (b_{JI} \mathcal{H}_{JI} + a_{I} k_{JI} \mathcal{K}_{JI})$ in Eq. (\ref{eq:surface_alpha_line})), 
\begin{equation}
\begin{aligned}
    \Theta_{,ij}(\textbf{x}) & = \sum_{I=1}^{N_I} -(\xi_I^0)_{i} \frac{\partial}{\partial x_j} \left[ \int_{S_I} \alpha \thinspace d\textbf{x}' \right] = \sum_{I = 1}^{N_I} -\frac{(\xi_I^0)_{i}}{2} \sum_{J=1}^{N_{JI}} \frac{\partial}{\partial x_{j}} \left[ b_{JI} \mathcal{H}_{JI} + a_{I} k_{JI} \mathcal{K}_{JI} \right] \\ 
    & = \sum_{I = 1}^{N_I} -(\xi_I^0)_{i} \sum_{J=1}^{N_{JI}} \left\{ (-\xi_I^0)_{j} \frac{\partial \mathcal{P}_{JI}}{\partial a_{I}} + (-\lambda_{JI}^0)_{j} \frac{\partial \mathcal{P}_{JI}}{\partial b_{JI}} + (-\eta_{JI}^0)_{j} \left[ \frac{\partial \mathcal{P}_{JI}}{\partial l_{JI}^+} + \frac{\partial \mathcal{P}_{JI}}{\partial l_{JI}^-} \right] \right\}
\end{aligned}
    \label{eq:theta_2nd}
\end{equation}
Because the surface area $A_I$ is independent of the field point, its derivatives vanish, which does not appear in Eq. (\ref{eq:theta_2nd}). The derivatives of $\Lambda$ can be obtained analogously by differentiating the surface integral $\Xi^\beta_{I}$ in Eq. (\ref{eq:beta_surface_final}). The third- and fourth-order partial differentiation chain rule are provided in \ref{sec:partial_deriv}. The partial derivatives of elementary integrals in Appendix B can be evaluated symbolically using Mathematica. Although the third- and fourth-order partial derivatives can be explicitly obtained, the resulting expressions are lengthy. In the numerical implementation, this paper utilizes the open-source automatic differentiation package ``autodiff'' to evaluate partial derivatives based on the original closed-form expressions, which is elaborated with the Supplemental Material and ``C++'' source code for readers to reproduce results. 

\section{Validation and numerical applications}

\subsection{Reproduction of analytical solutions}
Section 3 derives the closed-form formulae for two Boussinesq's displacement potentials. As mentioned earlier, their higher-order partial derivatives are essential for constructing the elastic and thermoelastic Eshelby tensors for polyhedral inclusions embedded in the bimaterial domain. The formulae allow the analytical evaluation of the disturbances caused by prescribed eigenstrain and eigen-temperature-gradient, respectively. In this section, the proposed formulae and their high-order partial derivatives are verified by reproducing three classic results. The three benchmarks are arranged progressively to verify the formulae at the level of the potential integrals, fourth-order partial derivatives, and the assembled Eshelby's tensors. Specifically,  

\begin{enumerate}
    \item[(1)] The domain integrals of $\Theta$ and $\Lambda$ for a spherical inclusion approximated with polyhedra are compared with the analytical solution by Walpole \cite{Walpole1997}. The field points are located in the lower phase, while the spherical inclusion is located in the upper phase. The radius of the sphere is $0.1$ m, which is located at $(0, 0, 0.15)$ m. The field point moves along a line parallel to the third axis, with $x_1 = -0.02$ m, $x_{2} = -0.05$ m, and $x_{3} \in [-0.25, 0]$ m.
    \item[(2)] The fourth-order partial derivatives of $\Theta$ and $\Lambda$ for a cuboidal inclusion are compared with the analytical solution by Liu et al. \cite{Liu2012}. The field points are located in the lower phase, while the cuboidal inclusion is located in the upper phase. The dimensions of the cuboid are $0.4 \times 0.3 \times 0.2$ m, whose center is located at $(0, 0, 0.15)$ m. The field point moves along a line parallel to the third axis, with $x_1 = -0.02$ m, $x_{2} = -0.05$ m, and $x_{3} \in [-0.25, 0]$ m.
    \item[(3)] The fourth-rank elastic and third-rank thermoelastic Eshelby's tensors for a spherical inclusion are compared with the analytical results derived in our recent work \cite{Wu2023}. The field point moves in both phases, while the spherical inclusion is located in the upper phase. The radius of the sphere is $0.1$ m, and it is located at $(0, 0, 0.15)$ m. The field point moves along a line parallel to the third axis, with $x_1 = -0.02$ m, $x_{2} = -0.05$ m, and $x_{3} \in [-0.3, 0.6]$ m. 
\end{enumerate}

The three verification examples validate the present formulae at three levels. Step (1) verifies the closed-form domain integrals derived in Section 3 before partial differentiation. Since tetrahedra approximate the spherical inclusion, this example examines the convergence of the polyhedral representation of the exact sphere. Step (2) verifies the fourth-order partial derivatives of two Boussinesq's displacement potentials and confirms that they can be reduced to the existing analytical solution for a cuboidal inclusion parallel to the bimaterial interface. Finally, Step (3) verifies the complete assembly of the potentials and their derivatives through the elastic and thermoelastic bimaterial Eshelby's tensors. 

\subsection{Two Boussinesq's displacement potentials over a spherical inclusion}
\begin{figure}
    \centering
    \includegraphics[width=1\linewidth]{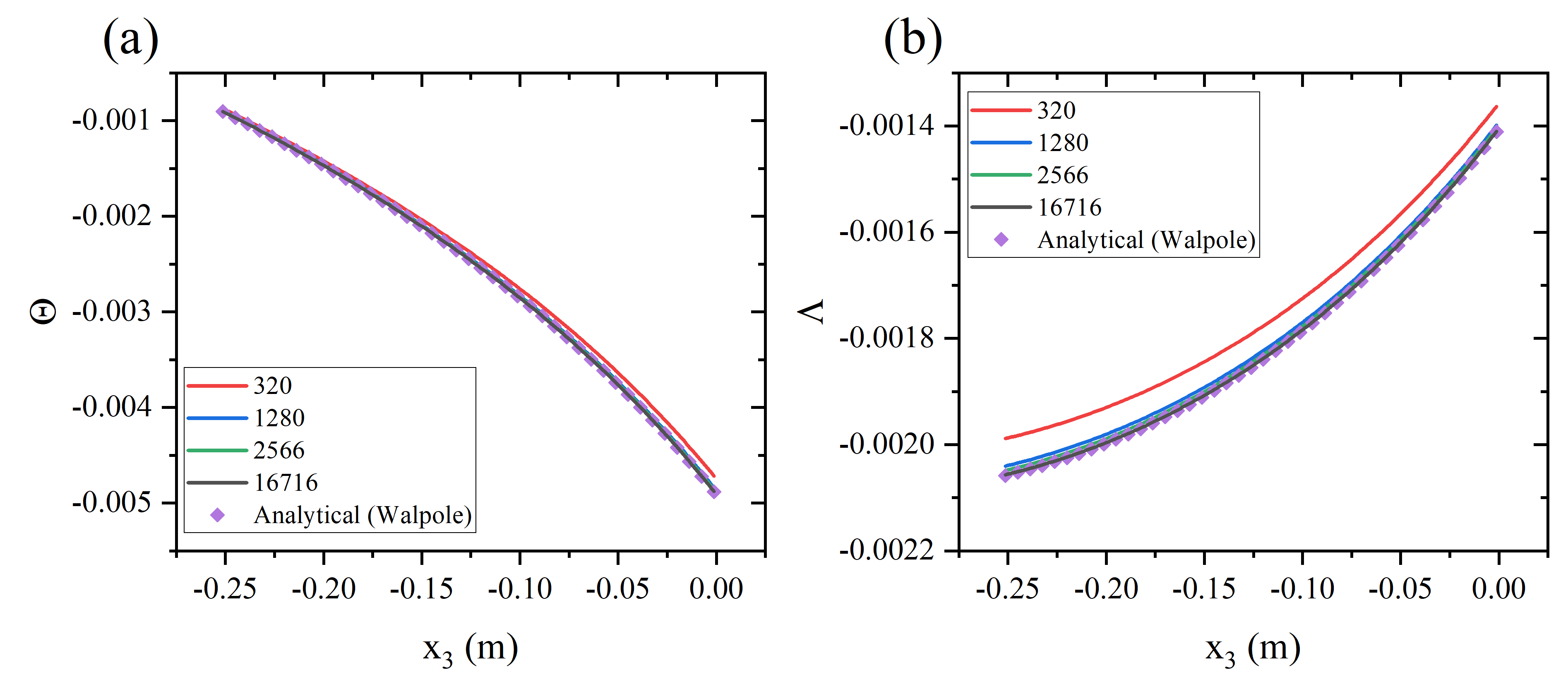}
    \caption{Comparison and variation of (a) $\Theta$ and (b) $\Lambda$ over a spherical inclusion using the polyhedra-approximated formulae and Walpole's solution \cite{Walpole1997}. The radius of the sphere is $0.1$ m, which is located at $(0, 0, 0.15)$ m. The field points move along the line parallel to the third axis, with $x_{1} = -0.02$, $x_{2} = -0.05$, and $x_{3} \in [-0.25, 0]$ m. The spherical inclusion is approximated by a polyhedra with \num{320}, \num{1280}, \num{2566}, \num{16716} faces. }
    \label{fig:sphere_potential}
\end{figure}

Figs. \ref{fig:sphere_potential} (a) and (b) compare the domain integrals of $\alpha$ and $\beta$ over a sphere approximated by a polyhedra with $N_I = $ \num{320}, \num{1280}, \num{2566}, and \num{16716} triangular faces, respectively. To avoid possible cancellation caused by spherical symmetry, this subsection intends to set field points deviated from the center line, with $x_{1} = -0.02, x_{2} = -0.05$, and $x_{3} \in [-0.25, 0]$ m. Specifically, Figs. \ref{fig:sphere_potential} (a) and (b) utilize Eqs. (\ref{eq:alpha_final}) and (\ref{eq:beta_final}) in Section 3, respectively. The analytical solution was first proposed by Walpole \cite{Walpole1997}. Note that the Boussinesq's displacement potentials originate from integration of the Newtonian potential with respect to the third component, which contains the infinite constant part. Because the additive constant in the logarithmic potential does not affect any further partial derivatives in Eshelby's tensors, only the regularized finite part is compared. 

As Fig. \ref{fig:sphere_potential} (a) indicates, the magnitude of $\Theta$ increases continuously as the field point $\textbf{x}$ approaches the bimaterial interface. This phenomenon can be interpreted as the decreasing distance between the source region and the field point, and the contribution of the logarithmic kernel gradually increases. Unlike the harmonic potential $\phi$ presented in \cite{Mura1987}, the distributions of $\Theta$ remain finite and exhibit $C^{1}$ continuity. Note that Boussinesq's displacement potentials require that the source and field points are separated in two phases; even for the image potentials, the source points are mirrored in another phase. Therefore, the field point and the source region are separated from the bimaterial interface by a finite distance. Consequently, there is no singularity during the integral process, which leads to a continuous distribution of $\Theta$. In contrast, Fig. \ref{fig:sphere_potential} (b) shows that the magnitude of $\Lambda$ decreases continuously as the field point $\textbf{x}$ approaches the bimaterial interface. The second Boussinesq's displacement potential involves competing contributions from the logarithmic kernel and the biharmonic potential. 

For both displacement potentials, all approximated solutions can provide good accuracy compared to the analytical solution. When the number of tetrahedra increases, the present formulae gradually converge to the analytical solution, as very minor discrepancies can be found in curves \num{2566}, \num{16716} and ``Analytical''. Because the polyhedra can never truly simulate an exact sphere, it is rational to predict some differences between curve $320$ and the analytical case. Note that the present formulae are evaluated analytically for each polyhedral approximation, and therefore, the remaining discrepancies should only be interpreted as the geometric approximation errors of the sphere. Hence, Figs. \ref{fig:sphere_potential} (a-b) confirm that the present formulae derived in Section 3 can provide acceptable accuracy, and verify the convergence to the exact spherical solution. Based on the numerical case study, Section 4.4 uses a polyhedra with \num{2566} triangular faces to approximate the sphere.

\subsection{Two Boussinesq's displacement potentials over a cuboidal inclusion}

\begin{figure}
    \centering
    \includegraphics[width=1\linewidth]{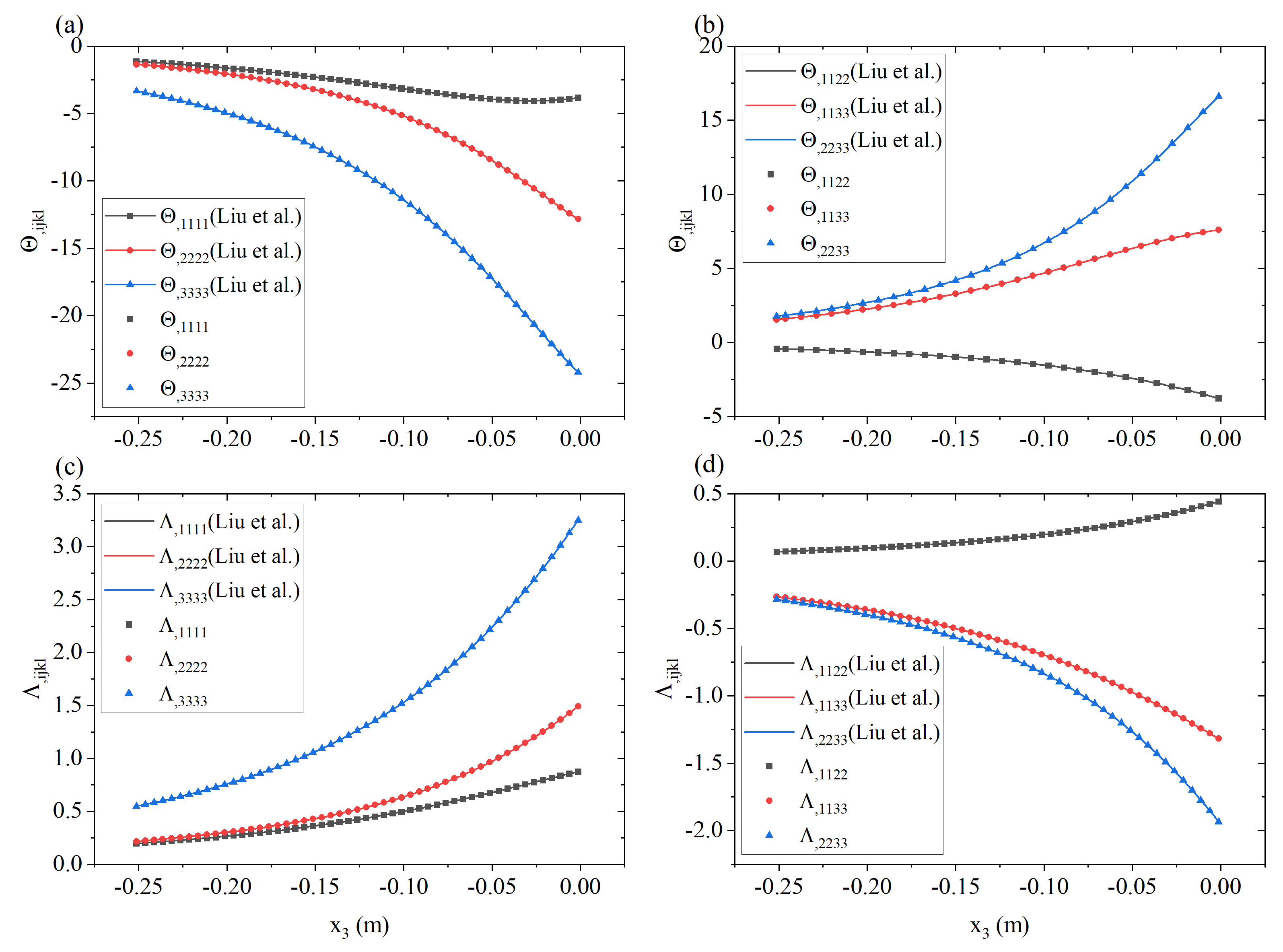}
    \caption{Comparison and variation of fourth-order partial derivatives of (a) $\Theta_{,1111}, \Theta_{,2222}, \Theta_{,3333}$, (b) $\Theta_{,1122}, \Theta_{,1133}, \Theta_{,2233}$, (c) $\Lambda_{,1111}, \Lambda_{,2222}, \Lambda_{,3333}$, and (d) $\Lambda_{,1122}, \Lambda_{,1133}, \Lambda_{,2233}$ over a cuboidal inclusion using the present formulae and Liu's solution \cite{Liu2012}. The dimensions of the cuboid are $0.4 \times 0.3 \times 0.2$ m, whose center is located at $(0, 0, 0.15)$ m. The field points move along the line parallel to the third axis, with $x_{1} = -0.02$, $x_{2} = -0.05$, and $x_{3} \in [-0.25, 0]$ m.}
    \label{fig:cube_potential}
\end{figure}

Figs. \ref{fig:cube_potential} (a-d) plot the variation of the fourth-order partial derivatives of $\Theta$ and $\Lambda$ for a cuboidal inclusion whose surfaces are parallel or perpendicular to the bimaterial interface. The dimensions of the cuboidal inclusion are $0.4 \times 0.3 \times 0.2$ m, and its center is located at $(0, 0, 0.15)$ m. To avoid possible cancellation caused by the symmetry, this subsection intends to set field points deviated from the center line, with $x_{1} = -0.02, x_{2} = -0.05$, and $x_{3} \in [-0.25, 0]$ m. The present results are compared with the closed-form formulae proposed by Liu et al. \cite{Liu2012}. Unlike the spherical inclusion case in the previous subsection, the geometric shape of the cuboid is exact in the present formulae. 

Fig. \ref{fig:cube_potential} (a) plots the components $\Theta_{,1111}, \Theta_{,2222}$, and $\Theta_{,3333}$. The numerical results obtained from the present method are consistent with the analytical solution by Liu et al. \cite{Liu2012}, as the curves overlap. Since the geometric shape is exact in the present formulae, only machine errors can be detected. While $\Theta_{,1111}, \Theta_{,2222}$, $\Theta_{,3333}$ are negative over the entire field-point path, the shearing components $\Theta_{,1122}, \Theta_{,1133}$, $\Theta_{,2233}$ exhibit different signs. The magnitude of partial derivatives increases as the field point approaches the bimaterial interface, which should be interpreted as a shorter distance from the source region to the field points. Components involving the normal direction, i.e., $\Theta_{,2233}$ and $\Theta_{,3333}$, exhibit the most intensive variation, since the field points move along the line parallel to the $x_{3}$ axis. Moreover, the unequal responses of $\Theta_{,1111}$ and $\Theta_{,2222}$ reflect the nonuniform dimensions in the $ x_1$ and $ x_2$ directions and the deviation of the field-point path from the symmetric centerline. 

Figs. \ref{fig:cube_potential} (c-d) plot the variation of the fourth-order partial derivative of $\Lambda$. Similar to the previous comparison of $\Theta_{,ijkl}$, all numerical results generated by the present formulae exhibit exact agreement with Liu's solution \cite{Liu2012}, which shows that the present formulae can successfully reproduce the reduced case. Specifically, when the surface normal vector is perpendicular to the bimaterial interface, the elementary integrals can simplify significantly. In contrast to $\Theta$, the normal components $\Lambda_{,1111}, \Lambda_{,2222}$, and $\Lambda_{,3333}$ are all positive, while the shearing components $\Lambda_{,1133}$, and $\Lambda_{,2233}$ are negative. Such sign reversal should be interpreted as the second Boussinesq's displacement potential mixing both the logarithmic term multiplied by the normal component of the distance vector and the biharmonic potential, which is different from the first Boussinesq's displacement potential. The present cuboid case confirms that the general polyhedral expressions have successfully recovered the existing reduced solution after fourth-order partial differentiation.

\subsection{Elastic and thermoelastic bimaterial Eshelby's tensors}
\begin{figure}
    \includegraphics[width=1\linewidth]{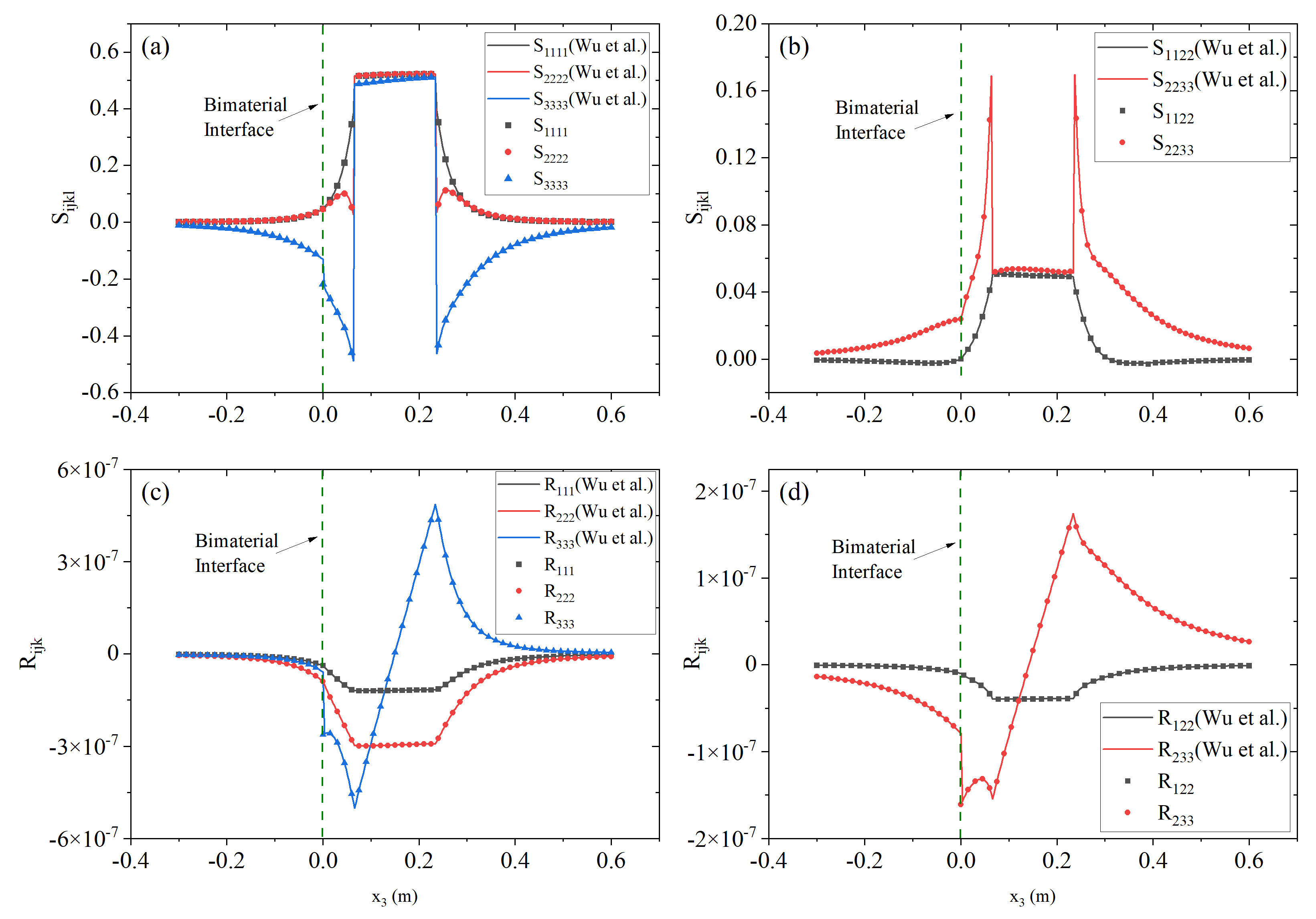}
    \caption{Comparison and variation of (a) normal and (b) shearing components of elastic bimaterial Eshelby's tensors, and (c) normal and (d) shearing components of thermoelastic bimaterial Eshelby's tensors over a spherical inclusion using the present formulae and our recent solution \cite{Wu2023}. The radius of the sphere is $0.1$ m, which is located at $(0, 0, 0.15)$ m. The field points move along the line parallel to the third axis, with $x_{1} = -0.02$, $x_{2} = -0.05$, and $x_{3} \in [-0.3, 0.6]$ m. The spherical inclusion is approximated by a polyhedra with \num{2566} triangular faces.}
    \label{fig:tensors}
\end{figure}

This subsection reproduces the complete components of the elastic and thermoelastic Eshelby's tensors. Without loss of any generality, the material properties of the upper and lower phases are specified as: (i) $\mu' = 7.673 \times 10^4 $ Pa, $\nu' = 0.3$, $\mathcal{A}' = 7.681$ Pa/K; and (ii) $\mu'' = 1.572 \times 10^5$ Pa, $\nu'' = 0.25$, $\mathcal{A}'' = 5.816$ Pa/K. The spherical inclusion with radius $0.1$ m is located at $(0, 0, 0.15)$ m, and the field point moves along a line parallel to the third axis, with $x_{1} = -0.02, x_{2} = -0.05$, and $x_{3} \in [-0.3, 0.6]$ m. Based on the results in Figs. \ref{fig:sphere_potential} (a-b), the numerical results are evaluated by the sphere approximated with \num{2566} triangular faces. Fig. \ref{fig:tensors} compares the elastic and thermoelastic Eshelby's tensors evaluated by the present formulae with the analytical solution proposed in our recent work \cite{Wu2023}. The dashed line refers to the bonded bimaterial interface. 

Figs. \ref{fig:tensors} (a) and (b) plot the normal and shearing components of the elastic Eshelby's tensor, which is shown in Eqs. (\ref{eq:ElasticEshelbyTensors_u}-\ref{eq:ElasticEshelbyTensors_l}). When field points are far from the spherical inclusion, all components gradually reduce to zero, which is consistent with the decaying feature of local disturbances by eigenstrain. In contrast, when the field point moves closer to the inclusion, the magnitude of components significantly increases. In particular, when the field point is inside the inclusion, the components $S_{3333}$ and $S_{2233}$ exhibit jumps. As Mura \cite{Mura1987} commented, the jumps originate from an integral singularity for interior field points, which exists for the full-space Eshelby's tensor as well. Unlike the case when the field points move along the symmetric line, the jump term ensures the continuity of the normal traction. Note that the jumps are jointly caused by two singular kernels $\phi_{,ij}$ and $\psi_{,ijkl}$, and the image kernels and two Boussinesq's displacement potentials have no contribution to them. Because the field-point path is deviated from the symmetric line, the jump is not exactly the unit, which can be observed in the curve ``$ S _ {2233} $ '' of Fig. \ref{fig:tensors} (b). 

For an ellipsoidal inclusion embedded in the infinite domain, the second-order partial derivatives of $\Phi$ and the fourth-order partial derivatives of $\Psi$ are invariants for interior field points. Consequently, the full-space Eshelby's tensor of the ellipsoidal inclusion remains constant for interior field points. As Figs. \ref{fig:tensors} (a) and (b) indicate, none of the components are constant within the spherical inclusion. This phenomenon should be interpreted as a bimaterial interfacial effect: the presence of the bimaterial interface breaks full-space isotropy and introduces interaction responses in the normal and tangential directions. Specifically, the component $S_{3333}$ exhibits apparent variation within the spherical inclusion, because the field points move along a line perpendicular to the bimaterial interface. Moreover, the difference between $S_{1111}$ and $S_{2222}$ arises from the off-axis location of the field-point axis, which is similar to $\Theta_{,1111}$ and $\Theta_{,2222}$ in Fig. \ref{fig:cube_potential} (a). 

Figs. \ref{fig:tensors} (c) and (d) show the normal and shearing components of the thermoelastic Eshelby's tensor $\mathcal{R}_{ijk}$, which describes the disturbed strain caused by the eigen-temperature-gradient, as shown in Eq. (\ref{eq:ThermoEshelbyTensors}). For components involving the normal direction, $\mathcal{R}_{333}, \mathcal{R}_{233}$, they exhibit the strongest spatial variation and change sign as the field point passes through the inclusion. Note that the full-space thermoelastic Eshelby's tensor is associated with the third-order partial derivative of $\Psi$. Hence, the components exhibit linear spatial variation within the spherical inclusion. As the curves $\mathcal{R}_{333}$ and $\mathcal{R}_{233}$ indicate, although the bimaterial interfacial effect causes discontinuity at the interface, the bimaterial thermoelastic Eshelby's tensor exhibits pseudo-linear distribution within the inclusion. The excellent agreement in Figs. \ref{fig:tensors} (a-d) not only verifies the closed-form Boussinesq's displacement potentials and their higher-order partial derivatives, but also confirms their assembly into the elastic and thermoelastic bimaterial Eshelby's tensors.

\subsection{Inclined cuboidal inclusion: FEM validation and EIM application}
The preceding subsections have shown that the closed-form formulae can reproduce classic solutions for spherical and specially aligned cuboidal inclusions. To further validate the general applicability of the present formulae, this subsection considers an inclined cuboid, whose faces are not parallel or perpendicular to the bimaterial interface. This subsection serves two purposes, (i) the closed-form solution is validated against finite element analysis for an inclined inclusion subjected to anisotropic eigenstrain; and (ii) the bimaterial Eshelby's tensor is further applied in the equivalent inclusion method (EIM) to solve an inhomogeneity problem with uniform eigenstrain approximation. The latter case primarily serves as a demonstration of application instead of an exact solution, since a uniform eigenstrain distribution is assumed. 

\begin{figure}
    \centering
    \includegraphics[width=0.5\linewidth]{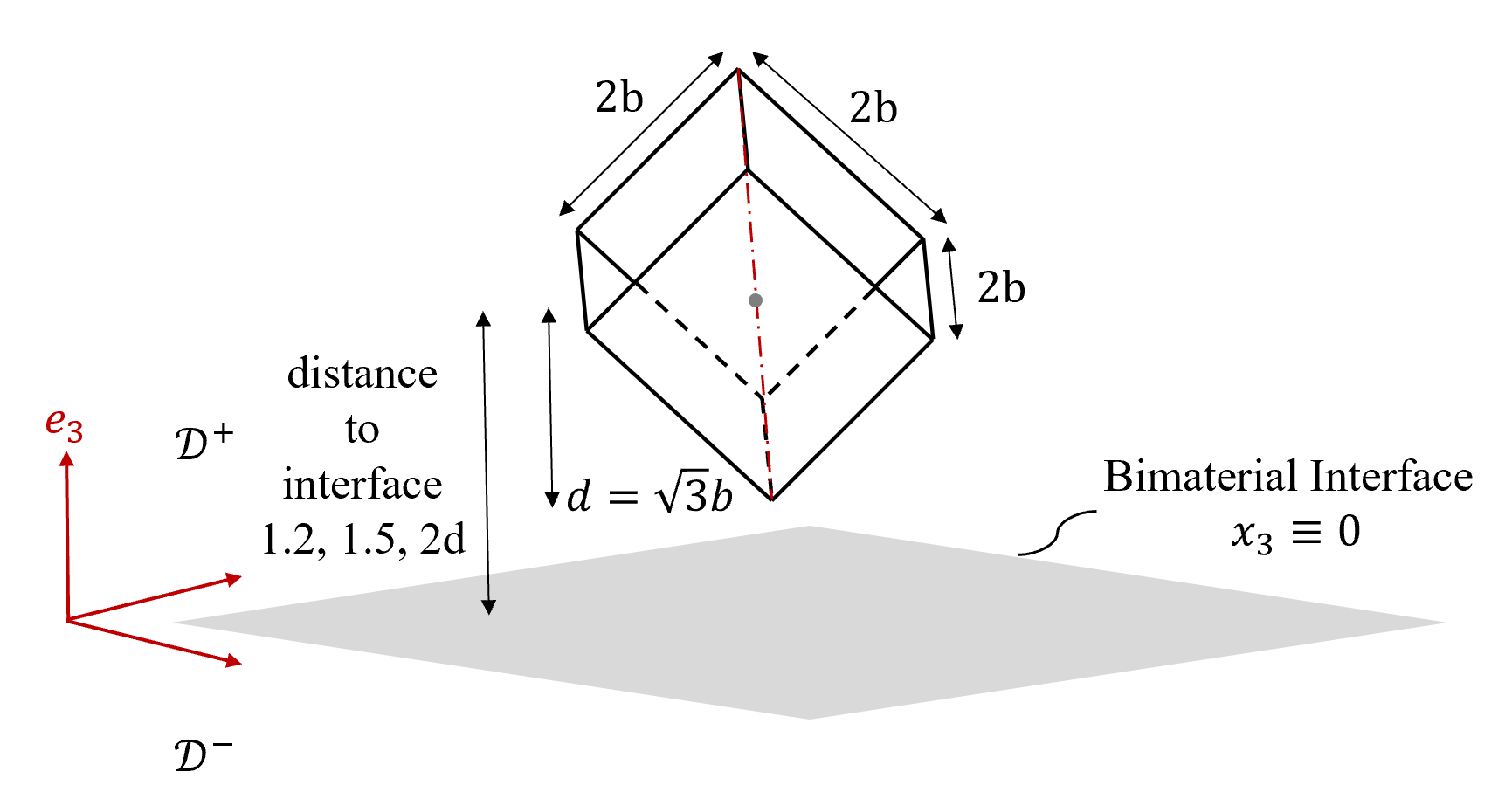}
    \caption{Schematic plot of an inclined cuboidal subdomain in the bimaterial domain, whose length $2b = 0.2$ m and the half length of the diagonal axis $d = \frac{\sqrt{3}}{10}$ m. Three distances between the center and the bimaterial interface are considered, $1.2 d, 1.5 d$ and $2 d$.}
    \label{fig:an_incline}
\end{figure}

Without loss of generality, Fig. \ref{fig:an_incline} plots an inclined cuboid with length $2b = 0.2$ m, whose diagonal axis is parallel to the $x_{3}$ axis. Three distances between the center of the cuboid and the bimaterial interface are considered: $1.2 d$, $1.5 d$, and $2 d$. Since the diagonal of the inclined cuboid is aligned with the $x_{3}$ axis, none of its edges are parallel to the $x_{3}$ axis. Hence, the edge cosines $\gamma_{JI} \neq \pm 1$, so that this configuration requires the general branch of the integral formulae in Appendix B. The material properties of the upper and lower half-spaces remain the same as in the previous subsection. Two numerical cases are considered: 

\begin{itemize}
    \item Inclusion case: the cuboid is subjected to anisotropic eigenstrain, $\varepsilon_{11}^* = 10^{-2}, \varepsilon^*_{22} = 2 \times 10^{-2}$, and $\varepsilon_{33}^* = 3 \times 10^{-2}$, and other components are zero. 
    \item Inhomogeneity case: the inclined cuboidal inhomogeneity has the elastic properties of the lower material phase, and the bimaterial domain is subjected to a far-field load of $\sigma_{33}^0 = 10^4$ Pa. 
\end{itemize}

Because the inclined cuboid is embedded in an infinite bimaterial domain, the infinite domain is truncated in the finite element simulation by two adjoining cuboidal domains, whose lengths are $1$ m. The mesh is locally refined within a $0.4$ m neighborhood of the inclined cuboid. For the distances of $1.2d, 1.5d$, and $2 d$, the finite element method uses \num{1527955}, \num{1991277}, \num{1992190} nodes and \num{980826}, \num{1187892}, \num{1188614} $10$-node quadratic tetrahedral elements, respectively. For both inclusion and inhomogeneity cases, the third component of the displacement is constrained on the bottom surface, and the lateral surfaces are free of traction. And the load $\sigma_{33}^0$ in the inhomogeneity case is imposed as a surface traction on the top surface of the adjoining cuboidal system. 

\begin{figure}
    \centering
    \includegraphics[width=1\linewidth]{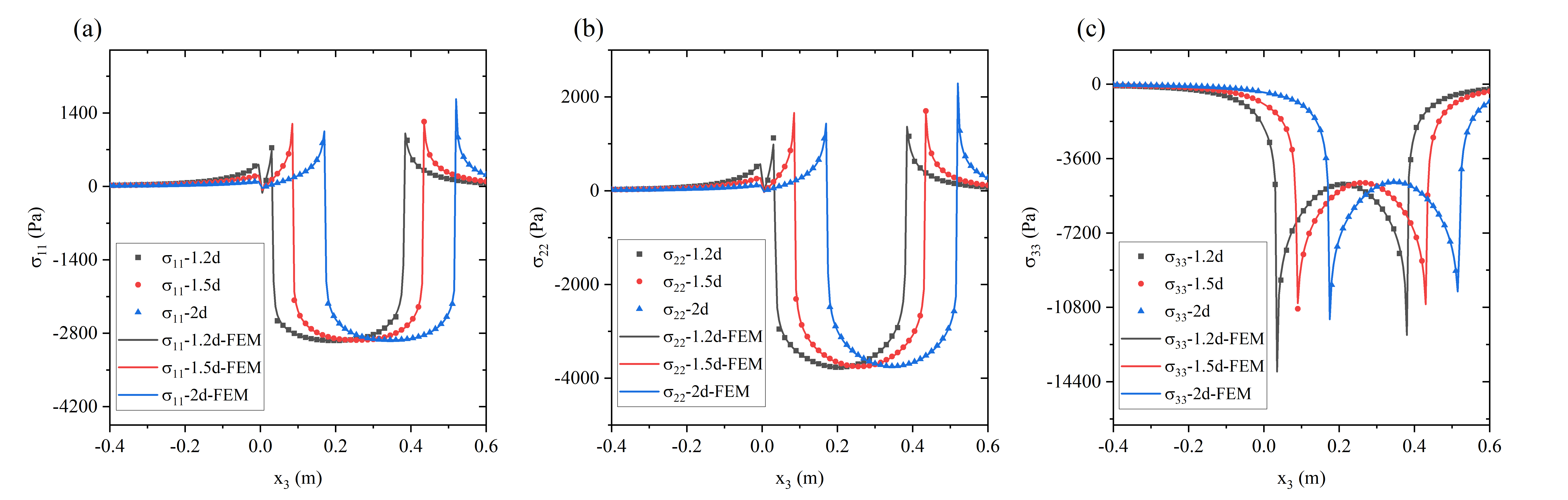}
    \caption{Comparison and variation of normal stresses (a) $\sigma_{11}$, (b) $\sigma_{22}$, (c) $\sigma_{33}$ along the vertical centerline of the inclined cuboidal inclusion, $x_{3} \in [-0.4, 0.6]$ m. The anisotropic eigenstrain, $\varepsilon_{11}^* = 10^{-2}$, $\varepsilon_{22}^* = 2 \times 10^{-2}$, and $\varepsilon_{33}^* = 3 \times 10^{-2}$ are considered.}
    \label{fig:inclusion}
\end{figure}

As Figs. \ref{fig:inclusion} (a-c) show, the closed-form results agree well with predictions by the finite element analysis for all three normal stress components. Note that there exists a stress/strain singularity at the vertices of inclusions, and Rodin \cite{Rodin1996} has stated the logarithmic singular behaviors. Hence, the curves in Figs. \ref{fig:inclusion} (a-c) cannot display the exact values at the vertices. Besides the singular stress, the present formulae accurately capture the variations of the stress magnitude, the discontinuity/continuity across the bimaterial interface, as well as the pronounced stress gradients in the neighborhood of the vertices. Minor discrepancies can be observed in the vicinity of the vertices, since the logarithmically singular stress fields are challenging to resolve with a finite mesh. These comparisons confirm that the present closed-form formulae are not restricted to polyhedra parallel/perpendicular to the bimaterial interface, but they can handle general geometric configurations with more complex eigenstrain settings. Hence, the excellent agreement provides independent validation of the present formulae for inclined polyhedral inclusions subjected to an anisotropic eigenstrain, beyond the special geometries considered in preceding analytical benchmarks. 

After verifying the closed-form formulae in Fig. \ref{fig:inclusion}, the remainder demonstrates their application in the EIM to solve an inhomogeneity problem. It should be emphasized that, unlike the inclusion problem with prescribed uniform eigenstrain, a non-ellipsoidal inhomogeneity generally requires a spatially varying eigenstrain to satisfy the equivalent stress condition, particularly in the presence of the bimaterial interfacial effects. We follow Walpole's strategy \cite{Walpole1997} to solve the cuboidal inhomogeneity embedded in the bimaterial domain, which requires the bimaterial elastic Eshelby's tensor in Eq. (\ref{eq:disturbed_strain}). Specifically, the equivalent stress condition using a uniform eigenstrain approximation can be written as follows, which is enforced at the center of the inhomogeneity, 

\begin{equation}
    C_{ijkl}' \left( \varepsilon_{kl}^0 + S_{klmn}(\textbf{x}^c) \thinspace \varepsilon_{mn}^* - \varepsilon_{kl}^* \right) = C''_{ijkl} \left( \varepsilon_{kl}^0 + S_{klmn}(\textbf{x}^c) \thinspace \varepsilon_{mn}^* \right)
    \label{eq:EIM}
\end{equation}
where $\textbf{x}^c$ represents the center of the cuboidal inhomogeneity; the upper branch of Eshelby's tensor in Eq. (\ref{eq:ElasticEshelbyTensors_u}) is adopted since the inhomogeneity is located in the upper phase; $\varepsilon_{kl}^0$ refers to the unperturbed strain field of the bimaterial domain under the same boundary conditions in the absence of the inhomogeneity, which is evaluated from the FEM model. Note that the unperturbed strain is used only as the incident field in Eq. (\ref{eq:EIM}), and the disturbed strain caused by the inhomogeneity is calculated using the solved eigenstrain $\varepsilon_{kl}^*$ and closed-form bimaterial Eshelby's tensor. 

\begin{figure}
    \centering
    \includegraphics[width=0.85\linewidth]{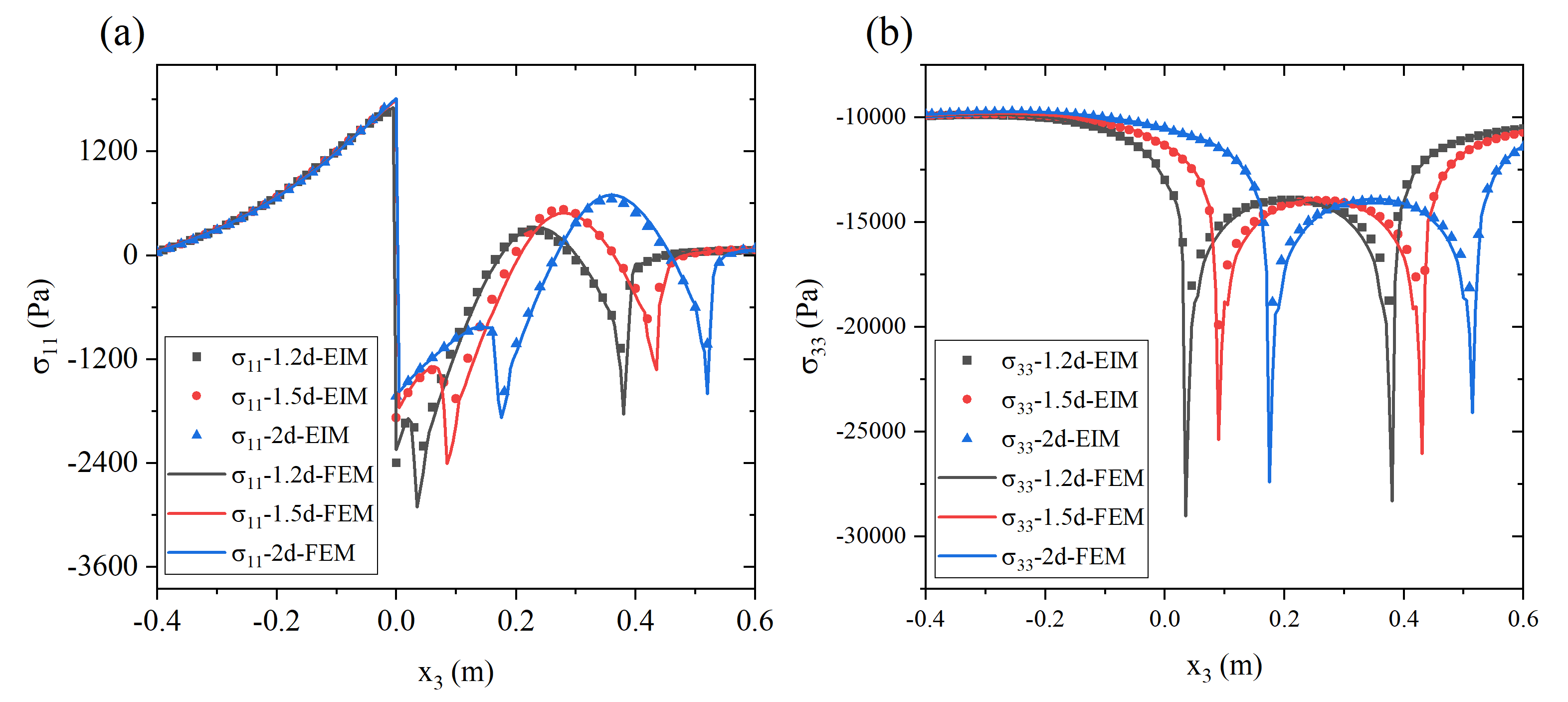}
    \caption{Comparison and variation of normal stresses (a) $\sigma_{11}$ and (b) $\sigma_{33}$ along the vertical centerline of the inclined cuboidal inhomogeneity, $x_{3} \in [-0.4, 0.6]$ m. The bimaterial domain is subjected to far-field stress $\sigma^{0}_{33} = 10^4$ Pa. The labels FEM and EIM refer to results predicted by the finite element method and Eshelby's EIM.}
    \label{fig:inhomogeneity}
\end{figure}

Figs. \ref{fig:inhomogeneity} (a) and (b) compare normal stress $\sigma_{11}$, $\sigma_{33}$ evaluated by the FEM and Eshelby's EIM with uniform eigenstrain approximation, respectively. The EIM can capture the overall stress magnitude and spatial distribution well, except for field points in the neighborhood of the vertices of the cuboidal inhomogeneity. Rodin \cite{Rodin1996} has demonstrated that Eshelby's tensors for cuboidal inclusions, even in the full space, are not constant for interior field points. Therefore, using the uniform eigenstrain only attempts to approximate the actual distribution of eigenstrain, which deviates from the actual stresses. Specifically, obvious discrepancies can be observed in two regions: (a) field points located between the lower vertex of the inhomogeneity and the bimaterial interface; and (b) the vicinity of the upper vertex of the inhomogeneity. The former should be interpreted as an increasing bimaterial interfacial effect, as the cuboidal inhomogeneity is closer to the bimaterial interface, resulting in rapid variation of stress and the corresponding eigenstrain. The latter primarily arises from the inaccurate approximation of eigenstrain, which has been discussed in our recent work \cite{Wu2021_polyhedral}. Despite the approximation strategy, the results demonstrate that the present formulae not only serve as an exact solution for arbitrarily shaped/oriented inclusions with prescribed eigenstrains, but can also be integrated with the inclusion-based methods.

\section{Singularity for a polyhedral inclusion touching the bimaterial interface}
The numerical results in the preceding section reveal significant changes in stress magnitude near the polyhedral vertices. For inclusions with uniform eigenstrain, Rodin \cite{Rodin1996} has proved that the disturbed strain and stress exhibit a logarithmic singular distribution in the neighborhood of vertices. However, the bimaterial Green's function contains additional image harmonic, biharmonic, and two Boussinesq's displacement potentials. Therefore, it is necessary to investigate whether these interface-related parts introduce additional singularities or change the classical singularity order. 

For this purpose, the bimaterial Eshelby's tensor is decomposed into three parts: (i) the full-space contribution associated with $\Psi$ and $\Phi$, which has been analyzed by Rodin \cite{Rodin1996}; (ii) the image-potential part related to $\overline{\Psi}$ and $\overline{\Phi}$; and (iii) the Boussinesq's displacement potentials with $\Theta, \Lambda$, $\overline{\Theta}$, and $\overline{\Lambda}$, which are investigated separately. 

When the polyhedral inclusion is separated from the bimaterial interface by a finite distance, the latter two parts remain regular. To show this, let $h$ denote the minimum vertical distance from the source region to the bimaterial interface as, 
\begin{equation}
    h = \min_{\textbf{x}'\in\Omega} x_3' > 0
\end{equation}
For a field point in the upper phase, the image source $\overline{\textbf{x}}'$ and the field point $\textbf{x}$ are separated at least $h$ in the third direction. When the field point is located in the lower phase, the source region and the field point are separated at least $h$ in the third direction. Therefore, the partial derivatives of the integral function and their domain integrals remain bounded. In conclusion, when $h > 0$, the strain and stress exhibit the same singularity order as in the full-space case without the bimaterial interface. 

However, the interface-related singularities may arise when the distance $h \to 0^+$, which indicates that the faces, edges or the vertices of the polyhedral inclusion touch the bimaterial interface. According to Eshelby's tensor in Eqs. (\ref{eq:ElasticEshelbyTensors_u}) and (\ref{eq:ElasticEshelbyTensors_l}), potentially singular contributions include $\overline{\Phi}_{,ij}$, $\overline{\Psi}_{,ijkl}$, $\overline{\Theta}_{,ijk}$, $\overline{\Lambda}_{,ijkl}$, and their higher-order partial derivatives multiplied by factors $x_{3}, x_{3}^2$. Note that when the edge or vertex lies on the bimaterial interface, the partial derivatives of the original and its image potentials can be related through the reflection vector $\bm{Q}$ (see Eq. (\ref{eq:image_diff})). Therefore, it is sufficient to investigate the singular behaviors of one branch of them. 

To quantify the leading singularity, consider a vertex $\bm{v}_{JI}$ located on the bimaterial interface and define, 
\begin{equation}
    \rho_{JI} = |\textbf{x} - \bm{v}_{JI}| = \sqrt{a_I^2 + b_{JI}^2 + l_{JI}^2}, \quad (a_I, b_{JI}, l_{JI}) = \rho (n_a, n_b, n_l) = \rho \textbf{n}, \quad |\textbf{n}| = 1
    \label{eq:scale}
\end{equation}
Since the vertex lies on the bimaterial interface, $(v_{JI})_{3} = 0$, the third component of the field point $\textbf{x}$ can be written as, 
\begin{equation}
    x_{3} = -\rho_{JI}\left[ n_a \chi_I + n_b \zeta_{JI} + n_l \gamma_{JI} \right] = -\rho_{JI} \hat{x}_{3} = \mathcal{O}(\rho_{JI}), \quad |\hat{x}_{3}| \leq 1
    \label{eq:x_3}
\end{equation}
The scaling will be combined with the local orders of higher-order partial derivatives multiplied by $x_{3}, x_{3}^2$ to determine whether the related terms remain bounded or not. 

\subsection{Singularity of partial derivatives of $\overline{\Phi}$ and $\overline{\Psi}$}
\label{sec:phi_psi}
The image harmonic and biharmonic potentials can be evaluated by mirroring the source regions about the bimaterial interface. For instance, $\overline{\phi}(\textbf{x}', \textbf{x}) = \phi(\overline{\textbf{x}}', \textbf{x})$. Therefore, the closed-form formulae derived for the original potentials can be applied by mirroring the source region. 

Based on our recent work \cite{Wu2021_polyhedral}, the first-order partial derivatives of $\overline{\Phi}$ and $\overline{\Psi}$ can be expressed, 
\begin{equation}
\begin{aligned}
    \overline{\Phi}_{,j}(\textbf{x}) = -Q_{J} \sum_{I=1}^{N_I} (\xi_I^0)_{j} \sum_{J=1}^{N_{JI}} \left[ \overline{F}(a_I, b_{JI}, l_{JI}^+) - \overline{F}(a_I, b_{JI}, l_{JI}^-) \right] \\ 
    \overline{\Psi}_{,j}(\textbf{x}) = -Q_{J} \sum_{I=1}^{N_I} (\xi_I^0)_{j} \sum_{J=1}^{N_{JI}} \left[ \overline{H}(a_I, b_{JI}, l_{JI}^+) - \overline{H}(a_I, b_{JI}, l_{JI}^-) \right]
\end{aligned}
\end{equation}
where closed-form expressions are listed in \ref{sec:FH_close}, which contains the algebraic, inverse trigonometric, and inverse hyperbolic tangent parts; in particular, the algebraic and inverse trigonometric parts remain bounded after partial differentiations to construct Eshelby's tensor. However, the inverse hyperbolic tangent part can cause a logarithmic singular distribution when the field point approaches the vertex or edge of the polyhedral inclusion. Therefore, it is sufficient to retain logarithmic parts to extract the dominant singularity.

The evaluation of $\overline{\Phi}_{,ij}$ involves the first-order partial derivative of $\overline{F}$. Among the terms in $\overline{F}$, only the inverse hyperbolic tangent part can provide the logarithmic contribution, which is defined as, $\overline{F}^\dagger(a_I, b_{JI}, l_{JI}) = b_{JI} \tanh^{-1}\left[\frac{l_{JI}}{\rho_{JI}} \right]$, and its first-order partial derivatives are, 
\begin{equation}
    \begin{aligned}
    & \frac{\partial \overline{F}^{\dagger}}{\partial a_I} = -\frac{a_I b_{JI} l_{JI}}{\rho_{JI}(a_I^2 + b_{JI}^2) }, \quad\frac{\partial \overline{F}^{\dagger}}{\partial b_{JI}} = -\frac{b_{JI}^2 l_{JI}}{\rho_{JI} (a_I^2 + b_{JI}^2) } + \tanh^{-1} \left[ \frac{l_{JI}}{\rho_{JI}} \right], \quad \frac{\partial \overline{F}^{\dagger}}{\partial l_{JI}} = \frac{b_{JI}}{\rho_{JI}}
    \end{aligned}
    \label{eq:phi_leading_order}
\end{equation}
Only $\frac{\partial \overline{F}^{\dagger}}{\partial b_{JI}}$ contains the inverse hyperbolic tangent function. First, to evaluate the edge limit, this subsection defines $R_{JI} = \sqrt{a_I^2 + b_{JI}^2}$. For fixed $l_{JI} \neq 0$ and $R_{JI} \to 0^+$, $\tanh^{-1} \left[ \frac{l_{JI}}{\sqrt{R_{JI}^2 + l_{JI}^2}} \right] = \tanh^{-1} \left[ \text{sgn}(l_{JI}) \right] + \mathcal{O}(R_{JI}^2)$, where $\text{sgn}(.)$ refers to the sign function. Although it may lead to the logarithmic singularity, the domain integrals should be evaluated as the difference between two edge endpoints. Therefore, the logarithmic part can be written as, $-[\text{sgn}(l_{JI}^+) -\text{sgn}(l_{JI}^-)] \ln R_{JI}$. (i) When $l_{JI}^+$ and $l_{JI}^-$ have the same signs, the divergent logarithmic terms cancel. This scenario indicates that a field point lies on the extension of the edge. (ii) When $l_{JI}^+$ and $l_{JI}^-$ have the different signs, the logarithmic terms remain. This case corresponds to the field point located on the edge between two endpoints. For the image contribution, it is possible only when the edge lies on the bimaterial interface. 

The vertex limit requires a slightly different interpretation. Consider the field point approaching one endpoint of the edge, i.e., $\textbf{x}$ approaching $\bm{v}_{JI}^+$. Using the scaling in Eq. (\ref{eq:scale}), the first-order partial derivatives in Eq. (\ref{eq:phi_leading_order}) remain bounded, 
\begin{equation}
\begin{split}
    & \frac{\partial \overline{F}^\dagger}{\partial a_I} = -\frac{n_a n_b n_l }{(n_a^2 + n_b^2) }, \quad \frac{\partial \overline{F}^\dagger}{\partial b_{JI}} = -\frac{n_b^2 n_l}{n_a^2 + n_b^2} + \tanh^{-1} \left[ n_{l} \right], \quad \frac{\partial \overline{F}^\dagger}{\partial l_{JI}} = n_b
\end{split}
\end{equation}
However, the complete domain integrals require to evaluate contributions from both endpoints. Thus, the contribution associated with the field point approaching the vertex remains bounded. Because $l_{JI}^+ \to 0$ and $l_{JI}^- \neq 0$, as Eq. (\ref{eq:phi_leading_order}) indicates, the contribution associated with the opposite endpoint exhibits same logarithmic singularity as $\tanh^{-1} \left[ \text{sgn}(l_{JI}^-) \right]$, which generates the logarithmic singularity of the complete endpoint difference. 

Note that the logarithmic part of the edge integral satisfies, $\Delta \overline{F}^\dagger = \overline{F}^\dagger(a_I, b_{JI}, l_{JI}^+) - \overline{F}^\dagger(a_I, b_{JI}, l_{JI}^-) = \mathcal{O}(\rho_{JI} \ln \rho_{JI})$. Each additional partial derivative can increase the singularity at most one order. Therefore, the dominant singularity orders are, $\overline{\Phi}_{,ij} = \mathcal{O}(\ln \rho_{JI})$, $\overline{\Phi}_{,ijk} = \mathcal{O}(\rho_{JI}^{-1})$, $\overline{\Phi}_{,ijkl} = \mathcal{O}(\rho_{JI}^{-2})$. As Eq. (\ref{eq:x_3}) indicates, when the vertex lies on the bimaterial interface, $x_{3} = \mathcal{O}(\rho_{JI})$. Hence, the dominant singularity orders are, 
\begin{equation}
    \overline{\Phi}_{,ij} = \mathcal{O}(\ln \rho_{JI}), \quad x_{3} \overline{\Phi}_{,ijk} = \mathcal{O}(1), \quad x_{3}^2 \overline{\Phi}_{,ijkl} = \mathcal{O}(1)
    \label{eq:Phi_sum}
\end{equation}

Consequently, among the three contributions considered in this subsection, only $\overline{\Phi}_{,ij}$ can generate the logarithmic singularity. The other two terms, $x_{3} \overline{\Phi}_{,ijk}$ and $x_{3}^2 \overline{\Phi}_{,ijkl}$, remain bounded, and they do not contribute to any additional singularity.

The evaluation of $\overline{\Psi}_{,ijkl}$ involves the third-order partial derivative of $\overline{H}$. Among the terms in $\overline{H}$, only the inverse hyperbolic tangent part can provide the logarithmic contribution, which is defined as, $\overline{H}^\dagger(a_I, b_{JI}, l_{JI}) = \frac{b_{JI}}{6} (3 a_I^2 + b_{JI}^2) \tanh^{-1}\left[\frac{l_{JI}}{\rho_{JI}} \right]$, and its third-order partial derivatives are listed in \ref{sec:der_psi}. Among them, only the $\frac{\partial^3 \overline{H}^\dagger}{\partial a_I^2 b_{JI}}$ and $\frac{\partial^3 \overline{H}^\dagger}{\partial b_{JI}^3}$ contain the logarithmic part. Let $\left(.\right)_{\text{log}}$ denote the logarithmic part, 
\begin{equation}
\begin{aligned}
    \left(\frac{\partial^3 \overline{H}^\dagger}{\partial a_I^2 b_{JI}}\right)_\text{log} = \tanh^{-1} \left[ \frac{l_{JI}}{\rho_{JI}} \right], \quad \left(\frac{\partial^3 \overline{H}^\dagger}{\partial b_{JI}^3}\right)_\text{log} = \tanh^{-1} \left[ \frac{l_{JI}}{\rho_{JI}} \right]
\end{aligned}
\label{eq:psi_leading_order}
\end{equation}
The third-order partial derivatives of $\overline{H}$ and the first-order partial derivative of $\overline{F}$ contain the same elementary logarithmic term. Hence, the same edge and endpoint analysis for $\overline{\Phi}_{,ij}$ applies. Consequently, $\overline{\Psi}_{,ijkl}$ exhibits the same singularity order as $\overline{\Phi}_{,ij}$. Moreover, since the edge primitive $\overline{H}^\dagger = \mathcal{O}(\rho_{JI}^3 \ln \rho_{JI})$ near a vertex, the dominant singularity for $\overline{\Psi}_{,ijkls} = \mathcal{O}(\rho_{JI}^{-1})$. Hence, the dominant singularity orders are, 
\begin{equation}
    \overline{\Psi}_{,ijkl} = \mathcal{O}(\ln \rho_{JI}), \quad x_{3} \overline{\Psi}_{,ijkls} = \mathcal{O}(1)
\end{equation}

Consequently, between the two contributions considered in this subsection, only $\overline{\Psi}_{,ijkl}$ can contribute to the logarithmic singularity. The other term, $x_{3} \overline{\Psi}_{,ijkls}$, remains bounded, and it does not contribute to any additional singularity.

\subsection{Singularity analysis of partial derivatives of $\overline{\Theta}$}

The evaluation of $\overline{\Theta}_{,ijk}$ requires the second-order partial derivatives of the surface integral $\Xi^\alpha_{I}$. As demonstrated in the preceding subsection, only terms containing an explicit logarithmic term contribute to the dominant singular behavior after second-order partial differentiation. Therefore, this subsection extracts the logarithmic parts of each second-order partial derivative of $\Xi^\alpha_{I}$. 

For $|\gamma_{JI}| < 1$, these logarithmic contributions are provided in \ref{sec:alpha_derivative}. For $R_{JI} \to 0^+$ and $l_{JI} \neq 0$, $\rho_{JI} = |l_{JI}|$, 
the dominant singularity order of edge limits are, 
\begin{equation}
\begin{aligned}
    \left(\frac{\partial^2 \Xi_{I}^{\alpha \dagger}}{\partial a_I^2} \right)_\text{log} & = -\frac{\zeta_{JI}}{1 - \gamma_{JI}} \ln [l_{JI} + |\rho_{JI}|] + \mathcal{O}(1) \notag \\  
    \left(\frac{\partial^2 \Xi_{I}^{\alpha \dagger}}{\partial a_I \partial b_{JI}} \right)_\text{log} & = \frac{\chi_I}{1 - \gamma_{JI}} \ln [l_{JI} + |l_{JI}|] + \mathcal{O}(1) \notag \\
    \left( \frac{\partial^2 \Xi_{I}^{\alpha \dagger}} {\partial b_{JI}^2} \right)_{\mathrm{log}} &= \frac{\zeta_{JI}}{1 - \gamma_{JI}} \ln [l_{JI} + |\rho_{JI}|] + \mathcal{O}(1) \\
    \left(\frac{\partial^2 \Xi_{I}^{\alpha \dagger}}{\partial a_I \partial l_{JI}} \right)_\text{log} & = \mathcal{O}(1), \quad \left(\frac{\partial^2 \Xi_{I}^{\alpha \dagger}}{\partial b_{JI} \partial l_{JI}} \right)_\text{log} = \mathcal{O}(1), \quad \left(\frac{\partial^2 \Xi_{I}^{\alpha \dagger}}{\partial l^2_{JI}} \right)_\text{log} = \mathcal{O}(1) 
\end{aligned}
\end{equation}
When the field point approaches the edge, $\ln [l_{JI} + |l_{JI}|]$ may generate the logarithmic singularity. Specifically, (i) When $l_{JI}^- > 0$ and $l_{JI}^+ > 0$, both endpoint primitives remain bounded; (ii) when $l_{JI}^- < 0$ and $l_{JI}^+ < 0$, both endpoint primitives contain the same logarithmic term $\ln [2 R_{JI}]$, which cancels in their difference; and (iii) when $l_{JI}^- l_{JI}^+ < 0$, only the endpoint with $l_{JI} < 0$ contains the divergent part, and therefore it results in $\mathcal{O}(\ln R_{JI})$ singularity. The cases (i) and (ii) indicate that the field point lies on the extension of the edge, while the case (iii) requires the field point is within the two endpoints of the edge, which lies on the bimaterial interface. Note that when $\gamma_{JI} = \pm 1$, the edge must be perpendicular to the bimaterial interface, therefore, the edge limit for $\gamma_{JI} = \pm 1$ does not involve any singularity. 

Subsequently, when the field point approaches the vertex, $\rho_{JI} \to 0^+$, $\ln [l_{JI} + \rho_{JI}] = \ln \rho_{JI} + \ln[1 + n_l]$, and $\ln [\rho_{JI} (1 + \hat{x}_3)] = \ln \rho_{JI} + \ln [1 + \hat{x}_{3}]$. The dominant singularity order of the second-order partial derivatives can be rewritten as, 
\begin{equation}
\begin{aligned}
    \left(\frac{\partial^2 \Xi_{I}^{\alpha \dagger}}{\partial a_I^2} \right)_\text{log} & = \left[ n_b (1 - n_a^2) - \frac{\zeta_{JI}}{1 - \gamma_{JI}^2} \right] \ln \rho_{JI} + \mathcal{O}(1)\\ 
    \left(\frac{\partial^2 \Xi_{I}^{\alpha \dagger}}{\partial a_I \partial b_{JI}} \right)_\text{log} & = \left[ n_a (1 - n_b^2) + \frac{\chi_I}{1 - \gamma_{JI}^2} \right] \ln \rho_{JI} + \mathcal{O}(1) \\ 
    \left(\frac{\partial^2 \Xi_{I}^{\alpha \dagger}}{\partial a_I \partial l_{JI}} \right)_\text{log} & = -\frac{n_a n_b n_l}{2} \ln \rho_{JI} + \mathcal{O}(1) \\
    \left(\frac{\partial^2 \Xi_{I}^{\alpha \dagger}}{\partial b^2_{JI}} \right)_\text{log} & = \left[ n_b (3 - n_b^2) + \frac{\zeta_{JI}}{1 + \gamma_{JI}} \right] \ln \rho_{JI}  + \mathcal{O}(1) \\ 
    \left(\frac{\partial^2 \Xi_{I}^{\alpha \dagger}}{\partial b_{JI} \partial l_{JI}} \right)_\text{log} & = \frac{n_l (1 - n_b^2)}{2} \ln \rho_{JI} + \mathcal{O}(1) \\ 
    \left(\frac{\partial^2 \Xi_{I}^{\alpha \dagger}}{\partial l^2_{JI}} \right)_\text{log} & = \frac{n_b (1 - n_l^2)}{2} \ln \rho_{JI} + \mathcal{O}(1)
\end{aligned}
    \label{eq:sing_alpha}
\end{equation}
It demonstrates that the second-order partial derivatives can generate $\ln \rho_{JI}$ singularities, but some of the coefficients can vanish. For instance, when $\zeta_{JI} = 0$ and $n_b = 0$, the coefficient of $\left(\frac{\partial^2 \Xi_{I}^{\alpha \dagger}}{\partial b^2_{JI}} \right)_\text{log}$ is zero. 

For the special case $\gamma_{JI} = \pm 1$, when the field point approaches the vertex, only the mixed second-order partial derivative with respect to $b_{JI}$ and $l_{JI}$ contains the singular part, which can be written as, ($\rho_{JI} \to 0^+$)
\begin{equation}
    \begin{aligned}
    \left(\frac{\partial^2 \Xi_{I}^{\alpha \dagger}}{\partial b_{JI} \partial l_{JI}} \right)_\text{log} = \frac{\text{sgn}[\gamma_{JI}]}{2} \ln \rho_{JI} + \mathcal{O}(1)
    \end{aligned}
\end{equation}

Based on the six components of the second-order partial derivatives, $\Theta_{,ijk}$ and $\overline{\Theta}_{,ijk}$ can exhibit logarithmic singularity when the field point is on the edge or the polyhedral vertices. Because the third-order partial derivatives of $\overline{\Theta}$ or $\Theta$ exhibit the logarithmic singularity, their subsequent fourth-order partial derivatives are at most $\rho_{JI}^{-1}$ singularity. Therefore, the dominant singularity orders for terms associated with $\overline{\Theta}$ are, 
\begin{equation}
    \overline{\Theta}_{,ijk} = \mathcal{O}(\ln \rho_{JI}) + \mathcal{O}(1), \quad x_{3} \overline{\Theta}_{,ijkl} = \mathcal{O}(1)
    \label{eq:Theta_sum}
\end{equation}

\subsection{Singularity analysis of partial derivatives of $\overline{\Lambda}$}

The evaluation of $\overline{\Lambda}_{,ijkl}$ requires the third-order partial derivatives of the surface integral $\Xi_I^\beta$. Following the same procedure in the preceding subsection, this subsection extracts the logarithmic parts of each third-order partial derivative of $\Xi_I^\beta$. 

For $|\gamma_{JI}| < 1$, these logarithmic contributions are provided in \ref{sec:beta_derivative}. For $R_{JI} \to 0^+$ and $l_{JI} \neq 0$, the dominant singularity order of edge limits are, 
\begin{equation}
\begin{aligned}
    \left( \frac{\partial^3 \Xi_{I}^{\beta \dagger}}{\partial a_I^3} \right)_{\text{log}} & = \frac{-2 (2 + \gamma_{JI}^2) \chi_I \zeta_{JI}}{(1 - \gamma_{JI}^2)^2} \ln \left[ l_{JI} + |l_{JI}| \right] \\ 
    \left( \frac{\partial^3 \Xi_{I}^{\beta \dagger}} {\partial a_I^2 \partial b_{JI}} \right)_{\text{log}} & = \frac{2}{3} \ln \left[ |l_{JI}| -l_{JI} \right] + \frac{\ln \left[ l_{JI} + |l_{JI}| \right]}{3 (1 - \gamma_{JI}^2 )^2} \left[ \gamma_{JI}^4 - \gamma_{JI}^2 (2 \chi_I^2 + 6 \zeta_{JI}^2 + 7) + 4 \chi_I^2 + 6 (1 - \zeta_{JI}^2) \right] \\
    \left( \frac{\partial^3 \Xi_{I}^{\beta \dagger}}{\partial a_I \partial b_{JI}^2} \right)_{\text{log}} & = \chi_I \frac{2 \zeta_{JI} (4 + \gamma_{JI}^2)}{3 (1 - \gamma_{JI}^2)^2} \ln [l_{JI} + \rho_{JI}] \\ 
    \left( \frac{\partial^3 \Xi_{I}^{\beta \dagger}}{\partial b^3_{JI}} \right)_{\text{log}} & = \frac{2 \zeta_{JI}^2 - \gamma_{JI}^2 + \gamma_{JI}^4}{(1 - \gamma_{JI}^2 )^2} \ln [l_{JI} + \rho_{JI}] \\ 
    \left( \frac{\partial^3 \Xi_{I}^{\beta \dagger}}{\partial a_I^2 \partial l_{JI}} \right)_{\text{log}} & = \mathcal{O}(1), \quad \left( \frac{\partial^3 \Xi_{I}^{\beta \dagger}}{\partial a_I \partial b_{JI} \partial l_{JI}} \right)_{\text{log}} = \mathcal{O}(1), \quad  \left( \frac{\partial^3 \Xi_{I}^{\beta \dagger}}{\partial a_I \partial l^2_{JI}} \right)_{\text{log}} = \mathcal{O}(1) \\
    \left( \frac{\partial^3 \Xi_{I}^{\beta \dagger}}{\partial b^2_{JI} \partial l_{JI}} \right)_{\text{log}} & = \mathcal{O}(1), \quad \left( \frac{\partial^3 \Xi_{I}^{\beta \dagger}}{\partial b_{JI} \partial l^2_{JI}} \right)_{\text{log}} = \mathcal{O}(1), \quad  \left( \frac{\partial^3 \Xi_{I}^{\beta \dagger}}{\partial l^3_{JI}} \right)_{\text{log}} = \mathcal{O}(1)
    \end{aligned}
\end{equation}
When the field point approaches the edge, the partial derivatives contain two types of logarithmic terms, $\ln [l_{JI} + |l_{JI}|]$ and $\ln [|l_{JI}|- l_{JI}]$. In analogy with $\ln [l_{JI} + |l_{JI}|]$, the same conclusion directly applies to $\ln [|l_{JI}| - l_{JI}]$.  The previous subsection has demonstrated that the occurrence of the logarithmic singularity requires that the field point be located between two endpoints and the edge lies on the bimaterial interface. 

Subsequently, when the field point approaches the vertex, $\rho_{JI} \to 0^+$, $\ln [l_{JI} + \rho_{JI}] = \ln \rho_{JI} + \ln[1 + n_l]$, $\ln [\rho_{JI} - l_{JI}] = \ln \rho_{JI} + \ln [1 - n_l]$, and $\ln [\rho_{JI} (1 + \hat{x}_3)] = \ln \rho_{JI} + \ln [1 + \hat{x}_{3}]$. The dominant singularity order of the third-order partial derivatives can be rewritten as, 
\begin{align}
    \left( \frac{\partial^3 \Xi_{I}^{\beta \dagger}}{\partial a_I^3} \right)_{\text{log}} & = \left( \frac{n_b (1 - n_a^2)^2}{2} - \frac{\zeta_{JI} (4 + \gamma_{JI})}{\left(1 + \gamma_{JI}\right)^2} \right) \ln \rho_{JI} + \mathcal{O}(1) \notag \\ 
    \left( \frac{\partial^3 \Xi_{I}^{\beta \dagger}}{\partial a_I^2 \partial b_{JI}} \right)_{\text{log}} & = \frac{1}{6 (1 - \gamma_{JI}^2)^2} \left[ 2 \left(8 + \gamma_{JI} [11 + 3 \gamma_{JI}] \right) + n_a \chi_I (1 + \gamma_{JI})^2 \right. \notag \\ & \left. \qquad \left( 2 n_a^4 + (5 n_a^2 - 3 n_l^2) (1 - n_a^2) \right) + 2 \chi_I^2 (4 + 3 \gamma_{JI}) \right]  \ln \rho_{JI} + \mathcal{O}(1) \notag  \\
    \left( \frac{\partial^3 \Xi_{I}^{\beta \dagger}}{\partial a_I^2 \partial l_{JI}} \right)_{\text{log}} & = -\frac{n_a n_b n_l (1 - n_a^2) \chi_I}{2} \ln \rho_{JI} + \mathcal{O}(1) \notag \\
    \left( \frac{\partial^3 \Xi_{I}^{\beta \dagger}}{\partial a_I \partial b_{JI}^2} \right)_{\text{log}} & = \frac{\chi_I}{6} \left[ n_b \left( (1 - n_a^2) (2 n_b^2 + 3 n_l^2) + n_a^2 (5 n_b^2 + 3 n_l^2) \right) + \frac{2 \zeta_{JI} (8 + \gamma_{JI})}{[1 + \gamma_{JI}]^2} \right] \ln \rho_{JI} + \mathcal{O}(1) \notag  \\
    \left( \frac{\partial^3 \Xi_{I}^{\beta \dagger}}{\partial a_I \partial b_{JI} \partial l_{JI}} \right)_{\text{log}} & = \frac{\chi_I}{6} \left[ 3 n_a^2 n_b^2 n_l + n_a^2 n_l^3 + n_b^2 n_l^3 + n_l^5 + 2  \right] \ln \rho_{JI} + \mathcal{O}(1)  \\ 
    \left( \frac{\partial^3 \Xi_{I}^{\beta \dagger}}{\partial a_I \partial l^2_{JI}} \right)_{\text{log}} & = \frac{n_b \chi_I}{6} \left[ n_b^2 (n_b^2 + n_l^2) + n_a^2 (n_b^2 + 3 n_l^2) \right]  \ln \rho_{JI} + \mathcal{O}(1) \notag \\ 
    \left( \frac{\partial^3 \Xi_{I}^{\beta \dagger}}{\partial b^3_{JI}} \right)_{\text{log}} & = \left( \frac{\gamma_{JI} ( 1 + \gamma_{JI}) + (2 + \gamma_{JI}) \zeta_{JI}^2}{[1 + \gamma_{JI}]^2} + \frac{n_a (1 - n_l^2)^2 \chi_I }{2} \right) \ln \rho_{JI} + \mathcal{O}(1) \notag \\ 
    \left( \frac{\partial^3 \Xi_{I}^{\beta \dagger}}{\partial b^2_{JI} \partial l_{JI}} \right)_{\text{log}} & = \left( \frac{2}{3} \zeta_{JI} -n_a n_b n_l (1 - n_b^2) \chi_I \right) \ln \rho_{JI} + \mathcal{O}(1) \notag \\
    \left( \frac{\partial^3 \Xi_{I}^{\beta \dagger}}{\partial b_{JI} \partial l^2_{JI}} \right)_{\text{log}} & = \left( \frac{\gamma_{JI}}{3} + n_a \left[ n_a^4 + 3 n_b^2 n_l^2 + n_a^2 (n_b^2 + n_l^2) \right] \chi_I \right) \ln \rho_{JI} + \mathcal{O}(1) \notag \\
    \left( \frac{\partial^3 \Xi_{I}^{\beta \dagger}}{\partial l^3_{JI}} \right)_{\text{log}} & = -\frac{n_a n_b n_l (1 - n_l^2) \chi_I }{2} \ln \rho_{JI} + \mathcal{O}(1) \notag
\end{align}
which exhibit logarithmic singularities. For the special case $\gamma_{JI} = \pm 1$, when the field point approaches the vertex, only three components of the third-order partial derivatives may cause logarithmic singularity, ($\rho_{JI} \to 0^+$)
\begin{equation}
    \begin{aligned}
    & \left(\frac{\partial^3 \Xi_{I}^{\beta \dagger}}{\partial a_I^2 \partial b_{JI}} \right)_\text{log} = \left( \frac{2}{3} + \frac{\text{sgn}[\gamma_{JI}]}{6} \right) \ln \rho_{JI} + \mathcal{O}(1) \\ 
    & \left(\frac{\partial^3 \Xi_{I}^{\beta \dagger}}{\partial b_{JI} \partial l_{JI}^2} \right)_\text{log} = \frac{\text{sgn}[\gamma_{JI}]}{2} \ln \rho_{JI} + \mathcal{O}(1), \qquad \left(\frac{\partial^3 \Xi_{I}^{\beta \dagger}}{\partial b^3_{JI}} \right)_\text{log} = \frac{\text{sgn}[\gamma_{JI}]}{3} \ln \rho_{JI} + \mathcal{O}(1)
    \end{aligned}
\end{equation}

Based on the ten components of the third-order partial derivatives, $\Lambda_{,ijkl}$ and $\overline{\Lambda}_{,ijkl}$ exhibit a logarithmic singularity when the field point is on the edge or the polyhedral vertices. Therefore, the image potentials and two Boussinesq's displacement potentials will not contribute to any higher-order singularity. 

\subsection{Summary of singularity cases}
The singularity orders investigated in the preceding subsections are summarized in Table  \ref{tab:sing}. When the polyhedral inclusion is separated from the bimaterial interface by a finite distance, the singularity structure remains the same as the full-space case, which is reported by Rodin \cite{Rodin1996}. Additional logarithmic singularities can only arise when an edge or vertex lies on the bimaterial interface. Although some higher-order partial derivatives may exhibit higher singularities, their multiplication by factors $x_{3}$ or $x_{3}^2$ remain bounded. The vector $\bm{Q}$ only changes the sign but it does not alter the singularity order. 

\begin{table}[htbp]
\centering
\renewcommand{\arraystretch}{1.35}
\caption{Summary of the singularity orders of the bimaterial contribution to elastic Eshelby's tensor.}
\label{tab:sing}
\begin{tabular}{C{0.35\textwidth} C{0.15\textwidth} C{0.18\textwidth} C{0.18\textwidth}}
\hline
Terms in Eshelby's tensor& $h>0$ & Interfacial edge & Interfacial vertex \\
\hline
$\overline{\Phi}_{,ij}$ & $\mathcal{O}(1)$ & $\mathcal{O}(\ln R_{JI})$ & $\mathcal{O}(\ln \rho_{JI})$ \\
$x_3\overline{\Phi}_{,ijk}$ & $\mathcal{O}(1)$ & $\mathcal{O}(1)$ & $\mathcal{O}(1)$ \\
$x_3^2\overline{\Phi}_{,ijkl}$ & $\mathcal{O}(1)$ & $\mathcal{O}(1)$ & $\mathcal{O}(1)$ \\
\hline
$\overline{\Psi}_{,ijkl}$ & $\mathcal{O}(1)$ & $\mathcal{O}(\ln R_{JI})$ & $\mathcal{O}(\ln \rho_{JI})$ \\
$x_3\overline{\Psi}_{,ijklm}$ & $\mathcal{O}(1)$ & $\mathcal{O}(1)$ & $\mathcal{O}(1)$ \\
\hline
$\Theta_{,ijk},\ \overline{\Theta}_{,ijk}$ & $\mathcal{O}(1)$ & $\mathcal{O}(\ln R_{JI})$ & $\mathcal{O}(\ln \rho_{JI})$ \\
$x_3\Theta_{,ijkl}, \thinspace x_3\overline{\Theta}_{,ijkl}$ & $\mathcal{O}(1)$ & $\mathcal{O}(1)$ & $\mathcal{O}(1)$ \\
\hline
$\Lambda_{,ijkl},\ \overline{\Lambda}_{,ijkl}$ & $\mathcal{O}(1)$ & $\mathcal{O}(\ln R_{JI})$ & $\mathcal{O}(\ln \rho_{JI})$ \\
\hline
\end{tabular}
\end{table}

\section{Conclusions}
This paper derives closed-form domain integrals of two Boussinesq's displacement potentials of polyhedral inclusions embedded in the three-dimensional bimaterial space. Using integration by parts and auxiliary potential functions, the original volume integrals have been reduced to simple surface integrals and elementary line integrals. The closed-form formulae apply to general polyhedral inclusions for all components of a general uniform eigenstrain, which breaks the previous limitations on dilatational eigenstrain or cuboids parallel to the bimaterial interface. The closed-form formulae are verified against classic solutions in the literature, including spheres and cuboids parallel to the bimaterial interface, as well as numerical solutions of an inclined cuboid by finite element analysis. With the complete bimaterial elastic Eshelby's tensor, the solution has been further applied in the equivalent inclusion method, which exhibits good agreement except in the vicinity of cuboid vertices. The singular behaviors of the closed-form formulae have been investigated. It is shown that the logarithmic singularity arises in the higher-order derivatives, such as the second-order partial derivatives of the harmonic potential, the third-order partial derivatives of the first Boussinesq's displacement potential, and the fourth-order partial derivatives of the biharmonic and the second Boussinesq's displacement potentials. When an edge or vertex of the polyhedral inclusion touches the bimaterial interface, these interface-related components will lead to additional logarithmic singularities. However, when the polyhedral inclusion is away from the bimaterial interface, the singularity is only caused by full-space components. The proposed formulae provide an exact and explicit tool for modeling polyhedral inclusions, defects, and residual strain in the presence of material interfaces.

\section*{CRediT Author Statement}
\textbf{Chunlin Wu}: Conceptualization, Methodology, Data Curation, Validation, Visualization, Writing-original draft, Funding Acquisition; \textbf{Huiming Yin}: Conceptualization, Writing Review \& Editing. 

\section*{Acknowledgment}
CW's work is supported by National Natural Science Foundation of China Grant No. 12302086. HY's work was sponsored by the US Department of Agriculture NIFA \#2021-67021-34201 and the National Science Foundation (NSF) (IIP \#1738802 and IIP \#1941244). These supports are gratefully acknowledged.
          
\appendix 

\section{Bimaterial thermoelastic Green's function and Eshelby's tensor}
\label{sec:thermoelastic}

Analogous to the elastic Green's function that provides the displacement field caused by a point force, thermoelastic Green's function describes the displacement field induced by a point heat source as follows \cite{Wu2023}:  
\begin{equation}
    G_{i} (\textbf{x}, \textbf{x}') = \frac{1}{2 \mu'}
    \begin{cases}
    \begin{aligned}
        & \mathcal{H}_5 \psi_{,i} + (\mathcal{H}_1 + L_B) \overline{\beta}_{,i} + L_B \overline{\psi}_{,i} \\ & + x_3 \Big[ \big( L_D - L_F \big) \overline{\alpha}_{,i} + L_C \big( \overline{\psi}_{,i3} + 2(1 - 2\nu') \delta_{i3} \overline{\phi} - x_3 \overline{\phi}_{,i} \big) \Big] \\ & + \delta_{i3} \big[- (3 - 4\nu') \big( \mathcal{H}_3 \overline{\alpha} + L_C \overline{\psi}_{,3} \big) + \big(4 (1-\nu') \mathcal{H}_6 - L_B \big) \overline{\alpha} \big]   
    \end{aligned}
     & x_3 \geq 0 \\
     \\
     \begin{aligned}
    & (\mathcal{H}_2 + L_G) \beta_{,i} - x_3 \Big[ \mathcal{H}_4 - L_G \Big] \alpha_{,i} + L_G \psi_{,i} \\ & - \delta_{i3} \alpha \Big[ L_G + (3-4\nu') \mathcal{H}_4 - 4(1-\nu') \mathcal{H}_7 \Big] \end{aligned} & x_3 < 0
    \end{cases}
    \label{eq:thermoelastic_Green}
\end{equation}
which is also expressed in terms of harmonic, biharmonic, and two Boussinesq's displacement potentials. The material coefficients $\mathcal{H}_1 \sim \mathcal{H}_7$ and $L_B \sim L_F$ are provided as follows:
\begin{equation}
    \begin{aligned}
     \mathcal{H}_1 = \mathcal{H}_5 - 2\mu' \big[ \frac{ \mathcal{H}_7 (1 - \nu'')}{(3 - 4\nu'')\mu' + \mu''} &+ \frac{(\mathcal{H}_5 + \mathcal{H}_6)(1 - \nu')}{(3 - 4\nu') \mu'' + \mu'} \big]\\ 
    \mathcal{H}_2 = -2\mu'' \big[ \frac{\mathcal{H}_7 (1-\nu'')}{(3-4\nu'') \mu' + \mu''} & + \frac{(\mathcal{H}_5 + \mathcal{H}_6) (1 - \nu')}{(3 - 4\nu') \mu'' + \mu'} \big] \\
     \mathcal{H}_3 =   \frac{4 \mathcal{H}_6 (1 - \nu') \mu'' - \mathcal{H}_5(\mu' - \mu'')}{(3 - 4\nu') \mu'' + \mu'}, & \quad \mathcal{H}_4 = 4 \mathcal{H}_7 \frac{(1 - \nu'') \mu'}{(3 - 4\nu'') \mu' + \mu''} \\
      \mathcal{H}_5 = \frac{1}{8 \pi K'} \frac{(1-2v')\mathcal{A}'}{1-v'}, \quad \mathcal{H}_6 = \frac{1}{8\pi K'} \frac{(1-2v') \mathcal{A}'}{1-v'} & \frac{K' - K''}{K' + K''}, \quad \mathcal{H}_7 = \frac{1}{4\pi (K' + K'')} \frac{(1-2v'')\mathcal{A}''}{1-v''} 
    \end{aligned}
    \label{eq:mathcal_H}
\end{equation}
\begin{equation}
\begin{aligned}
    L_B = \mathcal{H}_5 \frac{(3 - 4\nu')(\mu' - \mu'')}{(3 - 4\nu') \mu'' + \mu'}, \quad L_C = 2\mathcal{H}_5 \frac{(\mu' - \mu'')}{(3 - 4\nu') \mu'' + \mu'}\\
    L_D = \frac{4 \mathcal{H}_6 (1-\nu') \mu''}{(3-4\nu') \mu'' + \mu'}, \quad L_F = \frac{4 \mathcal{H}_5 (\mu' - \mu'') (1-\nu')}{(3-4\nu') \mu'' + \mu'} \notag
\end{aligned}
\label{eq:thermoelastic_coeff}
\end{equation}

Given a uniform eigen-temperature gradient (ETG), the thermoelastic Eshelby's tensors can be similarly defined by 
\begin{equation}
    \begin{aligned}
    \varepsilon_{ij}' = \int_{\Omega} \frac{K(\textbf{x}')}{2} \left[ G_{i,k'j}(\textbf{x}, \textbf{x}') + G_{j,k'i}(\textbf{x}, \textbf{x}') \right] \thinspace dV(\textbf{x}') \thinspace T_{k}^* = R_{ijk} T{k}^*
    \end{aligned}
    \label{eq:disturbed_strain_thermo}
\end{equation}
where they can be written in terms of the above integrals:
\begin{equation}
    R_{ijk} = \frac{K'}{4 \mu'} \begin{cases}
    \begin{aligned}
    & -2 \mathcal{H}_{5} \Psi_{,ijk} - 2 (\mathcal{H}_{1} + L_{B}) Q_K \overline{\Lambda}_{,ijk} - 2 L_B Q_K \overline{\Psi}_{,ijk} \\ 
    & - Q_K (L_D - L_F) (\delta_{j3} \overline{\Theta}_{,ik} + \delta_{i3} \overline{\Theta}_{,jk} + 2 x_3 \overline{\Theta}_{,ijk}) \\
    & - Q_K L_C \left[ \delta_{j3} \overline{\Psi}_{,i3k} + \delta_{i3} \overline{\Psi}_{,j3k} + 2 x_{3} \overline{\Psi}_{,ij3k} + 4 (1 - 2 \nu') \delta_{i3} \delta_{j3} \overline{\Phi}_{,k} \right. \\ & \left. \quad - 4 \nu' x_{3} (\delta_{i3} \overline{\Phi}_{,jk} + \delta_{j3} \overline{\Phi}_{,ik}) - 2 x_{3}^2 \overline{\Phi}_{,ijk} \right] - Q_K \left\{ \left[4 (1 - \nu') \mathcal{H}_6 - (3 - 4\nu') \mathcal{H}_{3} - L_B \right] \right. \\ & \left. \quad \times (\delta_{i3} \overline{\Theta}_{,jk} + \delta_{j3} \overline{\Theta}_{,ik}) - (3 - 4\nu') L_C (\delta_{i3} \overline{\Psi}_{,3jk} + \delta_{j3} \overline{\Psi}_{,3ik}) \right\}
    \end{aligned}
    & x_{3} \geq 0 \\
    \\
    \begin{aligned}
    & -2 \left[ (\mathcal{H}_{2} + L_G) \Lambda_{,ijk} + L_{G} \Psi_{,ijk} - x_{3} (\mathcal{H}_{4} - L_G) \Theta_{,ijk} \right. \\ & \left. \quad - 2 (1 - \nu') (\mathcal{H}_{4} - \mathcal{H}_7) (\delta_{i3} \Theta_{,jk} + \delta_{j3} \Theta_{,ik}) \right] \end{aligned} & x_{3} < 0 
    \end{cases}
    \label{eq:ThermoEshelbyTensors}
\end{equation}

Hence, once the closed-form domain integrals of two Boussinesq's displacement potentials are acquired, the bimaterial Eshelby's tensors can be obtained for both elastic and thermoelastic problems.

\section{Evaluation of elementary area/line integrals}
\label{sec:kernels}
The dimensional reductions in Section 3 require evaluating some elementary surface and edge integrals. This appendix section derives their closed-form formulae. Specifically, \ref{sec:KH} evaluates $\mathcal{K}_{JI}$ and $\mathcal{H}_{JI}$, which are utilized in $\Xi^\alpha_{I}$ and $\Xi^\beta_I$. \ref{sec:three_integrals} evaluates $\mathcal{A}_{I3}$, $\Phi_{I}$, and $\mathcal{L}_{JI}$, which are required for $\Xi^\beta_{I}$. \ref{sec:M_V} evaluates $\mathcal{M}_{JI}$ and $\mathcal{V}_{JI}$, which arises from the reduction of $D_{I}$. Finally, \ref{sec:FH_close} shows the integrals $\overline{F}$ and $\overline{H}$ for the singularity analysis in Section 5. 

\subsection{Evaluation of $\mathcal{K}_{JI}$ and $\mathcal{H}_{JI}$}
\label{sec:KH}
Based on the transformed coordinate, the distance and its third component can be expressed as, $r = \sqrt{a_{I}^2 + b_{JI}^2 + le^2}$ and $r_{3} = \textbf{e}_{3} \cdot \textbf{r} = a_{I} \chi_I + b_{JI} \zeta_{JI} + le \gamma_{JI}$ \cite{Wu2021_polyhedral}. Hence, the line integral in Eq. (\ref{eq:def_H_K_int}) can be written as, 
\begin{equation}
\begin{aligned}
    \mathcal{K}_{JI} = \int_{l_{JI}^-}^{l_{JI}^+} \frac{1}{\sqrt{a_I^2 + b_{JI}^2 + le^2} + a_{I} \chi_I + b_{JI} \zeta_{JI} + le \gamma_{JI}} \thinspace d \thinspace le \\ 
    \mathcal{H}_{JI} = \int_{l_{JI}^-}^{l_{JI}^+} \ln \left[ \sqrt{a_I^2 + b_{JI}^2 + le^2} + a_{I} \chi_I + b_{JI} \zeta_{JI} + le \gamma_{JI} \right] \thinspace d \thinspace le
\end{aligned}
\label{eq:line_integral_form}
\end{equation}

For more compact expressions of results, this subsection defines, 
\begin{equation}
\begin{aligned}
    R_{JI} = \sqrt{a_{I}^2 + b_{JI}^2}, \quad c_{JI} = a_{I} \chi_I + b_{JI} \zeta_{JI}, \quad t = \frac{le + r}{R_{JI}} \\ \Delta = R_{JI}^2 (1 - \gamma_{JI}^2) - c_{JI}^2, \quad
    P(t) = R_{JI} (1 + \gamma_{JI}) t^2 + 2 c_{JI} t + R_{JI} (1 - \gamma_{JI}) \\  Q(t) = \ln \left| \frac{P(t)}{2t} \right|, \quad F(t) = \frac{\tan^{-1} \left[ \frac{(1 + \gamma_{JI}) R_{JI} t + c_{JI}}{\sqrt{\Delta}} \right] }{\sqrt{\Delta}} 
\end{aligned}
    \label{eq:sym}
\end{equation}
where $\Delta = R_{JI}^2 (1 - \gamma_{JI}^2) - c_{JI}^2 = \left( a_{I} \zeta_{JI} - b_{JI} \chi_I \right)^2$ is non-negative. In particular, when $\gamma_{JI} = \pm 1$, the transformed coordinate provides $\chi_I = \zeta_{JI} = 0$, which leads to $\Delta = 0$. For these two degenerate cases, the line integrals are evaluated separately from their original integral forms, which are provided in Eq. (\ref{eq:explicit_KH_limits}). 

In the following, the line integrals are evaluated by replacing the integral variable $le$ with $t$. The replacement rationalizes the square-root dependence on the edge variable $le$. Therefore, both $r+r_{3}$ and $d le$ can be expressed in $t$, and the orientation dependence is collected into $c_{JI}$, $\gamma_{JI}$, and $\Delta$. The above transformation reduces the original integral in Eq. (\ref{eq:line_integral_form}) to elementary rational and logarithmic form, which can be evaluated as, 

\begin{equation}
    \int_{l_{JI}^-}^{l_{JI}^+} \mathcal{F}(le) \thinspace d \thinspace le = \frac{R_{JI}}{2} \int_{t_{JI}^-}^{t_{JI}^+} \mathcal{F}(t) \left( 1 + t^{-2} \right) \thinspace d \thinspace t, \qquad t_{JI}^\pm = \frac{l_{JI}^\pm + \sqrt{R_{JI}^2 + (l_{JI}^\pm)^2} }{R_{JI}}
    \label{eq:integral_sub}
\end{equation}
where $t_{JI}^\pm$ refer to the integral limits, and $d le = \frac{R_{JI}}{2} ( 1+ t^{-2}) dt$ has been used. The explicit integral results for $\mathcal{K}_{JI}$ and $\mathcal{H}_{JI}$ in Eq. (\ref{eq:line_integral_form}) are derived as below, ($\gamma_{JI} \neq \pm 1$)
\begin{equation}
    \begin{aligned}
    & \mathcal{K}_{JI} = \int_{t_{JI}^-}^{t_{JI}^+} \frac{R_{JI} \left(t + t^{-1}\right)}{R_{JI} (1 + \gamma_{JI}) t^2 + 2 c_{JI} t + R_{JI} (1 - \gamma_{JI})} \thinspace dt = K_{JI}(t_{JI}^+) - K_{JI}(t_{JI}^-) \\ & K_{JI}(t) = \frac{1}{1 - \gamma_{JI}} \ln [t] - \frac{\gamma_{JI}}{1 - \gamma_{JI}^2} \ln|P(t)| - \frac{2c_{JI}}{1 - \gamma_{JI}^2} F(t) 
    \end{aligned}
    \label{eq:explicit_K}
\end{equation}
and 
\begin{equation}
    \begin{aligned}
    & \mathcal{H}_{JI} = \int_{t_{JI}^-}^{t_{JI}^+} \ln \left[ \frac{R_{JI} (1 + \gamma_{JI}) t^2 + 2 c_{JI} t + R_{JI} (1 - \gamma_{JI})}{2t} \right] \frac{R_{JI}}{2} \left( 1 + t^{-2} \right) \thinspace dt = H_{JI}(t_{JI}^+) - H_{JI}(t_{JI}^-) \\ & H_{JI}(t) = \frac{R_{JI}}{2} \left[ \frac{2c_{JI}}{R_{JI} (1 - \gamma_{JI})} \ln[t] + (t - t^{-1}) (Q(t) - 1) - \frac{2 \gamma_{JI} c_{JI}}{R_{JI} (1 - \gamma_{JI}^2)} \ln|P(t)| + \frac{4 \Delta}{R_{JI} (1 - \gamma_{JI}^2)} F(t) \right]
    \end{aligned}
    \label{eq:explicit_H}
\end{equation}

The line integrals for the degenerate cases $\gamma_{JI} = \pm 1$ can be evaluated directly from their original integral forms, yielding, 
\begin{equation}
    K_{JI}(t) = \begin{cases} \frac{1}{2} \ln [t] - \frac{1}{4 t^2} & \gamma_{JI} = 1 \\ 
    \frac{1}{2} \ln [t] + \frac{1}{4} t^2 & \gamma_{JI} = -1
    \end{cases}, \qquad \mathcal{H}_{JI}(t) = \begin{cases} \frac{R_{JI}}{2t} \left[ (t^2 - 1) \ln[R_{JI} t] - (1 + t^2) \right] & \gamma_{JI} = 1 \\ \frac{R_{JI}}{2t} \left[ (t^2 - 1) \ln \left[ \frac{R_{JI}}{t} \right] + ( 1+ t^2) \right] & \gamma_{JI} = -1 \end{cases}
    \label{eq:explicit_KH_limits}
\end{equation}
The degenerate cases $\gamma_{JI} = \pm 1$ imply that the edges are perpendicular to the bimaterial interface. For instance, a cuboid is parallel to the bimaterial interface, and its lateral edges are perpendicular to the bimaterial interface. In such cases, $\chi_I = \zeta_{JI} = 0$ and $\Delta = 0$, so the general expressions in Eqs. (\ref{eq:explicit_H}) and (\ref{eq:explicit_K}) are more complicated, because they involve $1 - \gamma_{JI}^2$. As shown in Eq. (\ref{eq:explicit_KH_limits}), the integration results have been significantly simplified. This geometric degeneracy explains why pioneers \cite{Liu2012, Wang2016} derived compact, straightforward result formulae for cuboidal inclusions aligned with the bimaterial interface.

\subsection{Evaluation of $\mathcal{A}_{I3}, \Phi_{I}$ and $\mathcal{L}_{JI}$}
\label{sec:three_integrals}

\noindent (i) Evaluation of the surface integral: $\mathcal{A}_{I3} = \int_{S_I} r_{3} \thinspace dS(\textbf{x}')$ \\
The surface integral $\mathcal{A}_{I3}$ refers to the first surface moment of the third component $r_{3}$. Because $r_{3} = \rho \cos \theta \zeta_{JI} + \rho \sin \theta \gamma_{JI} + a_I \chi_I$ is linear in the local transformed coordinate, it can be evaluated as,  

\begin{equation}
\begin{split}
    \mathcal{A}_{I3} & = \sum_{J = 1}^{N_{JI}} \int_{\tan^{-1} \left[ \frac{l_{JI}^-}{b_{JI}} \right]}^{\tan^{-1} \left[ \frac{l_{JI}^+}{b_{JI}} \right]} \int_{0}^{b_{JI} \sqrt{1 + \tan^2 \theta}} \left[\rho \cos \theta \zeta_{JI} + \rho \sin \theta \gamma_{JI} + a_{I} \chi_I \right] \thinspace \rho d \rho \thinspace d\theta \\ 
    & = \sum_{J = 1}^{N_{JI}} \frac{1}{6} b_{JI} \left( l_{JI}^+ - l_{JI}^- \right) \left[ 3 a_{I} \chi_I + 2 b_{JI} \zeta_{JI} + (l_{JI}^+ + l_{JI}^-) \gamma_{JI} \right]
\end{split}
    \label{eq:r3_surface}
\end{equation}

\noindent (ii) Evaluation of the surface integral: $\Phi_{I} = \int_{S_I} \frac{1}{r} \thinspace d S(\textbf{x}')$ \\
The surface integral $\Phi_{I}$ is the Newtonian potential integral over the $\text{I}^\text{th}$ surface. The closed-form formulation was originally proposed by \cite{Rodin1996} and is provided below for the completeness of the derivation, 
\begin{equation}
    \begin{split}
    \Phi_{I} = \sum_{J = 1}^{N_{JI}} \int_{\tan^{-1} \left[ \frac{l_{JI}^-}{b_{JI}} \right]}^{\tan^{-1} \left[ \frac{l_{JI}^+}{b_{JI}} \right]} \int_{0}^{b_{JI} \sqrt{1 + \tan^2 \theta}} \frac{\rho}{\sqrt{a_{I}^2 + \rho^2}} \thinspace  d \rho \thinspace d\theta = \sum_{J=1}^{N_{JI}} \left[ F_{JI}(l_{JI}^+) - F_{JI} (l_{JI}^-)\right]
    \end{split}
    \label{eq:1/r_surface}
\end{equation}
because the distance $r = \sqrt{a_I^2 + b_{JI}^2 + le^2}$, 
\begin{equation}
\begin{split}
    F_{JI}(le) & = \left( a_{I} + |a_{I}| \right) \tan^{-1} \left[ \frac{a_{I} - le + r}{b_{JI}} \right] + \left( a_{I} - |a_{I}| \right) \tan^{-1} \left[ \frac{a_{I} + le -r}{b_{JI}} \right]  - b_{JI} \left( \ln \left[ r - le \right] \right)
\end{split}
\end{equation}

\noindent (iii) Evaluation of the line integral: $\mathcal{L}_{JI} = \int_{\Gamma_{JI}} r_{3} \ln[r + r_{3}] \thinspace d\textbf{x}' - \int_{\Gamma_{JI}} r \thinspace d\textbf{x}'$

The second Boussinesq's displacement potential, $\beta = r_{3} \ln[r + r_{3}] - r$, can be divided into two parts, the logarithmic term $r_{3} \ln[r + r_{3}]$ and the biharmonic term $r$. Therefore, this subsection evaluates them separately. Note that the line integral of $r$ has been reported by Gao et al. \cite{Gao2012}, however, the closed-form expression is provided here for completeness using the present notation. 

Following the derivation procedure in Eq. (\ref{eq:sym}), this subsection substitutes $le$ with $t$ as the integral variable, which yields, 
\begin{small}
\begin{equation}
\begin{split}
    \mathcal{L}_{JI}  &= \frac{R_{JI}}{2} \int_{t^-}^{t^+} \frac{\left[\gamma_{JI} R_{JI} t^2 + 2 c_{JI} t - \gamma_{JI} R_{JI}\right] (1 + t^{-2})}{2t} \ln\left[ \frac{R_{JI} (1+\gamma_{JI}) t^2 + 2c_{JI} t + R_{JI}(1-\gamma_{JI})}{2t} \right] \thinspace dt \\ 
    & - \int_{l_{JI}^-}^{l_{JI}^+} \sqrt{a_I^2 + b_{JI}^2 + le^2} \thinspace d \thinspace le = \mathcal{L}_{JI}(t^+) - \mathcal{L}_{JI}(t^-) 
\end{split}
    \label{eq:line_beta_L}
\end{equation}
\end{small}
and 
\begin{equation}
    \mathcal{L}_{JI}(t) = \frac{\gamma_{JI} R_{JI}^2}{4} \left[ \mathcal{I}_{1}(t) - \mathcal{I}_{-3}(t) \right] + \frac{R_{JI} c_{JI}}{2} \left[ \mathcal{I}_{0}(t) + \mathcal{I}_{-2}(t) \right] - \Psi_{JI}(t)
    \label{eq:mathcal_L_JI}
\end{equation}
where the explicit expression of $\mathcal{L}_{JI}$ is composed of four elementary integrals $\mathcal{I}_{n} = \int t^n \left(\ln[P(t)] - \ln[2 t]\right) \thinspace dt $ and $\Psi_{JI} = \int r \thinspace dle$ (n = -2, -3, 0, 1), when $\gamma_{JI} \neq \pm 1$
\begin{equation}
\begin{aligned}
    \mathcal{I}_{-3}(t) & = -\frac{2 \left( \ln[P(t)] - \ln[2 t] \right) - 1}{4 t^2} - \frac{c_{JI}}{R_{JI} (1 - \gamma_{JI}) t} + \frac{R_{JI}^2 (1 - \gamma_{JI}^2) - 2 c_{JI}^2}{R_{JI}^2 (1 - \gamma_{JI})^2 } \ln[t] \\ 
    & + \frac{2 c_{JI}^2 - R_{JI}^2 (1 - \gamma_{JI}^2)}{2 R_{JI}^2 (1 - \gamma_{JI})^2} \ln[P(t)] - \frac{2 c_{JI} \Delta}{R_{JI}^2 (1 - \gamma_{JI})^2} F(t) \\ 
    \mathcal{I}_{-2}(t) & = \frac{-\ln[P(t)] + \ln[2 t] + 1}{t} + \frac{2 c_{JI}}{R_{JI} (1 - \gamma_{JI})} \ln[t] - \frac{c_{JI}}{R_{JI} (1 - \gamma_{JI})} \ln [P(t)] + \frac{2 \Delta F(t)}{R_{JI} (1 - \gamma_{JI})} \\ 
    \mathcal{I}_{0}(t) & = t \left(\ln[P(t)] - \ln[2 t]  - 1 \right) + \frac{c_{JI}}{R_{JI} ( 1 + \gamma_{JI})} \ln[P(t)] + \frac{2 \Delta}{R_{JI} (1 + \gamma_{JI})} F(t) \\ 
    \mathcal{I}_1(t) & = \frac{\left( 2 \ln[P(t)] - 2 \ln[2 t] - 1\right)}{4} t^2 + \frac{c_{JI}}{R_{JI} (1 + \gamma_{JI})} t + \frac{R_{JI}^2 (1 - \gamma_{JI}^2) - 2 c_{JI}^2}{2 R_{JI}^2 (1 + \gamma_{JI})^2} \ln [P(t)] -\frac{2 c_{JI} \Delta}{R_{JI}^2 ( 1 + \gamma_{JI})^2} F(t) \\
\end{aligned}
    \label{eq:I_integrals}
\end{equation}
and 
\begin{equation}
    \Psi_{JI}(le) = \frac{1}{2} \left( le \thinspace r - R_{JI}^2 \ln [r - le] \right) \quad \text{or equivalently} \quad \Psi_{JI}(t) = \frac{R_{JI}^2}{2} \left( \frac{t^2 - t^{-2}}{4} + \ln [t] \right)
\end{equation}

When $\gamma_{JI} = \pm 1$, Eq. (\ref{eq:I_integrals}) can be reduced as, 
\begin{equation}
    \mathcal{L}_{JI}(t) = \begin{cases}
    \frac{R_{JI}^2}{4} \left[ \frac{t^2}{4} \left( 2 \ln [R_{JI} t] -1 \right) + \frac{1}{4 t^2} \left( 2 \ln[R_{JI} t] + 1 \right) \right] - \Psi_{JI}(t) & \gamma_{JI} = 1 \\ 
    -\frac{R_{JI}^2}{4} \left[ \frac{t^2}{4} \left( 2 \ln [R_{JI}] - 2 \ln [t] + 1 \right) + \frac{1}{4 t^2} \left( 2 \ln[R_{JI}] - 2 \ln[t] - 1 \right) \right] - \Psi_{JI}(t) & \gamma_{JI} = -1
    \end{cases}
    \label{eq:L_JI_special}
\end{equation}
Eqs. (\ref{eq:r3_surface} -\ref{eq:L_JI_special}) complete the evaluation of all integrals in Eq. (\ref{eq:def_int_beta}) except $D_{I}$, which is evaluated in the following subsection. 

\subsection{Evaluation of $\mathcal{M}_{JI}$ and $\mathcal{V}_{JI}$}
\label{sec:M_V}

The dimensional reduction of $D_{I}$ in Eq. (\ref{eq:beta_surface_manu}) requires evaluating two additional line integrals, $\mathcal{M}_{JI}$ and $\mathcal{V}_{JI}$, while $\mathcal{K}_{JI}$ has been evaluated in \ref{sec:KH}. Because both line integrals contain the same denominator $r + r_{3}$, the transformation in \ref{sec:KH} is utilized. 

\noindent(i) Evaluation of the line integral: $\mathcal{M}_{JI} = \int_{\Gamma_{JI}} \frac{r_{3}}{r + r_{3}} \thinspace d\textbf{x}'$
\begin{equation}
     \mathcal{M}_{JI} = \frac{R_{JI}}{2} \int_{t_{JI}^-}^{t_{JI}^+} \frac{\left[\gamma_{JI} R_{JI} t^2 + 2 c_{JI} t - \gamma_{JI} R_{JI} \right] (1 + t^{-2})}{R_{JI} (1 + \gamma_{JI}) t^2 + 2 c_{JI} t + R_{JI} (1 - \gamma_{JI})} \thinspace dt = M_{JI}(t^+) - M_{JI}(t^-)
\end{equation}
and when $\gamma_{JI} \neq \pm 1$
\begin{equation}
    \begin{split}
    M_{JI}(t) &= \frac{R_{JI} \gamma_{JI}}{2} \left( \frac{t}{1 + \gamma_{JI}} + \frac{1}{t (1 - \gamma_{JI})} \right) + \frac{c_{JI}}{(1 - \gamma_{JI})^2} \ln [t] - \frac{2 c_{JI} \gamma_{JI}}{(1 - \gamma_{JI}^2)^2} \ln|P(t)| \\ & + \frac{2\left[ \Delta (1 + \gamma_{JI}^2) - R_{JI}^2 (1 - \gamma_{JI}^2) \right]}{(1 - \gamma_{JI}^2)^2} F(t)
    \end{split}
    \label{eq:line_beta_M}
\end{equation}
when $\gamma_{JI} = \pm 1$, Eq. (\ref{eq:line_beta_M}) reduces as, 
\begin{equation}
    M_{JI} = \begin{cases}
    \frac{R_{JI}}{4} \left( t + \frac{1}{3t^3} \right) & \gamma_{JI} = 1 \\ 
    -\frac{R_{JI}}{4} \left( \frac{t^3}{3} + \frac{1}{t} \right) & \gamma_{JI} = -1
    \end{cases}
\end{equation}

\noindent(ii) Evaluation of the line integral: $\mathcal{V}_{JI} = \int_{\Gamma_{JI}} \frac{\chi_{I} r_{3} - a_I}{(r + r_{3})^2} \thinspace d\textbf{x}'$
\begin{equation}
    \mathcal{V}_{JI} = R_{JI} \int_{t_{JI}^-}^{t_{JI}^+} \frac{(t + t^{-1}) \left[ \chi_I \left( \gamma_{JI} R_{JI} t^2 + 2 c_{JI} t - \gamma_{JI} R_{JI} \right) - 2 a_I t \right]}{\left[ R_{JI} (1 + \gamma_{JI}) t^2 + 2 c_{JI} t + R_{JI} (1 - \gamma_{JI}) \right]^2 } \thinspace dt = V_{JI}(t^+_{JI}) - V_{JI}(t^-_{JI})
\end{equation}
and when $\gamma_{JI} \neq \pm 1$, 
\begin{equation}
\begin{split}
    V_{JI}(t) & = C_0 \ln[t] + \frac{C_{1}}{2 R_{JI} (1 + \gamma_{JI})} \ln[P(t)] + \left( C_{2} - \frac{C_{1} c_{JI}}{R_{JI} (1 + \gamma_{JI})} \right) F(t) - \frac{C_{3}}{2 R_{JI} (1 + \gamma_{JI}) P(t)} \\ & + \left( C_{4} - \frac{C_{3} c_{JI}}{R_{JI} (1 + \gamma_{JI})} \right) \left[ \frac{R_{JI} (1 + \gamma_{JI}) t + c_{JI}}{2 \Delta P(t)} + \frac{R_{JI} (1 + \gamma_{JI})}{2 \Delta} F(t) \right]
\end{split}
    \label{eq:line_beta_V}
\end{equation}
where the coefficients, 
\begin{equation}
    \begin{aligned}
    C_0 = \frac{-f_{JI}}{(1 - \gamma_{JI})^2}, \quad C_1 = \frac{2 f_{JI} R_{JI} (1 + \gamma_{JI}^2) }{(1 + \gamma_{JI}) (1 - \gamma_{JI})^2}, \quad C_{2} = 2 \left( \frac{4 c_{JI} f_{JI} \gamma_{JI}}{(1 - \gamma_{JI}^2)^2}  + \frac{d_{JI}}{1 + \gamma_{JI}} \right) \\ 
    C_{3} = 4 \left( \frac{f_{JI} \gamma_{JI} R_{JI}^2}{1 - \gamma_{JI}^2} + \frac{c^2_{JI} f_{JI}}{(1 + \gamma_{JI})^2} - \frac{c_{JI} d_{JI}}{1 + \gamma_{JI}} \right), \quad C_{4} = 4 \left( \frac{ d_{JI} \gamma_{JI} R_{JI}}{1 + \gamma_{JI}} + \frac{c_{JI} f_{JI} R_{JI} (1 + \gamma_{JI}^2)}{(1 + \gamma_{JI}) (1 - \gamma_{JI}^2) } \right)
    \end{aligned}
\end{equation}
where $d_{JI} = a_{I} \left( \chi_I^2 - 1 \right) + b_{JI} \chi_I \zeta_{JI} $ and $f_{JI} = \chi_I \gamma_{JI}$. When $\gamma_{JI} = \pm 1$, Eq. (\ref{eq:line_beta_V}) reduces as, 
\begin{equation}
    V_{JI} = \begin{cases}
        \frac{a_I}{2 R_{JI}} \left[ t^{-1} + \frac{1}{3} t^{-3} \right] & \gamma_{JI} = 1 \\ 
        -\frac{a_I}{2 R_{JI}} \left[ t + \frac{1}{3} t^3 \right] & \gamma_{JI} = -1
    \end{cases}
\end{equation}

Together with $\mathcal{K}_{JI}$ evaluated in \ref{sec:KH}, these results allow $D_I$ to be evaluated analytically, which completes the closed-form evaluation of $\Xi^\beta_{I}$ and $\Lambda$. 

\subsection{Expression of $\overline{F}$ and $\overline{H}$}
\label{sec:FH_close}

The following edge primitives have been reported in \cite{Wu2021_polyhedral}. Because their higher-order partial derivatives are utilized in the singularity analysis of Section 5, they are provided here for completeness. Note that their inverse hyperbolic tangent terms may cause logarithmic singularities in the vicinity of the polyhedral edges and vertices. 

\begin{align}
\begin{split}
    \overline{F}(a_I, b_{JI}, le) & =  a_I \frac{b_{JI}}{|b_{JI}|} \sin^{-1}\left[\frac{a_I le }{\sqrt{(a_I^2 + b_{JI}^2)(b_{JI}^2 + le^2)}}\right]\\
    &- |a_I| \tan^{-1}[\frac{le}{b_{JI}}] + b_{JI} \tanh^{-1}\left[\frac{le}{\sqrt{a_I^2 + b_{JI}^2 + le^2}} \right]
\end{split}\\
\begin{split}
    \overline{H}(a_I, b_{JI}, le) & = \frac{1}{6} \Big[  2|a_I|^3 \left( -\tan^{-1}\left[\frac{le}{b_{JI}} \right]  + \tan^{-1}\left[\frac{le |a_I|}{b_{JI} \sqrt{a_I^2 + b_{JI}^2 + le^2}}\right] \right) \\ 
    & + b_{JI} le \sqrt{a_I^2 + b_{JI}^2 + le^2} + b_{JI} (3a_I^2 + b_{JI}^2) \tanh^{-1}\left[\frac{le}{\sqrt{a_I^2 + b_{JI}^2 + le^2}} \right] \Big]
\end{split}
\end{align}

Although the algebraic and inverse trigonometric parts are required for evaluating the complete elastic fields, they do not alter the dominant singularity in Section 5. 

\section{Proof of the surface divergence identity in Eq. (\ref{eq:beta_surface_manu})}
\label{sec:proof_surface_grad_beta}
The remaining surface term in $D_{I}$ cannot be directly converted into line integrals. Eq. (\ref{eq:beta_surface_manu}) has introduced an auxiliary tangential vector whose surface divergence equals the original terms. To verify this identity, the auxiliary vector is decomposed into three parts, whose surface divergences are derived below. 
\begin{equation}
    \begin{split}
    2 \frac{\nabla^s r_{3} \cdot \bm{\mathcal{B}}_{I}}{r + r_{3}} = \nabla^s \cdot \left[ \frac{\chi_I r_{3} - a_{I}}{r+r_{3}} \bm{w}_{I} \right. \left.+ a_{I} \frac{(\chi_I^2 - 1) \bm{w}_{I} - a_{I} \chi_I \bm{m}_{I}}{r + r_{3}}  - a_{I} \frac{(\chi_I r_{3} - a_{I}) (\chi_I \bm{w}_{I} - a_{I} \bm{m}_{I})}{(r + r_{3})^2} \right]
    \end{split}
\end{equation}

\noindent (i) $\nabla^s \cdot \frac{\chi_I r_{3} - a_{I}}{r + r_{3}} \bm{w}_{I}$: 
Some results are given to facilitate the derivation, 
\begin{equation}
    \begin{aligned}
    \nabla^s \left( \chi_I r_{3} - a_I \right) = \left[ \delta_{ij} - (\xi_I^0)_{i} (\xi_I^0)_j \right] \left[ \chi_I \delta_{j3} - (\xi_I^0)_{j} \right] = \chi_I \delta_{i3} - \chi_I^2 (\xi_{I}^0)_{i} \\ 
    \nabla^s \left( \chi_I r_{3} - a_I \right) \cdot \bm{w}_I = \left[ \chi_I \delta_{i3} - \chi_I^2 (\xi_{I}^0)_{i} \right] \left[ r_i - a_I (\xi_I^0)_{i} \right] = a_I (1 - \chi_I^2) + \left( \chi_I r_{3} - a_I \right) \\
    \nabla^s (r + r_{3}) = \left[ \delta_{ij} - (\xi_I^0)_{i} (\xi_I^0)_j \right] \left[ r_{,j'} + \delta_{j3} \right] = \frac{r_i}{r} + \delta_{i3} - \frac{a_I}{r} (\xi_I^0)_{i} - \chi_I (\xi_I^0)_i = \frac{\bm{w}_I}{r} + \bm{m}_I \\ 
    \nabla^s (r + r_{3}) \cdot \bm{w}_I = \left[ \frac{(w_I)_i}{r} + (m_I)_i \right] \left[ r_i - a_I (\xi_I^0)_{i} \right] = r + r_{3} - a_I \left( \frac{a_I}{r} + \chi_I \right) 
    \end{aligned}
\end{equation}
Hence, the surface divergence of the first term provides, 
\begin{equation}
    \begin{split}
    \nabla^s \cdot \frac{\chi_I r_{3} - a_{I}}{r + r_{3}} \bm{w}_{I} & = \frac{\nabla^s \left( \chi_I r_{3} - a_I \right) \cdot \bm{w}_I}{r + r_{3}} + \frac{\left( \chi_I r_{3} - a_I \right) \nabla^s \cdot \bm{w}_I}{r + r_{3}} - \frac{\left( \chi_I r_{3} - a_I \right) \bm{w}_I \cdot \nabla^s (r + r_{3}) }{(r + r_{3})^2} \\ 
    & = \frac{\chi_{I} r_{3} - \chi_I^2 a_I}{r+r_{3}} + \frac{2 \left( \chi_I r_{3} - a_I \right)}{r + r_{3}} - \frac{a_I \left( \chi_I r_{3} - a_I \right)}{(r + r_3)^2} \left[ (r + r_{3}) - a_I \left( \frac{a_I}{r} + \chi_I \right)  \right] \\ 
    & = \frac{2 (\chi_I r_{3} - a_{I}) + a_I (1 - \chi_I^2)}{r + r_{3}} + \frac{a_I (\chi_I r_{3} - a_{I})}{(r + r_{3})^2} \left( \frac{a_I}{r} + \chi_I \right)
    \end{split}
    \label{eq:term_1_grad}
\end{equation}

\noindent(ii) $\nabla^s \cdot \left( a_I \frac{(\chi_I^2 - 1) \bm{w}_{I} - a_I \chi_{I} \bm{m}_I }{r + r_{3}} \right)$: 
Some results are given to facilitate the derivation, 
\begin{equation}
    \begin{aligned}
    \nabla^s \cdot \left[ (\chi_I^2 - 1) \bm{w}_{I} - a_I \chi_{I} \bm{m}_I \right] & = \nabla^s \cdot \bm{w}_I (\chi_I^2 - 1) - a_I \chi_I \nabla^s \cdot \bm{m}_I = 2 (\chi_I^2 - 1) \\ 
    \left[ (\chi_I^2 - 1) \bm{w}_{I} - a_I \chi_{I} \bm{m}_I \right] \cdot \nabla^s (r + r_{3}) & = (\chi_I^2 - 1) \frac{r^2 - a_I^2}{r} + (\chi_I^2 - 1) (r_{3} - a_I \chi_I) \\ & \quad - a_I \chi_I \frac{r_3 - a_I \chi_I}{r} - a_I \chi_I (1 - \chi_I^2)  \\ & \quad = (\chi_I^2 - 1) (r + r_{3}) - \frac{a_I \left( \chi_I r_{3} - a_I  \right)}{r} 
    \end{aligned}
\end{equation}
Hence, the surface divergence of the second term provides, 
\begin{equation}
    \begin{split}
    \nabla^s \cdot \left( a_I \frac{(\chi_I^2 - 1) \bm{w}_{I} - a_I \chi_{I} \bm{m}_I }{r + r_{3}} \right) & = a_I \left[ \frac{\nabla^s \cdot \left[ (\chi_I^2 - 1) \bm{w}_{I} - a_I \chi_{I} \bm{m}_I \right]}{r + r_{3}} - \frac{ \left[(\chi_I^2 - 1) \bm{w}_{I} - a_I \chi_{I} \bm{m}_I\right] \cdot \nabla^s (r + r_{3})}{(r + r_{3})^2} \right] \\ 
    & = \frac{2 a_I (\chi_I^2 - 1)}{r + r_{3}} - \frac{a_I \left[ (\chi_I^2 - 1) (r + r_{3}) - \frac{a_I \left( \chi_I r_{3} - a_I  \right)}{r} \right]}{(r + r_{3})^2} \\ & = \frac{a_I (\chi_I^2 - 1)}{r + r_{3}} + \frac{a_I^2 \left( \chi_I r_{3} - a_I  \right)}{r (r + r_{3})^2}
    \end{split}
    \label{eq:term_2_grad}
\end{equation}

\noindent (iii) $-\nabla^s \cdot \left( a_I \frac{(\chi_I r_{3} - a_I) (\chi_I \bm{w}_I - a_I \bm{m}_I)}{(r + r_{3})^2} \right)$: Some results are given to facilitate the derivation, 
\begin{equation}
    \begin{aligned}
     \nabla^s \left( \chi_I r_{3} - a_I  \right) \cdot \left( \chi_I \bm{w}_I - a_I \bm{m}_I \right)  & = (\chi_I \delta_{i3} - \chi_I^2 (\xi_I^0)_{i})  \left( \chi_I r_{i} - \chi_I a_I (\xi_I^0)_{i} - a_I \delta_{i3} + a_I \chi_I (\xi_I^0)_{i} \right) \\  
    & = \chi_I (\chi_I r_{3} - a_I) \\
    (\chi_I r_{3} - a_I) \nabla^s \cdot (\chi_I \bm{w}_I - a_I \bm{m}_{I}) & = 2 \chi_I (\chi_I r_{3} - a_I) \\
    (\chi_I r_{3} - a_I) (\chi_I \bm{w}_I - a_I \bm{m}_{I}) \cdot \nabla^s \left( \frac{1}{(r + r_{3})^2} \right) & = -\frac{2(\chi_I r_{3} - a_I)}{(r + r_{3})^3} (\chi_I \bm{w}_I - a_I \bm{m}_I) \cdot \nabla^s (r + r_{3}) \\ & = \frac{-2 (\chi_I r_{3} - a_I)}{(r + r_{3})^2} \left(\chi_I - \frac{a_I}{r} \right) 
    \end{aligned}
\end{equation}

Hence, the surface divergence of the third term provides, 
\begin{equation}
    \begin{split}   
    & -\nabla^s \cdot \left( a_I \frac{(\chi_I r_{3} - a_I) (\chi_I \bm{w}_I - a_I \bm{m}_I)}{(r + r_{3})^2} \right) = -\frac{3 a_I \chi_I (\chi_I r_{3} - a_I)}{(r + r_{3})^2} + \frac{2 a_I(\chi_I r_{3} - a_I)}{(r + r_{3})^2} \left(\chi_I - \frac{a_I}{r} \right) \\ 
    & = -\frac{a_I \left( \chi_I r_{3} - a_I  \right)}{(r+r_{3})^2} \left( \chi_I + \frac{2 a_I}{r} \right)
    \end{split}
    \label{eq:term_3_grad}
\end{equation}

Superposing Eqs. (\ref{eq:term_1_grad}), (\ref{eq:term_2_grad}), and (\ref{eq:term_3_grad}) reproduces the same result in Eq. (\ref{eq:beta_surface_manu}). Therefore, the surface integral in $D_I$ can be expressed through the surface divergence of the auxiliary vector, which can be subsequently converted into line integrals. 

\section{Partial differentiation results}
\subsection{Partial differentiation chain rule for the third- and fourth-order partial derivatives}
\label{sec:partial_deriv}

The second-order partial derivatives of the domain integrals have been discussed in Eq. (\ref{eq:theta_2nd}). The higher-order partial derivatives to construct Eshelby's tensor can be obtained through the partial differentiation chain rule. For completeness, the third- and fourth-order partial derivatives are provided below. 

(i) Third-order partial derivative: 
\begin{equation}
    \begin{split}
    \Theta_{,ijk} & = -\sum_{I=1}^{N_I} \sum_{J=1}^{N_{JI}} (\xi_I^0)_{i} \left[ (\xi_I^0)_j (\xi_I^0)_{k} \frac{\partial^2 \mathcal{P}_{JI}}{\partial a_I^2} + (\lambda^0_{JI})_j (\lambda^0_{JI})_k \frac{\partial^2 \mathcal{P}_{JI}}{\partial b_{JI}^2} + (\eta_{JI}^0)_j (\eta_{JI}^0)_k \left( \frac{\partial^2 \mathcal{P}_{JI}}{\partial (l_{JI}^+)^2} + \frac{\partial^2 \mathcal{P}_{JI}}{\partial (l_{JI}^-)^2} \right) \right. \\ & \left. 
    \left( (\xi_I^0)_j (\lambda_{JI}^0)_k  + (\xi_I^0)_k (\lambda_{JI}^0)_j \right) \frac{\partial^2 \mathcal{P}_{JI}}{\partial a_I \partial b_{JI}} + \left( (\xi_I^0)_j (\eta_{JI}^0)_k + (\xi_I^0)_k (\eta_{JI}^0)_j \right) \left( \frac{\partial^2 \mathcal{P}_{JI}}{\partial a_I \partial l_{JI}^+} + \frac{\partial^2 \mathcal{P}_{JI}}{\partial a_I \partial l_{JI}^-} \right) \right. \\ & \left. 
    + \left( (\lambda_{JI}^0)_j (\eta_{JI}^0)_k + (\lambda_{JI}^0)_k (\eta_{JI}^0)_j \right) \left( \frac{\partial^2 \mathcal{P}_{JI}}{\partial b_{JI} \partial l_{JI}^+} + \frac{\partial^2 \mathcal{P}_{JI}}{\partial b_{JI} \partial l_{JI}^-} \right)
    \right]
    \end{split}
    \label{eq:theta_3rd}
\end{equation}

(ii) Fourth-order partial derivative: 
\begin{small}
\begin{align}
    \begin{split}
    \Theta_{,ijkl} & = \sum_{I=1}^{N_I} \sum_{J=1}^{N_{JI}} (\xi_I^0)_{i} \left[ (\xi_I^0)_j(\xi_I^0)_k(\xi_I^0)_l \frac{\partial^3\mathcal{P}_{JI}}{\partial a_I^3} + (\lambda_{JI}^0)_j(\lambda_{JI}^0)_k(\lambda_{JI}^0)_l \frac{\partial^3\mathcal{P}_{JI}}{\partial b_{JI}^3} \right. \\ &  \left.
    + (\eta_{JI}^0)_j(\eta_{JI}^0)_k(\eta_{JI}^0)_l \left(\frac{\partial^3\mathcal{P}_{JI}}{\partial(l_{JI}^+)^3}
    +3\frac{\partial^3\mathcal{P}_{JI}}{\partial(l_{JI}^+)^2\partial l_{JI}^-}
    +3\frac{\partial^3\mathcal{P}_{JI}}{\partial l_{JI}^+\partial(l_{JI}^-)^2}+ \frac{\partial^3\mathcal{P}_{JI}} {\partial(l_{JI}^-)^3}\right) \right. \\ & \left.
    +\left((\xi_I^0)_j(\xi_I^0)_k(\lambda_{JI}^0)_l+(\xi_I^0)_j(\xi_I^0)_l(\lambda_{JI}^0)_k+(\xi_I^0)_k(\xi_I^0)_l(\lambda_{JI}^0)_j\right) \frac{\partial^3\mathcal{P}_{JI}}{\partial a_I^2\partial b_{JI}}
    \right. \\ & \left. 
    +\left((\xi_I^0)_j(\xi_I^0)_k(\eta_{JI}^0)_l+(\xi_I^0)_j(\xi_I^0)_l(\eta_{JI}^0)_k+
    (\xi_I^0)_k(\xi_I^0)_l(\eta_{JI}^0)_j\right)\left(\frac{\partial^3\mathcal{P}_{JI}}{\partial a_I^2\partial l_{JI}^+}+\frac{\partial^3\mathcal{P}_{JI}}{\partial a_I^2\partial l_{JI}^-}\right)
    \right. \\ & \left.
    +\left((\xi_I^0)_j(\lambda_{JI}^0)_k(\lambda_{JI}^0)_l+(\xi_I^0)_k(\lambda_{JI}^0)_j(\lambda_{JI}^0)_l
    +(\xi_I^0)_l(\lambda_{JI}^0)_j(\lambda_{JI}^0)_k\right)\frac{\partial^3\mathcal{P}_{JI}}{\partial a_I\partial b_{JI}^2}
    \right.\\ & \left.
    +\left((\lambda_{JI}^0)_j(\lambda_{JI}^0)_k(\eta_{JI}^0)_l+(\lambda_{JI}^0)_j(\lambda_{JI}^0)_l(\eta_{JI}^0)_k+(\lambda_{JI}^0)_k(\lambda_{JI}^0)_l(\eta_{JI}^0)_j \right) \left(\frac{\partial^3\mathcal{P}_{JI}}{\partial b_{JI}^2\partial l_{JI}^+}+ \frac{\partial^3\mathcal{P}_{JI}}{\partial b_{JI}^2\partial l_{JI}^-}\right)
    \right. \\ & \left.
    +\left((\xi_I^0)_j(\eta_{JI}^0)_k(\eta_{JI}^0)_l+(\xi_I^0)_k(\eta_{JI}^0)_j(\eta_{JI}^0)_l+
    (\xi_I^0)_l(\eta_{JI}^0)_j(\eta_{JI}^0)_k\right)\left(\frac{\partial^3\mathcal{P}_{JI}} {\partial a_I\partial(l_{JI}^+)^2}+2\frac{\partial^3\mathcal{P}_{JI}}{\partial a_I\partial l_{JI}^+\partial l_{JI}^-}+\frac{\partial^3\mathcal{P}_{JI}}{\partial a_I\partial(l_{JI}^-)^2}\right)
    \right.\\ & \left.
    +\left((\lambda_{JI}^0)_j(\eta_{JI}^0)_k(\eta_{JI}^0)_l+(\lambda_{JI}^0)_k(\eta_{JI}^0)_j(\eta_{JI}^0)_l
    +(\lambda_{JI}^0)_l(\eta_{JI}^0)_j(\eta_{JI}^0)_k\right)
    \left(\frac{\partial^3\mathcal{P}_{JI}}{\partial b_{JI}\partial(l_{JI}^+)^2}+2\frac{\partial^3\mathcal{P}_{JI}}{\partial b_{JI}\partial l_{JI}^+\partial l_{JI}^-}+\frac{\partial^3\mathcal{P}_{JI}}{\partial b_{JI}\partial(l_{JI}^-)^2}\right)
    \right. \\ & \left.
    +
    \big((\xi_I^0)_j(\lambda_{JI}^0)_k(\eta_{JI}^0)_l+(\xi_I^0)_j(\lambda_{JI}^0)_l(\eta_{JI}^0)_k
    +(\xi_I^0)_k(\lambda_{JI}^0)_j(\eta_{JI}^0)_l\right.\\ & \left. \qquad
    +(\xi_I^0)_k(\lambda_{JI}^0)_l(\eta_{JI}^0)_j+ (\xi_I^0)_l(\lambda_{JI}^0)_j(\eta_{JI}^0)_k
    +(\xi_I^0)_l(\lambda_{JI}^0)_k(\eta_{JI}^0)_j
    \big) \left(
    \frac{\partial^3\mathcal{P}_{JI}}
    {\partial a_I\partial b_{JI}\partial l_{JI}^+}+
    \frac{\partial^3\mathcal{P}_{JI}}
    {\partial a_I\partial b_{JI}\partial l_{JI}^-}
    \right)
    \right]
    \end{split}
\end{align}
\end{small}

Eq. (D.1) shows that the third-order partial derivatives require the second-order partial derivatives of the edge primitive $\mathcal{P}_{JI}$. Subsequently, Eq. (D.2) indicates that the fourth-order partial derivatives require the third-order partial derivatives of $\mathcal{P}_{JI}$. Although Eqs. (D.1) and (D.2) are lengthy, no additional integration procedure is required, as they can be obtained by differentiating the integrals in Appendix B. 

\subsection{Partial derivatives of the inverse-hyperbolic-tangent part of $\overline{H}$}
\label{sec:der_psi}

As discussed in Section \ref{sec:phi_psi}, the logarithmic singularity of the biharmonic potential arises from the inverse hyperbolic tangent part of the edge primitive $\overline{H}$. Although the algebraic and inverse trigonometric parts are required to evaluate the complete elastic fields, they remain bounded in the vicinity of polyhedral edges and vertices. Therefore, it is sufficient for the singularity analysis to differentiate the inverse hyperbolic tangent part, 

\begin{equation}
\begin{split}
    \frac{\partial^3 \overline{H}^\dagger}{\partial a_I^3} & = \frac{a_I b_{JI} l_{JI}}{\left(a_I^2 + b_{JI}^2 \right)^3 \left(a_I^2 + b_{JI}^2 + l_{JI}^2\right)^\frac{5}{2}} \left[ 3 b_{JI}^2 (a_I^2 - 9 b_{Ji}^2) (a_I^2 + b_{JI}^2)^2 \right. \\ & \left. + (a_I^2 + b_{JI}^2) (3 a_I^4 - 14 a_I^2 b_{JI}^2 - 57 b_{JI}^4) l_{JI}^2 -2 (3 a_I^4 + 10 a_I^2 b_{JI}^2 + 15 b_{JI}^4) l_{JI}^4 \right]
\end{split}
\end{equation}
\begin{equation}
\begin{split}
    \frac{\partial^3 \overline{H}^\dagger}{\partial a_I^2 \partial b_{JI}}
    &=-\frac{l_{JI}}{\left(a_I^2+b_{JI}^2\right)^3\left(a_I^2+b_{JI}^2+l_{JI}^2\right)^{\frac{5}{2}}}
    \left[3\left(a_I^2+b_{JI}^2\right)^2\left(2a_I^6+7a_I^4b_{JI}^2-3a_I^2b_{JI}^4+2b_{JI}^6\right)\right.\\
    &\quad+\left(a_I^2+b_{JI}^2\right)\left(15a_I^6+33a_I^4b_{JI}^2-9a_I^2b_{JI}^4+13b_{JI}^6\right)l_{JI}^2\\
    &\quad\left.+\left(3a_I^2+7b_{JI}^2\right)\left(3a_I^4+b_{JI}^4\right)l_{JI}^4\right]
    +6\tanh^{-1}\left[\frac{l_{JI}}{\sqrt{a_I^2+b_{JI}^2+l_{JI}^2}}\right]
\end{split}
\end{equation}

\begin{equation}
    \frac{\partial^3 \overline{H}^\dagger}{\partial a_I^2 \partial l_{JI}} = \frac{b_{JI} \left[ -a_I^2 (b_{JI}^2 + 3 l_{JI}^2) + (b_{JI}^2 + l_{JI}^2) (5 b_{JI}^2 + 6 l_{JI}^2) \right]}{\left(a_I^2+b_{JI}^2+l_{JI}^2\right)^{\frac{5}{2}}}
\end{equation}
\begin{equation}
\begin{split}
    \frac{\partial^3 \overline{H}^\dagger}{\partial a_I \partial b_{JI}^2}
    &=\frac{a_I b_{JI} l_{JI}}{\left(a_I^2+b_{JI}^2\right)^3\left(a_I^2+b_{JI}^2+l_{JI}^2\right)^{\frac{5}{2}}}
    \left[3a_I^2\left(a_I^2-9b_{JI}^2\right)\left(a_I^2+b_{JI}^2\right)^2 \right.\\
    &\quad\left.-\left(a_I^2+b_{JI}^2\right)\left(3a_I^4+46a_I^2b_{JI}^2+3b_{JI}^4\right)l_{JI}^2
    -2\left(3a_I^4+14a_I^2b_{JI}^2+3b_{JI}^4\right)l_{JI}^4\right]
\end{split}
\end{equation}
\begin{equation}
    \frac{\partial^3 \overline{H}^\dagger}{\partial a_I \partial b_{JI} \partial l_{JI}} = \frac{3 \left[ a_I^5 + 3 a_I^3 (b_{JI}^2 + l_{JI}^2) + a_I l_{JI}^2 (b_{JI}^2 + 2 l_{JI}^2) \right] }{\left(a_I^2+b_{JI}^2+l_{JI}^2\right)^{\frac{5}{2}}}
\end{equation}

\begin{equation}
    \frac{\partial^3 \overline{H}^\dagger}{\partial a_I \partial l_{JI}^2} = \frac{a_I b_{JI} l_{JI} (a_I^2 - b_{JI}^2 - 2 l_{JI}^2)}{2 \left(a_I^2+b_{JI}^2+l_{JI}^2\right)^{\frac{5}{2}}}
\end{equation}

\begin{equation}
\begin{split}
    \frac{\partial^3 \overline{H}^\dagger}{\partial b_{JI}^3}
    &=-\frac{l_{JI}}{\left(a_I^2+b_{JI}^2\right)^3\left(a_I^2+b_{JI}^2+l_{JI}^2\right)^{\frac{5}{2}}}
    \left[3\left(a_I^2+b_{JI}^2\right)^2\left(3a_I^6-3a_I^4b_{JI}^2+6a_I^2b_{JI}^4+2b_{JI}^6\right)\right.\\
    &\quad\left.+2\left(a_I^2+b_{JI}^2\right)\left(9a_I^6+18a_I^2b_{JI}^4+7b_{JI}^6\right)l_{JI}^2
    +\left(9a_I^6+9a_I^4b_{JI}^2+27a_I^2b_{JI}^4+11b_{JI}^6\right)l_{JI}^4\right]\\
    &\quad+6\tanh^{-1}\left[\frac{l_{JI}}{\sqrt{a_I^2+b_{JI}^2+l_{JI}^2}}\right]
\end{split}
\end{equation}

\begin{equation}
    \frac{\partial^3 \overline{H}^\dagger}{\partial b_{JI}^2 \partial l_{JI}} =\frac{b_{JI}\left[-3a_I^4+5a_I^2b_{JI}^2+2b_{JI}^4+\left(3a_I^2+5b_{JI}^2\right)l_{JI}^2+6l_{JI}^4\right]}{\left(a_I^2+b_{JI}^2+l_{JI}^2\right)^{\frac{5}{2}}}
\end{equation}

\begin{equation}
    \frac{\partial^3 \overline{H}^\dagger}{\partial b_{JI}\partial l_{JI}^2}
    =-\frac{3l_{JI}\left[a_I^4-a_I^2b_{JI}^2+\left(a_I^2+b_{JI}^2\right)l_{JI}^2\right]}{\left(a_I^2+b_{JI}^2+l_{JI}^2\right)^{\frac{5}{2}}}
\end{equation}

\begin{equation}
    \frac{\partial^3 \overline{H}^\dagger}{\partial l^3_{JI}} = -\frac{b_{JI} (3 a_I^2 + b_{JI}^2) (a_I^2 + b_{JI}^2 - 2 l_{JI}^2)}{\left(a_I^2+b_{JI}^2+l_{JI}^2\right)^{\frac{5}{2}}} 
\end{equation}

Among Eqs. (D.3-D.12), only Eqs. (D.4) and (D.9) exhibit inverse hyperbolic tangent terms. Therefore, the two derivatives in Eqs. (D.4) and (D.9) provide the logarithmic contributions reported in Eq. (\ref{eq:psi_leading_order}). The remaining algebraic and inverse trigonometric parts are bounded for the singularity analysis. 

\subsection{Partial derivatives of logarithmic parts of $\Xi^\alpha_I$}

Based on Eq. (D.1), the third-order partial derivatives of $\Theta$ requires to evaluate the second-order partial derivatives of $\Xi_I^\alpha$. In analogy with the partial derivatives discussed about $\overline{\Psi}$ and $\overline{\Phi}$, only the logarithmic and inverse hyperbolic tangent terms determine the dominant singularity. Therefore, the other bounded terms are omitted here, and the logarithmic parts $(.)_{\text{log}}$ are differentiated below,  
\label{sec:alpha_derivative}
\begin{align}
    \left(\frac{\partial^2 \Xi_{I}^{\alpha \dagger}}{\partial a_I^2} \right)_\text{log} & = \frac{1}{2} \left\{ \left[ \frac{b_{JI}(b_{JI}^2 + l_{JI}^2)}{\rho_{JI}^3} - \frac{2 \zeta_{JI}}{1 - \gamma_{JI}} \right] \ln \left[ l_{JI} + \rho_{JI} \right] \right. \notag \\ 
    & \left. + \left[ \frac{b_{JI}(b_{JI}^2 + l_{JI}^2)}{\rho_{JI}^3} + \frac{2 \zeta_{JI} \gamma_{JI} }{1 - \gamma_{JI}^2} \right] \ln \left[ \rho_{JI} (1 + \hat{x}_{3}) \right] \right\} \notag \\ 
    \left(\frac{\partial^2 \Xi_{I}^{\alpha \dagger}}{\partial a_I \partial b_{JI}} \right)_\text{log} & = \frac{1}{2} \left\{ \left[ \frac{a_I(a_{I}^2 + l_{JI}^2)}{\rho_{JI}^3} + \frac{2 \chi_I}{1 - \gamma_{JI}} \right] \ln \left[ l_{JI} + \rho_{JI} \right] \right. \notag\\ 
    & \left. + \left[ \frac{a_I (a_{I}^2 + l_{JI}^2)}{\rho_{JI}^3} - \frac{2 \chi_I \gamma_{JI} }{1 - \gamma_{JI}^2} \right] \ln \left[ \rho_{JI} (1 + \hat{x}_{3}) \right] \right\} \notag\\ 
    \left(\frac{\partial^2 \Xi_{I}^{\alpha \dagger}}{\partial a_I \partial l_{JI}} \right)_\text{log} & = -\frac{a_I b_{JI} l_{JI}}{2 \rho_{JI}^3} \ln \left[ \rho_{JI} (1 + \hat{x}_{3}) \right] \\ 
    \left( \frac{\partial^2 \Xi_{I}^{\alpha \dagger}} {\partial b_{JI}^2} \right)_{\mathrm{log}} &= \left( \frac{b_{JI}(3 a_{I}^2 + 2 b_{JI}^2 + 3 l_{JI}^2)}{2 \rho_{JI}^3} + \frac{\zeta_{JI}}{1 - \gamma_{JI}} \right) \ln \left[ l_{JI} + \rho_{JI} \right] \notag\\ 
    &  + \left[ \frac{b_{JI} (3 a_{I}^2 + 2 b_{JI}^2 + 3 l_{JI}^2)}{2 \rho_{JI}^3} - \frac{2 \zeta_{JI} \gamma_{JI} }{1 - \gamma_{JI}^2} \right] \ln \left[ \rho_{JI} (1 + \hat{x}_{3}) \right]  \notag\\
    \left(\frac{\partial^2 \Xi_{I}^{\alpha \dagger}}{\partial b_{JI} \partial l_{JI}} \right)_\text{log} & = \frac{l_{JI}\left(a_I^2+l_{JI}^2\right)}{2 \rho_{JI}^3 } \ln \left[  \rho_{JI} (1 + \hat{x}_{3}) \right] \notag\\
    \left(\frac{\partial^2 \Xi_{I}^{\alpha \dagger}}{\partial l^2_{JI}} \right)_\text{log} & = \frac{b_{JI}\left(a_I^2+b_{JI}^2\right)}{2 \rho_{JI}^3} \ln \left[ \rho_{JI} ( 1 + \hat{x}_{3}) \right] \notag   
\end{align}

The expressions in Eq. (D.13) are endpoint primitives. Hence, the contribution by the complete edge is evaluated by taking the difference between the values at $l_{JI}^+$ and $l_{JI}^-$. Note that the endpoint subtraction is necessary to discuss the logarithmic singularities on the polyhedral edges and vertices. 

\subsection{Partial derivatives of logarithmic parts of $\Xi^\beta_I$}
\label{sec:beta_derivative}

Similarly, the fourth-order partial derivatives of $\Lambda$ requires to evaluate the third-order partial derivatives of $\Xi_I^\beta$. The complete derivatives contain many bounded terms, therefore, only the logarithmic parts are differentiated below, 
\begin{align}
    \left( \frac{\partial^3 \Xi_{I}^{\beta \dagger}}{\partial a_I^3} \right)_{\text{log}} & = \frac{-2 (2 + \gamma_{JI}^2) \chi_I \zeta_{JI}}{(1 - \gamma_{JI}^2)^2} \ln \left[ l_{JI} + \rho_{JI} \right] \notag \\ & 
    + \frac{1}{2} \chi_I \left( \frac{b_{JI}(b_{JI}^2 + l_{JI}^2)^2}{\rho_{JI}^5} + \frac{2 \zeta_{JI} \gamma_{JI} \left(7- \gamma_{JI}^3 \right) }{(1 - \gamma_{JI}^2)^2} \right) \ln \left[ \rho_{JI} (1 + \hat{x}_{3}) \right] \notag \\
    \left( \frac{\partial^3 \Xi_{I}^{\beta \dagger}} {\partial a_I^2 \partial b_{JI}} \right)_{\text{log}} & = \frac{2}{3} \ln \left[ \rho_{JI} -l_{JI} \right] + \frac{1}{3(1 - \gamma_{JI}^2)^2} \left[ \gamma_{JI}^4 - \gamma_{JI}^2 (2 \chi_I^2  \right. \notag \\ &  \left. \quad + 6 \zeta_{JI}^2 + 7) + 4 \chi_I^2 - 6 \zeta_{JI}^2 + 6 \right] \ln \left[ l_{JI} + \rho_{JI} \right] \notag \\
    & + \frac{1}{6 \rho_{JI}^5} \left[ a_I \chi_I \left( 2 a_I^4 + 5 a_I^2 (b_{JI}^2 + l_{JI}^2) + 3 l_{JI}^2 (b_{JI}^2 + l_{JI}^2) \right) \right] - \frac{\gamma_{JI}}{3(1 - \gamma_{JI}^2)^2} \notag \\ 
    & \times \left[ 5 (1 + \chi_I^2) - 12 \zeta_{JI}^2 - \gamma_{JI}^2 ( 5 + 3 \chi_I^2)  \right] \ln \left[ \rho_{JI} (1 + \hat{x}_{3}) \right] \notag \\
    \left( \frac{\partial^3 \Xi_{I}^{\beta \dagger}}{\partial a_I^2 \partial l_{JI}} \right)_{\text{log}} & = -\frac{a_I b_{JI} l_{JI} (b_{JI}^2 + l_{JI}^2) \chi_I }{2 \rho_{JI}^5} \ln \left[ \rho_{JI} (1 + \hat{x}_{3}) \right] \notag \\
    \left( \frac{\partial^3 \Xi_{I}^{\beta \dagger}}{\partial a_I \partial b_{JI}^2} \right)_{\text{log}} & = \frac{\chi_I}{6} \left\{ \frac{4 \zeta_{JI} (4 + \gamma_{JI}^2)}{(1 - \gamma_{JI}^2)^2} \ln [l_{JI} + \rho_{JI}] + 
    \left[ \frac{b_{JI}}{\rho_{JI}^5} \left( (b_{JI}^2 + l_{JI}^2) (2 b_{JI}^2 + 3 l_{JI}^2)  \right. \right. \right. \notag \\ & \left. \left. \left. \qquad + a_I^2 (5 b_{JI}^2 + 3 l_{JI}^2) \right)  + \frac{2 \zeta_{JI} \gamma_{JI} [3 \gamma_{JI}^2 - 13] }{(1 - \gamma_{JI}^2)^2} \right] \ln [\rho_{JI} (1 + \hat{x}_{3})]
    \right\}  \notag \\ 
    \left( \frac{\partial^3 \Xi_{I}^{\beta \dagger}}{\partial a_I \partial b_{JI} \partial l_{JI}} \right)_{\text{log}} & = \frac{\chi_I}{6 \rho_{JI}^5} \left[ 3 a_I^2 b_{JI}^2 l_{JI} + a_{I}^2 l_{JI}^3 + b_{JI}^2 l_{JI}^3 + l_{JI}^5 + 2 \rho_{JI}^5 \right] \ln [\rho_{JI} (1 + \hat{x}_{3})] \notag \\
    \left( \frac{\partial^3 \Xi_{I}^{\beta \dagger}}{\partial a_I \partial l^2_{JI}} \right)_{\text{log}} & = \frac{b_{JI} \chi_I}{6 \rho_{JI}^5} \left[ b_{JI}^2 (b_{JI}^2 + l_{JI}^2) + a_I^2 (b_{JI}^2 + 3 l_{JI}^2) \right] \ln [\rho_{JI} (1 + \hat{x}_{3})] \\
    \left( \frac{\partial^3 \Xi_{I}^{\beta \dagger}}{\partial b^3_{JI}} \right)_{\text{log}} & = \frac{2 \zeta_{JI}^2 - \gamma_{JI}^2 + \gamma_{JI}^4}{(1 - \gamma_{JI}^2 )^2} \ln [l_{JI} + \rho_{JI}] + \frac{1}{2} \left( \frac{a_I (\xi_{JI}^0)_{3} (a_I^2 + l_{JI}^2)^2 }{\rho_{JI}^5} \right. \notag \\ & \left. \quad \frac{2 \gamma_{JI} \left[1 - 3 \zeta_{JI}^2 - \gamma_{JI}^2 (1 - \zeta_{JI}^2) \right]}{(1 - \gamma_{JI}^2 )^2} \right) \ln [\rho_{JI} (1 + \hat{x}_{3})] \notag \\ 
    \left( \frac{\partial^3 \Xi_{I}^{\beta \dagger}}{\partial b^2_{JI} \partial l_{JI}} \right)_{\text{log}} & = \left( \frac{2}{3} \zeta_{JI} - \frac{a_I b_{JI} l_{JI} (a_I^2 + l_{JI}^2) \chi_I}{2 \rho_{JI}^5} \right) \ln [\rho_{JI} (1 + \hat{x}_{3})] \notag \\
    \left( \frac{\partial^3 \Xi_{I}^{\beta \dagger}}{\partial b_{JI} \partial l^2_{JI}} \right)_{\text{log}} & = \left( \frac{\gamma_{JI}}{3} + \frac{a_I \left[ a_I^4 + 3 b_{JI}^2 l_{JI}^2 + a_I^2 (b_{JI}^2 + l_{JI}^2) \right] \chi_I}{\rho_{JI}^5} \right) \ln [\rho_{JI} (1 + \hat{x}_{3})] \notag \\
    \left( \frac{\partial^3 \Xi_{I}^{\beta \dagger}}{\partial l^3_{JI}} \right)_{\text{log}} & = -\frac{a_I b_{JI} (a_I^2 + b_{JI}^2) l_{JI} \chi_I }{2 \rho_{JI}^5} \ln [\rho_{JI} (1 + \hat{x}_{3})] \notag
\end{align}

The expressions in Eq. (D.14) are endpoint primitives. Hence, the contribution by the complete edge is evaluated by taking the difference between the values at $l_{JI}^+$ and $l_{JI}^-$. Note that the endpoint subtraction is necessary to discuss the logarithmic singularities on the polyhedral edges and vertices.

\end{document}